\documentclass[aps, preprint, amsmath, showkeys, a4paper, byrevtex]{revtex4-1}
\usepackage{subfig, multirow, rotating,xcolor,bm,mathtools}% Include figure files
\usepackage{amssymb}
\usepackage[]{graphicx}
\usepackage[section]{placeins}
\usepackage[printfigures]{figcaps}
\usepackage[percent]{overpic}
\usepackage[final]{changes}
\definechangesauthor[name={EA}, color=red]{ea}
\definechangesauthor[name={OS}, color=blue]{os}
\figcapsoff % enable to keep floats in their positions
\usepackage[font=small,justification=justified,singlelinecheck=false]{caption}
\usepackage{mathrsfs}
\usepackage{xr-hyper}
\usepackage{hyperref}
\usepackage{cleveref}
\usepackage{accents}
\renewcommand\eqref[1]{Eq.~(\ref{#1})}
\newcommand\figref[1]{Figure~\ref{#1}}

\begin{document}
\title[]{Open Quantum System Theory of Muon Spin Relaxation in Materials}% Force line breaks with \\
\author{Elvis F. Arguelles}
\affiliation{Institute for Solid State Physics, The University of Tokyo}
\author{Osamu Sugino}
\affiliation{Institute for Solid State Physics, The University of Tokyo}

\date{\today}
\begin{abstract}
We present a non-Markovian theory of muon spin relaxation that treats the
implanted muon as an open quantum spin coupled to a temporally correlated local
magnetic environment.
Using a Schwinger-Keldysh influence-functional
formulation, we derive a stochastic equation of motion for the muon spin, in
which the fluctuation kernel is fixed by the local-field correlation tensor,
while the retarded memory torque is introduced through an effective
phenomenological backaction kernel.
In the appropriate limits, the theory 
reduces to standard Kubo-Toyabe descriptions.
This enables quantitative, global analysis of zero-field (ZF) and weak
longitudinal-field (LF) $\mu$SR spectra beyond the strong-collision approximation.
Applied to $\mathrm{Li}_{0.73}\mathrm{CoO}_2$, 
the model supports a decomposition into a quenched width and a Li-driven dynamical component 
within the adopted parametrization, and yields fluctuation rates approximately consistent 
with activated behavior over the intermediate-temperature window. 
The fitted memory parameter is most visible in the crossover between quasi-static and fast-fluctuation limits.
%the approach separates quenched broadening from Li-driven fluctuations and 
%extracts a thermally activated fluctuation rate over the intermediate-temperature
%window. It also reveals a distinct non-Markovian line-shape signature 
%captured by a retarded backaction (memory) kernel that is most evident in the crossover 
%between quasi-static and fast-fluctuation limits.

\end{abstract}
%\pacs{75.47.Lx, 75.50.Pp}% insert suggested PACS numbers in braces on next line
\maketitle
\section{Introduction}
Muon spin relaxation ($\mu$SR) is a sensitive local probe of
magnetic dynamics and ordering in condensed matter.
In a typical $\mu$SR experiment, a spin-polarized positive
muon ($\mu^{+}$) is implanted in the sample. $\mu^{+}$ rapidly thermalizes
and localizes at a local electrostatic minimum, commonly adjacent
to an anion\cite{Sulaiman1994,Moller2013,Blundell2023} in the crystal.
Local magnetic fields predominantly generated by dipolar 
interactions with neighboring nuclei and
electrons induce the muon spin precession at a rate set by its gyromagnetic
ratio $\gamma_{\mu}=2\pi\times135.539~$MHz/T.
The vector-axial (V-A) weak decay of $\mu^{+}$ imprints a forward-backward
asymmetry in the emitted positron angular distribution relative to the
muon spin direction. Monitoring this temporal asymmetry 
directly maps the spin polarization of the muon ensemble, yielding insights to 
the amplitude and dynamics of local fields. This has enabled
studies of magnetism\cite{Uemura1985,Williams2016}, superconductivity\cite{Blundell2004}
and ionic transport in materials
such as layered cathodes $\mathrm{Li}_x\mathrm{CoO}_2$\cite{Sugiyama2009,Sugiyama2013,Ohishi2022} and
related compounds\cite{Sugiyama2011,Sugiyama2012,Ashton2014,Laveda2018,Okabe2024}.

From the standpoint of ion dynamics, $\mu$SR complements established probes
such as NMR\cite{heitjans_03,chandran_16} and quasielastic neutron
scattering (QENS)\cite{hempelmann_2000}.
NMR detects ionic motion via the nuclear spin-lattice relaxation
rate $1/T_1$, but paramagnetic transition-metal ions in common possitive-electrode materials
induce dominant magnetic relaxation pathways, making the interpretation highly complex\cite{mizushima_80,amatucci_2003,padhi_1997}.
QENS is in principle insensitive to nuclear magnetic moments, but 
typically requires elevated temperatures to yield practical diffusion signals, or conditions under which many
charged cathode materials become thermally unstable\cite{hempelmann_2000}.
A distinctive advantage of $\mu$SR is that a nearly 100\% spin-polarized muon beam is
available due to parity violation in the muon-production 
process \cite{kalvius,yaouanc}
enabling measurements in true zero field (ZF) and weak longitudinal fields (LF).

A faithful quantitative analysis of $\mu$SR spectra faces several significant challenges.
Firstly, the ionic motions that induce field fluctuations are often correlated giving 
rise to long-lived temporal structure in the local field.
Moreover, the implanted $\mu^{+}$ probes a complex magnetic environment containing both 
slowly varying nuclear magnetic fields and rapidly fluctuating sources, often
electronic in origin. In practice, the measured spectra relaxation is controlled mainly 
by the stochastic fluctuations of the slow sector, while the fast sector may leave only weak 
direct noise signature within the $\mu$SR window owing to strong motional narrowing.
Furthermore, there is growing evidence that implanted $\mu^{+}$ is not an inert spectator
in $\mu$SR. During the final thermalization stage it can undergo charge-exchange processes 
and form muonium\cite{Vilao2017,Vilao2025}. In addition, recent first-principles and
experimental studies have shown that implanted muons can induce substantial lattice
relaxation\cite{Blundell2023}, self-trapping\cite{ohta2025}, and 
even charge-neutral muon-polaron complexes that measurably affect
the $\mu$SR frequencies and relaxation rates\cite{Dehn2020,Dehn2021}.
These considerations suggest that the local environment relevant to the measured $\mu$SR
signal need not be fully characterized by the visible stochastic fluctuations alone. In
particular, rapidly fluctuating electronic or probe-induced degrees of freedom may contribute
only weakly to the resolved noise component while still exerting an appreciable causal
influence on the muon spin dynamics through the effective response of the environment.

The standard framework for analyzing ZF and LF $\mu$SR spectra is the
Kubo-Toyabe (KT) theory\cite{Kubo-Toyabe} and its dynamic extensions\cite{Hayano1979}.
These approaches assume a Gaussian distribution of local fields and commonly
incorporate field fluctuations through the strong-collision (SC) approximation,
leading to closed-form polarization functions that describe many $\mu$SR spectra
successfully. In its usual form, however, the dynamic KT framework represents the
field dynamics through a Markovian renewal process and does not explicitly encode
the above-mentioned effects. To go beyond this, open quantum system approaches, 
such as the hierarchical equations of motion (HEOM), have been adapted to $\mu$SR
to incorporate non-Markovian fluctuations and dissipation at finite temperatures nonperturbatively\cite{Takahashi2020}.
In their usual form, however, HEOM starts from a specified thermal bath correlation function,
equivalently a spectral density and temperature, so that the fluctuation and dissipative sectors 
are linked at the bath-model level by equilibrium statistics.
Although such a framework may be enlarged to include additional fast channels,
this raises a basic question of observability: a channel that does not appear as 
a clearly resolved stochastic component in the spectra cannot be inferred reliably 
from the visible noise alone.
This motivates an effective formulation in which the slow stochastic sector is fixed by the local-field correlator, 
while unresolved fast degrees of freedom are encoded separately through an independent retarded backaction kernel.

Here we propose a unified and efficient framework that facilitates this paradigm shift 
by integrating stochastic and open-system viewpoints.
Starting from a Schwinger-Keldysh spin coherent state path integral, we
represent the muon as an SU(2) coherent state coupled to an effective magnetic
environment composed of local background and ion-modulated fields.
Tracing over the bath degrees of freedom yields an influence
functional specified by effective nonlocal retarded kernel
which generates a causal backaction (memory) torque, and a
Keldysh kernel, which fixes the correlations of the effective
colored magnetic noise. The formulation fixes the stochastic structure 
of the spin dynamics and relates the colored-noise kernel to the symmetrized local-field correlator. 
By contrast, we treat the retarded kernel phenomenologically rather
than imposing an equilibrium fluctuation-dissipation theorem (FDT) constraint
\emph{a priori}. 
By decoupling the retarded response from the visible noise, this construction allows unresolved 
fast environmental processes to be encoded through an independent backaction parameter 
$\Lambda$, while incorporating quenched and dynamical components of the local field on equal footing.

We apply this formalism to $\mathrm{Li}_{0.73}\mathrm{CoO}_2$, where Li motion
induces slow, temporally correlated local-field fluctuations.
We set $\nu_\mu=0$, so the background contribution
enters as a quenched Gaussian width $\Delta_\mu$, while Li-driven
fluctuations are characterized by a dynamical width $\Delta_{\mathrm{Li}}$,
a fluctuation rate $\nu_{\mathrm{Li}}$, and a backaction scale $\Lambda$ that
controls the retarded torque.
Global fits of ZF and weak-LF (5~G and 10~G) spectra reproduce the field-dependent
line shapes with a single parameter set at each temperature. 
Within the adopted parametrization, including the static-muon assumption $\nu_\mu=0$,
the fits support a decomposition into a quenched width
$\Delta_\mu$, a Li-driven dynamical component
$(\Delta_{\mathrm{Li}},
\nu_{\mathrm{Li}})$, and an effective temperature-dependent memory parameter
$\Lambda(T)$ that is most evident in the crossover regime between quasi-static
and motional-narrowing limits.
\begin{figure}
  \centering
	  \includegraphics[scale=0.42]{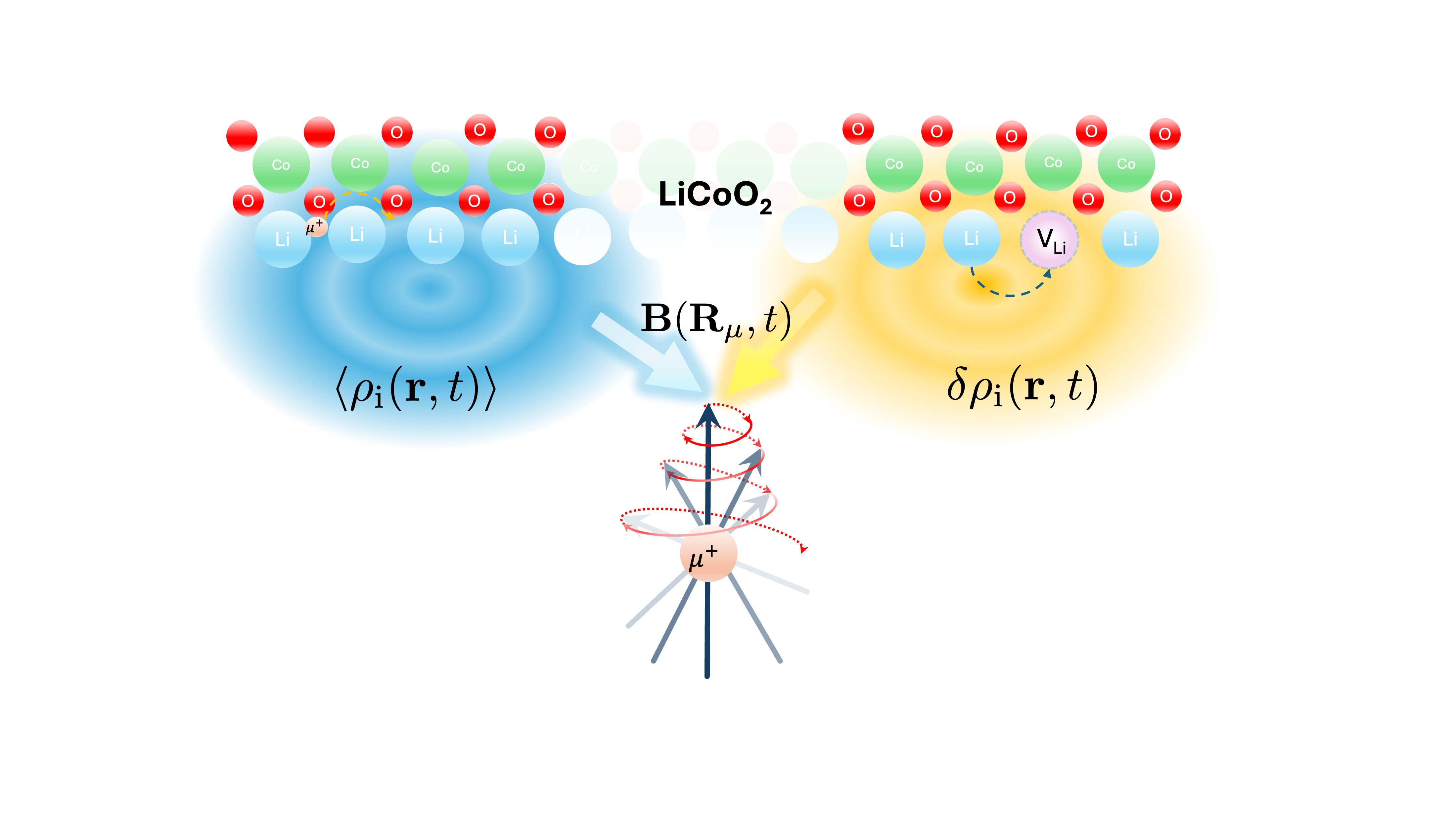}\label{fig:system}
  \captionsetup{justification=raggedright,singlelinecheck=false}
  \caption{Schematic representation of the muon spin relaxation as an open quantum
	  system. By linearizing the magnetic density at the muon site,
	  the effective magnetic field $\bm{B}(\bm{R}_{\mu},t)$ that couples 
	  to muon spin $\bm{S}$ is composed of a background field averaged over ion density ($\langle \rho_{\mathrm{i}} \rangle$) and
  the ion density fluctuation ($\delta \rho_{\mathrm{i}}$)-modulated field.
  }
  \label{fig:system}
\end{figure}

\section{Local Field Correlation Tensor Model \label{sec:model}}
The subsequent open-system formulation involves a fluctuation kernel and a
retarded memory kernel. The fluctuation kernel is determined by the
symmetrized autocorrelation tensor of the local magnetic field operator $\hat{\bm{B}}$
at the muon site,
\begin{equation}
	C_{\alpha\beta}(t)\equiv \frac{1}{2}\big\langle \left\{\hat B_\alpha(t),\hat B_\beta(0)\right\}\big\rangle,
  \label{eq:sym_corr_tens}
\end{equation}
with $\alpha,\beta\in\{x,y,z\}$.
In this section, we derive an effective form
of $C_{\alpha\beta}(t)$ appropriate for the stochastic sector of the theory.
Formally, if $\hat{\bm B}$ is identified with the full microscopic local-field
operator, and the expectation value is taken in the exact stationary state of
the environment, then \eqref{eq:sym_corr_tens} defines the exact symmetrized
local-field correlator. This object governs the fluctuations experienced by
the muon and, in principle, includes the effects of correlated muon and ion
dynamics.
In the present work, however, our goal is not to evaluate
\eqref{eq:sym_corr_tens} microscopically from first principles.
Rather, we use it as the formal starting point from which we derive the
effective correlation model employed in the fits.
To this end, we introduce a sequence of physically motivated approximations
that reduce the exact correlator to a tractable 
phenomenological two-component correlation function proposed by
Ito and Kadono~\cite{Ito2024}.
Let us assume the thermalized muon is localized at some potential minimum in the crystal and its 
instantaneous position is denoted by $\bm{R}_{\mu}(t)$.
Possible hopping of the muon between such sites will be incorporated
later through the muon intermediate scattering function with the static-muon limit
is recovered by setting $\nu_\mu=0$.
The muon spin operator $\hat{\bm S}$ couples to the effective local field 
through the Zeeman interaction
\begin{equation}
	\hat H_{\mathrm{int}} = -\gamma_{\mu}\hat{\bm{S}}\cdot\hat{\bm{B}}.
	\label{eq:H_SB}
\end{equation}
In specific materials such as $\mathrm{Li}_{x}\mathrm{CoO_{2}}$, $\hat{\bm{B}}$
is primarly generated by $^{7}\mathrm{Li}$ and $^{59}\mathrm{Co}$, nuclear
moments and possibly electronic $\mathrm{Co}$ moments from paramagnetic $\mathrm{Co}^{4+}$
defects. We focus on dipolar interaction for simplicity; 
additional hyperfine terms can be included similarly.
We denote $i\in\{1,2,3,\ldots\}$ as the index of magnetic moments residing on
transition-metal (TM) sites, ion nuclei, and other atoms, and let
$\hat{\bm\mu}_i$ be the corresponding Schr\"odinger-picture magnetic
dipole moment operator at lattice site $\bm R_i$. Its time dependence is
generated by the autonomous bath Hamiltonian $\hat H_B$, so that the associated
free Heisenberg operator is $
  \hat{\boldsymbol\mu}_i(t)
  =
  e^{\,i\hat H_B (t-t_0)}
  \hat{\boldsymbol\mu}_i
  e^{-\,i\hat H_B (t-t_0)}$.
This generates a field operator
\[
	\hat{\bm{B}}_{i}(t)=\frac{\mu_{0}}{4\pi r_{i}^{3}}\
	\left[3\left(\hat{\bm\mu}_{i}(t)\cdot\hat{\bm{n}}_{i}\right)\hat{\bm{n}}_{i}- \boldsymbol\mu_{i}(t)\right]
	\label{eq:Bj}
\]
at $\bm{R}_{\mu}$, where $\bm{r}_{i}=\bm{R}_{\mu}-\bm{R}_{i}$, 
$r_{i}=|\bm{r}_{i}|$, and $\hat{\bm{n}}_{i}=\frac{\bm{r}_{i}}{r_{i}}$.
The total local field operator at the muon site is therefore, $\hat{\bm{B}}(t)=\sum_{i}\hat{\bm{B}}_{i}(t)$.
Defining a continuous magnetization density operator
from the discrete dipoles $\hat{\bm{M}}(\bm{r},t)=\sum_{i}\hat{\bm\mu}_{i}(t)\delta(\bm{r}-\bm{R}_{i})$,
the local magnetic field operator at $\bm{R}_{\mu}$ reads 
\begin{equation}
	\hat B_{\alpha}(t)=\int d^{3} r \mathcal{K}_{\alpha\beta}
	\left(\bm{R}_{\mu}(t) - \bm{r}\right)\
	\hat M_{\beta}(\bm{r},t),
	\label{eq:B_tot}
\end{equation}
where $\mathcal{K}_{\alpha\beta}\left(\bm{r}\right)=\
\frac{\mu_{0}}{4\pi r^{3}}\
\left(3\hat{n}_{\alpha}\hat{n}_{\beta}- \delta_{\alpha\beta}\right)$
is the dipolar tensor. 
In principle, $\hat{\bm M}(\bm r,t)$ contains both slow nuclear contributions
and faster electronic components of the local magnetic environment. In the
present section we focus on the slow part that remains visible in the
symmetrized local-field correlator within the $\mu$SR time window. By
contrast, sufficiently rapid electronic fluctuations
are motionally narrowed in the stochastic sector, so that their contribution to
$C_{\alpha\beta}(t)$ is reduced to a weak relaxation rate and may
be neglected at this level.
From here onwards, we use Einstein summation convention only over repeated Cartesian indices.
Microscopically, the random occupation of ionic site affects the magnetic moment 
at the TM site by changing the valence charge and
the local crystal field. Furthermore, the ionic (nuclear) spin moments only exist when the 
$l$ site is actually occupied.
Therefore, $\hat{\bm\mu}_{i}$ depends functionally on the entire ion density field operator driven by local stoichiometry
and nearby ion occupancy. It follows that $\hat{\bm\mu}_{i}=\hat{\bm\mu}_{i}[\hat\rho_{\mathrm{i}}]$, where
$\hat\rho_{\mathrm i}=\hat\rho_{\mathrm{i}}(\bm{r},t)$ is the coarse-grained ion density operator.
We can decompose the ion density as $\hat\rho_{\mathrm{i}}(\bm{r},t)=\bar{\rho}_{\mathrm{i}}(\bm{r})\
+\delta\hat\rho_{\mathrm{i}}(\bm{r},t)$, where
$\bar{\rho}_{\mathrm{i}}(\bm{r})$ is the c-number stationary equilibrium 
ensemble-averaged density,
and $\delta\hat\rho_{\mathrm{i}}(\bm{r},t)$ is the zero-mean fluctuations operator
arising from time-dependent changes in site
occupations (ion hopping).
Assuming the magnetization responds linearly to these fluctuations, the magnetization density can be linearized as
\begin{equation}
	\hat M_{\beta}[\bar{\rho}_{\mathrm{i}}+\delta\hat\rho_{\mathrm{i}}](t)\approx\
	\hat M_{\beta}[\bar{\rho}_{\mathrm{i}}](t) + \int d^{3}r'\int dt'\
	\chi_{\beta}(\bm{r},\bm{r}',t,t')\delta\hat\rho_{\mathrm{i}}(\bm{r}',t'),
	\label{eq:M_FT}
\end{equation}
where $\chi_{\beta}(\bm{r},\bm{r}',t,t')=\
\frac{\delta \hat M_{\beta}(\bm{r},t)}{\delta\hat\rho_{\mathrm{i}}(\bm{r}',t')}\Big|_{\bar{\rho}_{\mathrm{i}}}$
is magneto-ionic response (susceptibility) kernel. 
We assume that the magnetic bath 
degrees of freedom relax on a time scale
$\tau_{\mathrm{bath}}$ that is short compared to the characteristic ion
hopping time $\tau_{\mathrm{ion}}\sim \nu_{\mathrm{i}}^{-1}$.
Then the magneto-ionic susceptibility can be taken as local in time,
$\chi_{\beta}(\bm{r},\bm{r}',t,t')=\
\chi_{\beta}(\bm{r},\bm{r}',t)\delta(t-t')$.
Inserting \eqref{eq:M_FT} allows one to write
the local field at the muon site as the sum 
\[
	\hat B_{\alpha}(t)=\hat B_{\alpha}^{(\mu)}(t)+ \hat B_{\alpha}^{(\mathrm i)}(t),
\]
where 
\[
	\hat B_{\alpha}^{(\mu)}(t)
	=\int d^{3} r \mathcal{K}_{\alpha\beta}\left(\bm{R}_{\mu} - \bm{r}\right)\
	\hat M_{\beta}^{(\mu)}(\bm{r},t) 
\]
with $\hat M_{\beta}^{(\mu)}(\bm{r},t)\equiv \hat M_{\beta}[\bar{\rho}_{\mathrm{i}}](t)$
is the background internal field that would remain even if the ion sublattice were frozen at its
coarse-grain density, and
\[
\hat B_{\alpha}^{(\mathrm i)}(t)=\int d^{3} r\int d^{3} r'\
\mathcal{K}_{\alpha\beta}\left(\bm{R}_{\mu} - \bm{r}\right)\
\chi_{\beta}(\bm{r},\bm{r}',t)\delta\hat\rho_{\mathrm{i}}(\bm{r}',t)
\]
is the ion density fluctuation-modulated contribution.
To simplify the formulation, we neglect the mixed symmetrized correlations
between the background and ion-modulated field operators, so that the local
field symmetrized autocorrelation tensor may be approximated as
\begin{equation}
  C_{\alpha\beta}(t-t')
  \simeq
  C_{\alpha\beta}^{(\mu)}(t-t') + C_{\alpha\beta}^{(\mathrm i)}(t-t'),
  \label{eq:autocorr_C}
\end{equation}
where
\[
  C_{\alpha\beta}^{(\mu)}(t-t')
  \equiv
  \frac{1}{2}
  \Big\langle
    \big\{
      \hat B_{\alpha}^{(\mu)}(t),
      \hat B_{\beta}^{(\mu)}(t')
    \big\}
  \Big\rangle,
  \qquad
C_{\alpha\beta}^{(\mathrm i)}(t-t')
  \equiv
  \frac{1}{2}
  \Big\langle
    \big\{
      \hat B_{\alpha}^{(\mathrm i)}(t),
      \hat B_{\beta}^{(\mathrm i)}(t')
    \big\}
  \Big\rangle.
\]
Here and below,
$\langle \cdots \rangle \equiv \mathrm{Tr}_{B}[\rho_{B}\,(\cdots)]$
denotes the expectation value over the bath state.
We treat these two channels in turn.

\paragraph*{Background field}
Let us first consider the magnetic background field at the muon site.
Inserting the expression for $\hat B_{\alpha}^{(\mu)}(t)$ into
\eqref{eq:autocorr_C}, the background-field symmetrized
autocorrelation tensor in Fourier space becomes
\begin{equation}
  C_{\alpha\beta}^{(\mu)}(t)
  =
  \sum_{\bm q}
  K_{\alpha\gamma}(\bm q)\,
  K_{\beta\delta}(-\bm q)\,
  F_{\mu}(\bm q,t)\,
  S^{(\mu)}_{\gamma\delta}(\bm q,t)
   .
  \label{eq:B0_corr1}
\end{equation}
where, $F_{\mu}(\bm{q},t)=\langle e^{i\bm{q}\cdot(\bm{R}_{\mu}(t)-\bm{R}_{\mu}(0))}\rangle$
is the muon self intermediate scattering function (ISF) and
$S^{(\mu)}_{\gamma\delta}(\bm q,t)=\frac{1}{2}
  \Big\langle
    \big\{
	    \hat M^{(\mu)}_{\gamma}(\bm q,t),
	    \hat M^{(\mu)}_{\delta}(-\bm q,0)
    \big\}
    \Big\rangle$.
To this end, we factorized the joint average 
\[
	\frac{1}{2}\Big\langle e^{i\bm{q}\cdot\bm{R}_{\mu}(t)}e^{-i\bm{q}'\cdot\bm{R}_{\mu}(0)}\
\left\{ \hat M_{\gamma}^{(\mu)}(\bm{q},t),\hat M_{\delta}^{(\mu)}(\bm{q}',0)\right\} \Big\rangle\approx\
\Big\langle e^{i\bm{q}\cdot\bm{R}_{\mu}(t)}e^{i\bm{q}'\cdot\bm{R}_{\mu}(0)}\Big\rangle\
S^{(\mu)}_{\gamma\delta}(\bm q,t)
\]
by invoking that the stochastic muon motion and the host background magnetization fluctuations
are statistically independent processes. Further, we assume
that the system is spatially homogeneous where all crystal lattice sites are equivalent
so that the equilibrium thermal average is translationally invariant.
This implies that the displacement $\Delta\bm{R}_{\mu}=\bm{R}_{\mu}(t)-\bm{R}_{\mu}(0)$
is independent of the absolute initial position $\bm{R}_{\mu}(0)$,
and $\langle e^{i\bm{q}\cdot\bm{R}_{\mu}(t)}e^{i\bm{q}'\cdot\bm{R}_{\mu}(0)}\rangle\
\approx F_{\mu}(\bm{q},t)\langle e^{i(\bm{q}+\bm{q}')\cdot\bm{R}_{\mu}(0)}\rangle$. 
In this case, the initial muon site is equally likely to be any of the lattice sites, with
$\langle e^{i(\bm{q}+\bm{q}')\cdot\bm{R}_{\mu}(0)}\rangle
\propto\delta_{\bm{q}+\bm{q}',0}$, giving \eqref{eq:B0_corr1}.
The strong attraction with the anions in the lattice restricts muon jumps 
to neareast neighbor sites and in the spirit of homogeneous lattice approximation,
the jump rates $\nu_{\mu}$ can be taken as symmetric. Therefore, the probability that the muon made 
$n$ jumps has a Poisson distribution $\frac{(\nu_{\mu}t)^{n}}{n!}e^{-\nu_{\mu}t}$. 
In Appendix \ref{app:ISF_QME_CME}, we derived the expressions
for the muon and ion ISF in this limit. 
For the muon, the result is $F_{\mu}(\bm{q},t)=\exp\left[-\nu_{\mu}t(1-\alpha(\bm{q}))\right]$,
where $\alpha(\bm{q})=\langle e^{i\bm{q}\Delta\bm{R}_{\mu}}\rangle$ is
the single jump characteristic function. We specialize in a case in which after a single jump
the phase is randomized and cancels out after averaging over all possible jump directions. 
This random phase approximation entails
$\alpha(\bm{q})=0$ and $F_{\mu}(t) = e^{-\nu_{\mu}t}$. 

In the typical $\mu$SR time window,
the predominantly nuclear dipolar fields evolve slowly and are essentially
considered quasi-static\cite{Willwater2022}. We therefore approximate 
$S^{(\mu)}_{\gamma\delta}(\bm q,t)\approx S^{(\mu)}_{\gamma\delta}(\bm q,0)$
so that
\eqref{eq:B0_corr1} becomes
\begin{equation}
	C_{\alpha\beta}^{(\mu)}(t) = \Delta_{(\mu)\alpha\beta}^{2}e^{-\nu_{\mu}t},
	\label{eq:B0_corr2}
\end{equation}
where we introduce the second moment tensor of the background local field distribution as
\begin{equation}
	\Delta_{(\mu)\alpha\beta}^{2}=C_{\alpha\beta}^{(\mu)}(0) =\
	\sum_{\bm{q}}\mathcal{K}_{\alpha\gamma}(\bm{q})\
	\mathcal{K}_{\beta\delta}(-\bm{q})S^{(\mu)}_{\gamma\delta}(\bm q,0)
	\label{eq:Delta_mu}
\end{equation}
For an isotropic background, one has $\Delta_{(\mu)\alpha\beta}^{2}=\Delta_{(\mu)}^{2}\delta_{\alpha\beta}$,
so that the scalar parameter $\Delta_{(\mu)}^{2}$ can be interpreted as the root-mean-square width
of the quasi-static local background field distribution at the muon site. 

\paragraph*{Ion-Modulated Field }

Employing the same decoupling approximations as for the muon background
field case the ion-modulated field in Fourier space reads
\begin{equation}
	C_{\alpha\beta}^{(\mathrm{i})}(t)=\
	\sum_{\bm{q}}\mathcal{K}_{\alpha\gamma}(\bm{q})\
	\mathcal{K}_{\beta\delta}(-\bm{q}) F_{\mu}(\bm{q},t)\
	\langle\chi_{\gamma}(\bm{q},t)\chi_{\delta}(-\bm{q},0) \rangle\
	S_{\rho\rho}^{(i)}(\bm q,t),
	\label{eq:dB_corr}
\end{equation}
where $S_{\rho\rho}^{(i)}(\bm q,t)=\frac{1}{2}
\langle\left\{\delta\hat\rho_{\mathrm{i}}(\bm{q},t),
\delta\hat\rho_{\mathrm{i}}(-\bm{q},0)\right\}\rangle$. 
Here,
$\delta\hat\rho_{\mathrm{i}}(\bm{q},t)=\hat\rho_{\mathrm{i}}(\bm{q},t)-\bar{\rho}_{\mathrm{i}}(\bm{q})$,
is the ion density fluctuation, with
$\hat\rho_{\mathrm{i}}(\bm{q},t)=\sum_{n}e^{-i\bm{q}\cdot\bm{R}_{n}}\hat n_{n}(t)$,
and 
$\bar{\rho}_{\mathrm{i}}(\bm{q})=c\sum_{n}e^{-i\bm{q}\cdot\bm{R}_{n}}$
where $c$ is the ion concentration.
At this point we adopt an incoherent-hopping, classical-stochastic approximation for the ion occupations
in order to obtain an explicit closed form for the symmetrized density correlator
$S_{\rho\rho}^{(i)}$. We do not use this reduction to determine the retarded response kernel, 
which remains an independent effective input unless additional equilibrium assumptions are imposed.
Within this approximation, the occupation operators may be replaced
by c-number variables $\hat n_{n}(t)\to n_{n}(t)=\sum_{l}^{N_{l}}\delta_{\bm{R}_{n},\bm{R}_{l}(t)}$,
where $N_{l}$ is the number of lattice sites. The total number of ion is
$N_i=\sum_n\langle n_n\rangle=cN_l$ with $c=\langle n_{n}\rangle$.
This implies that $S_{\rho\rho}^{(i)}(\bm q,t) \to
\langle\delta\rho_{\mathrm{i}}(\bm{q},t)
\delta\rho_{\mathrm{i}}(-\bm{q},0)\rangle$. 

We focus on density fluctuations at nonzero
wavevector $\bm{q}\neq 0$ since the $\bm{q}=0$ mode corresponds to conserved total ion number for which
$\delta\rho_{\mathrm{i}}(\bm{0},t)=0$ identically and therefore does not contribute to the fluctuation
correlator. For a periodic lattice,  
$\sum_{n} e^{-i\bm q\cdot\bm R_n}=0$ for $\bm q\neq 0$, so
$\bar{\rho}_{\mathrm{i}}(\bm{q})=0$ and thus 
$\delta\rho_{\mathrm{i}}(\bm{q},t)=\rho_{\mathrm{i}}(\bm{q},t)=\sum_{l}e^{-i\bm{q}\cdot\bm{R}_{l}(t)}$.
Using this the density correlator splits as $S_{\rho\rho}^{(i)}(\bm q,t)=S_{\mathrm i}^{\mathrm{self}}(\bm q,t)
+ S_{\mathrm i}^{\mathrm{dist}}(\bm q,t)$, where
the self (incoherent) part is given by 
$S_{\mathrm i}^{\mathrm{self}}(\bm q,t)
\equiv
\Big\langle \sum_{l}
e^{\,i\bm q\cdot\left[\bm R_l(t)-\bm R_l(0)\right]}
\Big\rangle$, and distinct (coherent) component reads 
$S_{\mathrm i}^{\mathrm{dist}}(\bm q,t)
\equiv
\Big\langle \sum_{l\neq l'}
e^{\,i\bm q\cdot\left[\bm R_l(t)-\bm R_{l'}(0)\right]}
\Big\rangle$.
In the following, we assume
weak ion-ion interactions and neglect the distinct term. 
For independent, identical hopping dynamics, the self correlator becomes
$S_{\mathrm i}^{\mathrm{self}}(\bm q,t)
=N_{\mathrm i}F_{\mathrm i}(\bm q,t)$, where
$F_{\mathrm{i}}(\bm q,t)=\exp\left[-\nu_{\mathrm i}t(1-\lambda(\bm q))\right]$
is the ion ISF with
$\lambda(\bm{q})=\langle e^{i\bm{q}\Delta\bm R_{\mathrm i}} \rangle$
being the ion single jump characteristic function, and $\Delta\bm R_{\mathrm i}$
is the displacement. As in the muon background field case, we adopt the random-phase 
approximation $\lambda(\bm{q})=0$ so that
$F_{\mathrm{i}}(\bm q,t)\to F_{\mathrm{i}}(t)=e^{-\nu_{\mathrm i}t}$.
We further assume that the magnetic environment equilibrates rapidly compared
to ion motion so that the magneto-ionic susceptibility kernel may be taken
effectively static on the ion time scale. With these approximations \eqref{eq:dB_corr}
reduces to a single exponential form
\begin{equation}
	C_{\alpha\beta}^{(\mathrm{i})}(t)=\
	\Delta^{2}_{(\mathrm i)\alpha\beta}e^{-\left(\nu_{\mu}+\nu_{\mathrm i} \right)t},
	\label{eq:dB_corr2}
\end{equation}
where
\begin{equation}
	\Delta^{2}_{(\mathrm i)\alpha\beta}=\
	N_{\mathrm i}\sum_{\bm{q}}\mathcal{K}_{\alpha\gamma}(\bm{q})\
	\mathcal{K}_{\beta\delta}(-\bm{q}) \
	\langle\chi_{\gamma}(\bm{q})\chi_{\delta}(-\bm{q}) \rangle
	\label{eq:Delta_i}
\end{equation}
is the second-moment tensor of ion-modulated field distribution.

The above derivations determine the correlator for $t>0$. Since
\[
C_{\alpha\beta}(t)=C_{\beta\alpha}(-t)
\]
for a stationary bath, the negative-time branch is fixed by symmetry.
This implies that the correlator is
even in time, so we extend the exponential form as
$e^{-\nu t}\to e^{-\nu |t|}$.
Under the above assumptions and using \eqref{eq:B0_corr2} and \eqref{eq:dB_corr2} the total autocorrelation
tensor is 
\begin{equation}
	C_{\alpha\beta}(t)=\
	\Delta_{(\mu)\alpha\beta}^{2}e^{-\nu_{\mu}|t|} + 
	\Delta^{2}_{(\mathrm i)\alpha\beta}e^{-\left(\nu_{\mu}+\nu_{\mathrm i} \right)|t|}.
	\label{eq:tot_corr}
\end{equation}
In the static-muon limit $\nu_\mu\to 0$ with dynamically fluctuating ions ($\nu_{\mathrm i}>0$),
\eqref{eq:tot_corr} reduces to
$C_{\alpha\beta}(t)=\Delta^{2}_{(\mu)\alpha\beta}+\Delta^{2}_{(\mathrm i)\alpha\beta}e^{-\nu_{\mathrm i}|t|}$.
It exhibits nonzero long-time behavior, formally analogous to Edwards-Anderson form of glassy
correlation functions\cite{Edwards1976}.
Ito and Kadono\cite{Ito2024} previously proposed the isotropic version of this 
correlation tensor phenomenologically. 
In the next subsection we construct an effective
spin-boson representation whose spectral densities reproduce \eqref{eq:tot_corr}.

\section{Schwinger-Keldysh Path Integral}

In $\mu$SR the measured quantity is the normalized longitudinal polarization,
\begin{equation}
  G_z(t)\equiv \big\langle n_z(t)\big\rangle,
  \label{eq:Gz_def}
\end{equation}
where $\bm n(t)$ is the spin-direction field in the coherent-state
representation satisfying
$\bm n(t)\cdot \bm n(t)=1$.
To derive $G_z(t)$ we employ the Schwinger-Keldysh coherent-state path integral and
integrate out the bath degrees of freedom to obtain an influence functional for the spin.
The corresponding real-time partition function is
\begin{equation}
  \mathcal{Z}
  =
  \mathrm{Tr}\!\left[
    \mathcal{T}_{\mathcal{C}}
    e^{-i\int_{\mathcal{C}}dt\,\hat{H}}\,
    \rho_{0}
  \right],
  \label{eq:z}
\end{equation}
where $\mathcal{T}_{\mathcal{C}}$ denotes the time ordering over the contour $\mathcal C$,
and the trace is over the system and bath degrees of freedom.
Here, $\hat H = \hat H_{\mathrm S} + \hat H_{\mathrm{B}} + \hat H_{\mathrm{int}}$ where
\begin{equation}
	\hat H_{\mathrm S} = -\gamma_\mu \hat{\bm S}\cdot \bm B_{\mathrm L}
  \label{eq:H_S}
\end{equation}
is the system Hamiltonian for muon spin the presence of a c-number external longitudinal field
$\bm B_{\mathrm L}$, $\hat H_{\mathrm{B}}$ is the local bath field Hamiltonian 
and the interaction Hamiltonian $\hat H_{\mathrm{int}}$ is given by \eqref{eq:H_SB}.
In the present formulation, $\hat H_{\mathrm{B}}$ serves only as the abstract generator of
the autonomous environment dynamics that defines the bath field correlators. The muon-spin
theory depends only on those correlators, not on unique closed-form expression for $\hat H_{\mathrm{B}}$.
From the linearization of magnetization density in Section \ref{sec:model}, the bath field operator
in $\hat H_{\mathrm{int}}$ is a sum $\hat{\bm B}=\hat{\bm B}^{(\mu)}+\hat{\bm B}^{(\mathrm i)}$
of contributions from the background  
and the ion density fluctuations modulated fields.
For notational clarity we carry out the Schwinger-Keldysh formulation in terms of the total local
field operator $\hat{\bm B}$.
We assume a factorized initial state $\rho_{0}=\rho_{\mathrm S}(0)\otimes\rho_{\mathrm B}$
so that there are no initial correlations or entanglement between the muon spin
and the environment.
We do not assume the bath state to be thermal. Instead, we assume that
the environment is stationary with respect to its autonomous bath dynamics,
$[\rho_{\mathrm B},\,H_{\mathrm B}]=0,$
so that the corresponding bath correlation functions are invariant under time
translations and depend only on the time difference $t-t'$.
We further assume that the environmental field statistics are Gaussian
(or, equivalently, that the influence functional is truncated at second
cumulant order), so that the reduced spin dynamics is fully determined by the
one- and two-point functions of the bath fields.

Using the Euler-angle parametrization of an SU(2) rotation, we define the
spin coherent state as
\begin{equation*}
  |g(\theta,\phi)\rangle
  \equiv
  e^{-i\phi \hat S_{z}}
  e^{-i\theta \hat S_{y}}
  |\uparrow\rangle,
\end{equation*}
where $\hat S_{x},\hat S_{y},\hat S_{z}$ are the Cartesian components of the
spin operator $\hat{\bm S}$, and $|\uparrow\rangle$ denotes the highest-weight
state with quantization axis along $z$. The third Euler angle $\psi$ contributes
only an overall phase and is therefore omitted.
We fix this phase convention and parametrize the spin state by
the direction $(\theta,\phi)$, i.e. by a unit vector $\bm n(t)$.
Following the standard spin coherent-state path-integral construction
over the real-time closed contour
\cite{AltlandSimons2010}, 
we arrive at the continuum representation
\begin{equation}
  \mathcal{Z}
  =
  \int\mathcal{D}[\bm{n}^{+},\bm{n}^{-}]\,
  e^{i\left(
		  S_{\mathrm S}[\bm n^{+}] - S_{\mathrm S}[\bm n^{-}]
  \right)}\mathscr F[\bm{n}^{+},\bm{n}^{-}]
  \label{eq:Z2}
\end{equation}
where $\bm n^{+}$ and $\bm n^{-}$ denote the c-number spin fields on the forward
($C_{+}$) and backward ($C_{-}$) branches of the Schwinger-Keldysh contour,
respectively. The spin action reads
\begin{equation*}
	S_{\mathrm S}[\bm n^{\pm}] =  S_{\mathrm{Berry}}[\bm n^{\pm}]-\int dt\,\gamma_{\mu}S\,\bm n^{\pm}\cdot\bm B_{\mathrm L}
\end{equation*}
with
\[
  S_{\mathrm{Berry}}
  =
  S\int dt\,\dot{\bm n}\cdot\bm A 
\]
being the spin Berry-phase (solid-angle) contribution of the coherent state path integral\cite{Radcliffe1971,AltlandSimons2010}.
$\bm A(\bm n)$ is the gauge-dependent Berry-connection (geometric vector potential) of the
spin coherent states on the unit sphere that satisfies
$\boldsymbol\nabla_{\bm n}\times\bm A(\bm n)=\bm n$.
In the North-pole (Dirac) gauge one may take
$\bm A=(1-\cos\theta)\,\hat{\bm e}_\phi/\sin\theta$.

The influence functional appearing in \eqref{eq:Z2} reads
\begin{equation}
	\mathscr F[\bm{n}^{+},\bm{n}^{-}] =\mathrm{Tr}_{B}\left\{
		\hat\rho_{B}\mathcal T_{\mathcal C} \exp\left[-i\int_{\mathcal C}dt\left(\hat H_{B}
-\gamma_\mu S \bm n^{\mathcal C}(t)\cdot \hat{\bm{B}}\right) \right] \right\},
  \label{eq:IF}
\end{equation}
where $\bm n^{\mathcal C} = \bm n^{+}$ on $C^{+}$ and
$\bm n^{\mathcal C} = \bm n^{-}$ on $C^{-}$. On a closed real time contour,
the free bath evolution satisfies $\mathcal T_{\mathcal C}e^{-i\int_{\mathcal C}dt \hat H_{B}}=1$. 
However the noncommutativity of the terms inside the contour exponential in \eqref{eq:IF} does not
permit for naive factorization. We therefore use the exact contour Dyson identity 
\[
	\mathcal T_{\mathcal C}\exp\left[-i\int_{\mathcal C} dt (\hat X(t) + \hat Y(t)) \right]
= 
U_{X,\mathcal C}
\mathcal T_{\mathcal C}\exp\left(-i\int_{\mathcal C} ds \hat{Y}_{I}(t) \right),
\]
where $U_{X,\mathcal C}=\mathcal T_{\mathcal C}\exp\left(-i\int_{\mathcal C} dt \hat X(t) \right)$
and $\hat Y_{I}(t) = U_{X}^{-1}(t,t_0)\hat Y(t) U_{X}(t,t_0)$. Applying this to our case yields
\begin{equation}
	\mathscr F[\bm{n}^{+},\bm{n}^{-}] =\mathrm{Tr}_{B}\left\{\rho_{B}\mathcal T_{\mathcal C} \exp\left[
i\gamma_\mu S \int_{\mathcal C} dt \bm n^{\mathcal C}(t)\cdot \hat{\bm{B}}(t) \right] \right\},
  \label{eq:IF1}
\end{equation}
where $\hat{\bm{B}}(t)=U_{B}^{-1}(t,t_0)\hat{\bm B}U_{B}(t,t_0)$ with
$U_{B}(t,t_0)=\exp\left[-i\hat H_B(t-t_0)\right]$
is the interaction picture local field operator.
We now perform cumulant expansion of \eqref{eq:IF1}. As mentioned above,
we assume a Gaussian bath so that the expansion is terminated up to the second order.
The first order cumulant
	\[
		i\gamma_\mu S \int_{\mathcal C} dt n^{\mathcal C}_{\alpha}(t)\bar{B}_{\alpha}
	\]
with $\bar{B}_{\alpha}\equiv\langle \hat{B}_{\alpha}(t) \rangle$ 
represents a deterministic internal mean field generated by the environment. 
For a centered bath, this term vanishes and if nonzero, renormalizes the 
deterministic applied longitudinal field. 
After absorbing the mean bath field into the deterministic spin Hamiltonian,
the remaining influence functional is governed entirely by the second cumulant.
Therefore, \eqref{eq:IF1} can be written as
\begin{equation}
	\mathscr F[\bm{n}^{+},\bm{n}^{-}] =\exp\left[-\frac{\gamma_\mu^2 S^2}{2}
	\int_{\mathcal C} dt\int_{\mathcal C} dt' n^{\mathcal C}_{\alpha}(t)
G^{\mathcal C}_{\alpha\beta}(t,t') n^{\mathcal C}_{\beta}(t')\right],
  \label{eq:IF2}
\end{equation}
where $G^{\mathcal C}_{\alpha\beta}(t,t')=\Big<\mathcal T_{\mathcal C}\hat{B}_{\alpha}(t)\hat{B}_{\beta}(t') \Big>$
is the contour-ordered connected correlator.
In the following, it is convenient to introduce the influence functional action 
$\ln\mathscr F =iS_{\mathrm{IF}}$. Performing Keldysh rotation\cite{Kamenev2011}
yields
\begin{equation}
	S_{\mathrm{IF}}
  =
  \int dt \int dt' n_{\alpha}^{q}(t)\Gamma_{\alpha\beta}^{R}(t,t')n_{\beta}^{cl}(t') 
  + \frac{i}{2}\int dt \int dt' n_{\alpha}^{q}(t)K_{\alpha\beta}(t,t')n_{\beta}^{q}(t'),
  \label{eq:S_IF}
\end{equation}
where 
\begin{equation}
	\Gamma_{\alpha\beta}^{R}(t,t') = i\theta(t-t')\gamma_{\mu}^{2}S^{2}\Big<\left[\hat B_{\alpha}(t),\hat B_{\beta}(t')\right] \Big>
  \label{eq:res_ker}
\end{equation}
is the retarded response kernel and
\begin{equation}
	K_{\alpha\beta}(t,t') = \frac{\gamma_{\mu}^{2}S^{2}}{2}\Big<\left\{\hat B_{\alpha}(t),\hat B_{\beta}(t')\right\}\Big>
	=\gamma_{\mu}^{2}S^{2}C_{\alpha\beta}(t,t')
  \label{eq:noise_ker}
\end{equation}
is the stochastic noise kernel which is fixed by the correlation tensor
\eqref{eq:sym_corr_tens} in Section \ref{sec:model}. 
Here the variables
$X^{cl}=\frac{1}{2}\left(X^{+}+ X^{-}\right)$ and $X^{q}=\left(X^{+}- X^{-}\right)$ 
represent the classical and quantum
fields obtained by rotation of components residing on the forward ($+$) and
backward ($-$) branches of $\mathcal C$.
Similarly after Keldysh rotation, the spin action reads
\begin{equation}
  S_{\mathrm S}
  =
  S\int dt
  \left[
    \bm n^{q}\cdot\left(\bm n^{cl}\times\dot{\bm n}^{cl}\right)
    +
    \gamma_{\mu}\bm n^{q}\cdot\bm B_{\mathrm L}
  \right].
  \label{eq:S_S}
\end{equation}
To this end, we expanded the Berry connection $\bm A$ to linear
order in the quantum component $\bm n^{q}$ and discarded a total time
derivative that depends on the gauge choice. The Berry action therefore
reduces to the standard gauge-invariant form giving the first term in \eqref{eq:S_S}
which enforces the kinematic precession structure of a unit spin.

We next decouple the quadratic $qq$ term in the influence action
by means of a Hubbard-Stratonovich (HS) transformation. Introducing
an auxiliary Gaussian field $\bm{\xi}(t)$, the stochastic representation
of the influence functional becomes
\begin{equation}
  e^{iS_{\mathrm{IF}}[\bm n^{cl},\bm n^{q}]}
  =
  \int \mathcal D\bm{\xi}\, P[\bm{\xi}]\,
  \exp\!\left\{
    i \int dt\, n_\alpha^{q}(t)
    \left[
      \int dt'\,\Gamma_{\alpha\beta}(t,t')\,n_\beta^{cl}(t')
      - \xi_\alpha(t)
    \right]
  \right\},
  \label{eq:IF_HS}
\end{equation}
where the Gaussian weight is
  $P[\bm{\xi}]
  \propto
  \exp\!\left[
    -\frac{1}{2}
    \int dt\,dt'\,
    \xi_\alpha(t)\,
    N^{-1}_{\alpha\beta}(t,t')\,
    \xi_\beta(t')
  \right]$,
so that
\begin{equation}
  \langle \xi_\alpha(t)\rangle_{\xi}=0,
  \qquad
  \langle \xi_\alpha(t)\xi_\beta(t')\rangle_{\xi}
  =
  K_{\alpha\beta}(t,t').
  \label{eq:xi_corr}
\end{equation}
Using \eqref{eq:IF_HS}, the full stochastic effective action reads
\begin{equation}
	S_{\mathrm{eff}}
  =
  \int dt\bm n^{q}(t)\cdot
  \left[
	  S\left(\bm n^{cl}(t)\times\dot{\bm n}^{cl}(t)\right)
    +
    \gamma_{\mu}S\bm B_{\mathrm L}
    + \int dt' \bm \Gamma(t,t')\cdot \bm n^{cl}(t') -\bm{\xi}(t)
  \right].
  \label{eq:S_eff}
\end{equation}
At this stage the derivation is written entirely in terms of the total local
field $\hat{\bm B}(t)$. We only now resolve it into the background and
ion-modulated contributions,
  $\hat B_\alpha(t)
  =
  \hat B_\alpha^{(\mu)}(t)
  +
  \hat B_\alpha^{(i)}(t)$.
Adopting the same decoupling assumptions introduced in Section \ref{sec:model},
we drop the $\mu-\mathrm i$ cross correlation terms so that 
the retarded and stochastic kernels decompose as
$\Gamma_{\alpha\beta} \approx \Gamma_{\alpha\beta}^{(\mu)}+\Gamma_{\alpha\beta}^{(\mathrm i)}$
and 
$K_{\alpha\beta} \approx K_{\alpha\beta}^{(\mu)}+K_{\alpha\beta}^{(\mathrm i)}$.
Here, 
\[
	\Gamma_{\alpha\beta}^{(\zeta)}(t,t') =  i\theta(t-t')\gamma_{\mu}^{2}S^{2}\Big<\left[\hat B_{\alpha}^{(\zeta)}(t),\hat B_{\beta}^{(\zeta)}(t')\right] \Big>
	\qquad
	K_{\alpha\beta}^{(\zeta)}(t,t') =  \frac{\gamma_{\mu}^{2}S^{2}}{2}\Big<\left\{\hat B_{\alpha}^{(\zeta)}(t),\hat B_{\beta}^{(\zeta)}(t')\right\}\Big>
\]
in the $\zeta\in \{\mu,\mathrm i\}$ channel. 
Consequently, for the slow channels retained explicitly in the noise model,
the stochastic noise is a sum of statistically independent Gaussian
components,
$\bm \xi(t)=\sum_{\zeta}\bm \xi^{(\zeta)}(t)$, with correlations
$K_{\alpha\beta}^{(\zeta)}$. From here onwards we absorb $\gamma_\mu$ into the field width,
$\Delta_{\alpha\beta}\to\gamma_\mu\Delta_{\alpha\beta}$, so that $\Delta_{\alpha\beta}$ is 
expressed in inverse-time units, e.g. $\mu s^{-1}$. The 
stochastic noise kernel from \eqref{eq:noise_ker} reads 
\begin{equation}
  K^{(\zeta)}_{\alpha\beta}(t)
  = S^{2}\Delta_{(\zeta)\alpha\beta}^{2}\,e^{-\nu_{\zeta}|t|}.
  \label{eq:Noise_zeta}
\end{equation}
where
\begin{equation}
  \nu_{\zeta} =
  \begin{cases}
    \nu_{\mu}, & \zeta=\mu,\\
    \nu_{\mu}+\nu_{\mathrm i}, & \zeta=\mathrm i.
  \end{cases}
  \label{eq:nu_zeta_def}
\end{equation}
As discussed in Section \ref{sec:model}, \eqref{eq:Noise_zeta}
only describes the slow local-field components that remain
visible in the symmetrized correlator on the $\mu$SR timescale. 
In contrast, the effective retarded response may encode residual
backaction from environmental sources, including fast fluctuations. 
We therefore model the
channel-resolved retarded kernel as
\begin{equation}
  \Gamma_{\alpha\beta}^{R,(\zeta)}(t)
  =
  \Lambda_{\alpha\beta}^{(\zeta)}
  e^{-\nu_\zeta t}\theta(t),
  \label{eq:GammaR_exp}
\end{equation}
where the channel index $\zeta$ is inherited from the field decomposition, and
$\Lambda_{\alpha\beta}^{(\zeta)}$ sets the corresponding strength of the
effective retarded memory torque.
A useful formal parametrization of
\eqref{eq:GammaR_exp} is obtained by writing the retarded kernel in terms
of an effective susceptibility from the explicit expressions for the local-field operator
$\hat B_\alpha^{(\zeta)}(t)$
\begin{equation}
  \Gamma_{\alpha\beta}^{R,(\zeta)}(t)
  =
  \gamma_\mu^2 S^2
  \sum_{\bm q}
  \mathcal K_{\alpha\gamma}(\bm q)\,
  \mathcal K_{\beta\delta}(-\bm q)\,
  \Pi_{\gamma\delta}^{R,(\zeta)}(\bm q,t),
  \label{eq:Gamma_ker}
\end{equation}
where $\Pi_{\gamma\delta}^{R,(\zeta)}$ is an effective retarded
susceptibility for channel $\zeta$, namely the causal response kernel
associated with the commutator sector of the local field. For the ion channel, this response is
not identified with the bare commutator of the classical hopping variables
used above to construct the symmetrized density correlator. Instead, if one assumes
a passive field-following static response in the projected channel,
$\Pi^{R,(\zeta)}(\bm q)$ is positive semidefinite and the corresponding
scalar backaction amplitude is nonnegative.
For simplicity, one may approximate the susceptibility a single-pole form,
\begin{equation*}
  \Pi_{\gamma\delta}^{R,(\zeta)}(\bm q,t)
  \approx
  \theta(t)\,
  \nu_\zeta e^{-\nu_\zeta t}\,
  \Pi_{\gamma\delta}^{R,(\zeta)}(\bm q),
  \qquad
  \Pi_{\gamma\delta}^{R,(\zeta)}(\bm q)
  =
  \int_0^\infty dt\,
  \Pi_{\gamma\delta}^{R,(\zeta)}(\bm q,t).
  \label{eq:PiR_singlepole}
\end{equation*}
This recovers \eqref{eq:GammaR_exp} with
\begin{equation}
  \Lambda_{\alpha\beta}^{(\zeta)}
  =
  \nu_\zeta\gamma_\mu^2 S^2
  \sum_{\bm q}
  \mathcal K_{\alpha\gamma}(\bm q)\,
  \mathcal K_{\beta\delta}(-\bm q)\,
  \Pi_{\gamma\delta}^{R,(\zeta)}(\bm q).
  \label{eq:Lambda}
\end{equation}
In the isotropic reduction used below,
$\Lambda_{\alpha\beta}^{(\zeta)}\to \Lambda^{(\zeta)}\delta_{\alpha\beta}$.
In the numerical simulations, $\Lambda^{(\zeta)}$ is treated as a
phenomenological parameter controlling the strength of the retarded
backaction. The exponential form of $\Gamma^{R,(\zeta)}$ is therefore
used as an effective memory kernel, rather than as a microscopic
consequence of the classical hopping approximation introduced above for
the symmetrized noise correlator.

Varying \eqref{eq:S_eff} with respect to $\bm n^{q}$ and then setting
$\bm n^{q}\!\to 0$ and $\bm n^{c}\to \bm n$  yields the spin stochastic (integro-)differential equation (SDE)
\begin{equation}
  \dot{\bm n}(t) =
  \bm n(t)\times\!\left\{
	  \gamma_{\mu}\bm B_{\mathrm L}
     + \sum_{\zeta}\left[\frac{1}{S}\!\int\!dt'\;
     \boldsymbol{\Gamma}^{(\zeta)}(t-t')\cdot\bm n(t')
     - \frac{\boldsymbol{\xi}^{(\zeta)}(t)}{S}\right]
  \right\}.
  \label{eq:LLG}
\end{equation}
\eqref{eq:LLG} is the central result of this work. Its Monte Carlo (MC)
sampling provides a fit-ready non-Markovian description of ZF/LF
$\mu$SR spectra.

\subsection{Static Kubo-Toyabe limit}
\label{sec:static_KT}
The static KT limit follows from the spin SDE [\eqref{eq:LLG}] by
neglecting the retarded damping kernel,
$\boldsymbol{\Gamma}^{(\zeta)}(t-t')$, and
freezing the stochastic field,
$\boldsymbol{\xi}^{(\zeta)}(t)\to \boldsymbol{\xi}^{(\zeta)}$ (time independent).
We define the static effective precession vector  
$\bm{\Omega}
  \equiv
\omega_{0}\hat{\bm{z}}-\bm{\xi}$,
with $\bm{\xi}\equiv\sum_{\zeta}\frac{\bm{\xi}^{\zeta}}{S}$ and $\omega_{0}=\gamma_\mu  B_{\mathrm L}$.
The spin direction obeys uniform precession via 
$\dot{\bm n}(t)=-\boldsymbol{\Omega}\times \bm n(t),$
whose formal solution is
\begin{equation}
	\bm{n}(t)=\bm{R}(t)\bm{n}(0).
	\label{eq:stat_sol}
\end{equation}
Here, $\bm{R}=\exp\left(-t[\boldsymbol{\Omega}]_{\times}\right)$
is the $SO(3)$ rotation propagator with 
\begin{equation*}
[\boldsymbol{\Omega}]_{\times}\equiv
\begin{pmatrix}
0 & -\Omega_{z} & \Omega_{y}\\
\Omega_{z} & 0  & -\Omega_{x}\\
-\Omega_{y} & \Omega_{x} & 0
\end{pmatrix},
\end{equation*}
being the antisymmetric cross-product matrix obeying
$[\boldsymbol{\Omega}]_{\times}\bm v
=\boldsymbol{\Omega}\times\bm v$ for any vector $\bm v$.
For an initially polarized ensemble along $+\hat{z}$,
$\bm n(0)=\hat{z}$,
Rodrigues' formula with $\hat{\boldsymbol{\Omega}}=\boldsymbol{\Omega}/\Omega$
and $\Omega=|\boldsymbol{\Omega}|$ yields
\begin{equation}
  n_z(t) = 
  \hat{\Omega}_z^{\,2}
  +
  \bigl(1-\hat{\Omega}_z^{\,2}\bigr)\cos(\Omega t).
  \label{eq:nz_fixedOmega}
\end{equation}
For a randomly oriented polycrystal (powder), the angular
average at fixed $\Omega$ is the usual $\langle n_z(t)\rangle
=\frac{1}{3}+\frac{2}{3}\cos(\Omega t)$. 
The standard static KT model further assumes that the components of
$\bm{\xi}$ are independent, isotropic Gaussians,
\begin{equation}
	P(\bm{\xi}) =
  \frac{1}{(2\pi\Delta^{2})^{3/2}}
  \exp\!\left[
    -\frac{\xi_x^{2}+\xi_y^{2}+\xi_z^{2}}{2\Delta^{2}}
  \right],
  \label{eq:gaussian_prob}
\end{equation}
where $\Delta^{2}=\langle \xi_\alpha^{2}\rangle$.
Using \eqref{eq:gaussian_prob} the statistical average of \eqref{eq:nz_fixedOmega} 
over $\bm{\xi}$ then yields the familiar
static KT polarization function\cite{Hayano1979}
\begin{equation}
	G_{\mathrm{stat}}(t) =\langle n_z(t) \rangle_{\bm{\xi}}= 1 -\frac{2\Delta^2}{\omega_{0}^{2}}\left[
	1-e^{-\frac{1}{2}\Delta^{2}t^{2}}\cos(\omega_0 t)\right]
	+ \frac{2\Delta^4}{\omega_{0}^{3}}\int_{0}^{t}d\tau 
	e^{-\frac{1}{2}\Delta^{2}\tau^{2}}\sin(\omega_0 \tau).
  \label{eq:G_z_kt}
\end{equation}

\subsection{Analytical Function}
The spin SDE [\eqref{eq:LLG}] may be reduced to an analytical function under some controlled approximations.
Firstly, for simplicity, let us assume that the muon is static ($\nu_\mu=0$) so that the stochastic noise
becomes $\sum_{\zeta} \boldsymbol\xi^{\zeta}(t)=\boldsymbol\xi^{\mu}+\boldsymbol\xi^{\mathrm i}(t)$.
\eqref{eq:LLG} with $t'\to t-\tau$ therefore becomes 
\begin{equation}
\dot{\bm{n}}(t)
=
\bm{n}(t)\times
\left[
\bm{\Omega}_\mathrm s
+
\frac{1}{S}\int_0^\infty d\tau\,\boldsymbol\Gamma^{\mathrm i}(\tau)\,\cdot\bm{n}(t-\tau)
-
\frac{1}{S}\bm{\xi}^{\mathrm i}(t)
\right],
\label{eq:SDE_n_split}
\end{equation}
where we defined the static frequency vector as
$\bm{\Omega}_\mathrm s 
\;\equiv\;
\gamma_\mu \bm{B}_L - \frac{1}{S}\bm{\xi}^{\mathrm \mu}$.
Let us rewrite \eqref{eq:SDE_n_split} in the rotating spin frame by defining
$\bm{m}(t)\equiv \bm{R}_\mathrm s(t)^{-1}\bm{n}(t)=\bm{R}_\mathrm s(t)^\mathsf{T}\bm{n}(t)$, 
$\bm{\eta}\equiv\bm{R}_\mathrm s(t)^{-1}\bm{\xi}^{\mathrm i}(t)$,
$\tilde{\bm{\Gamma}}^{\mathrm i}(t,\tau)\equiv\bm{R}_\mathrm s(t)^{-1}\bm{\Gamma}^{\mathrm i}(\tau)\bm{R}_\mathrm s(t-\tau)$
where
$\bm{R}_\mathrm{s}(t)\equiv \exp\!\left(t[\bm{\Omega}_\mathrm s]_\times\right)$
is the rotation operator generated by the antisymmetric vector
$[\bm{\Omega}_\mathrm s]_\times$. The resulting spin SDE reads
\begin{equation}
\dot{\bm{m}}(t)
=
\bm{m}(t)\times
\left[
	\frac{1}{S}\int_0^\infty d\tau\,\tilde{\bm{\Gamma}}^{\mathrm i}(t,\tau)\cdot\bm{m}(t-\tau)
-\frac{1}{S}\bm{\eta}(t)
\right],
\label{eq:SDE_rot}
\end{equation}
Consequently, the measured polarization can be written as
\begin{equation}
	G_{z}(t)=\langle n_z(t) \rangle =\big< [\bm{R}_{\mathrm s}(t)\bm{m}(t)]_{z} \big>_{\bm{\xi}^{\mu}},
\label{eq:G_z_rot}
\end{equation}
where the statistical average is over the static muon field $\bm{\xi}^{\mu}$.
As before we assume that the muon spin direction 
$\bm{m}(t) = (m_x(t),m_y(t),m_z(t))$ is initially polarized along $+\hat{z}$. 
We consider small transverse deviations $|m_x|,|m_y|\ll 1$ so that
$m_{z}=\sqrt{1-m_x^{2}-m_y^{2}} \approx 1-\frac{1}{2}\left(|m_x|^{2}+|m_y|^{2} \right)$,
and to linear order, we set $m_z\approx 1$. 
The linearized equations of motion in this small-angle approximation are
\[\dot{m}_x= -\Phi_{y}(t)+\Phi_{z}(t)m_{y}\qquad\dot{m}_y= \Phi_{x}(t)-\Phi_{z}(t)m_{x}\]
where
\[
\Phi_{a}(t)=
\frac{1}{S}\int_0^\infty d\tau\,\tilde{\Gamma}_{a,b}^{\mathrm i}(t,\tau)m_{j}(t-\tau)
-\frac{1}{S}\eta_{a}(t).
\]
Here $a,b\in\{x,y,z \}$ are the Cartesian indices in rotating spin frame.
The causality of $\Gamma_{a,b}^{\mathrm i}(\tau)$ implies
that $m_{b}(t-\tau)$ is only defined for $\tau<t$, allowing us to replace
$\int_{0}^{\infty}\to\int_{0}^{t}$.
We specialize in the isotropic case so that the rotating frame backaction kernel
$\tilde{\Gamma}^{\mathrm i}_{a,b}(t,\tau)=\Gamma^{\mathrm i}(\tau)[\bm{R}_\mathrm s(-\tau)]_{a,b}$
and the noise correlator [with \eqref{eq:Noise_zeta}]
$\langle\eta_{a}(t)\eta_{b}(t')\rangle=2S^{2}\Delta_{\mathrm i}^{2}\delta_{\alpha\beta}e^{-\nu_{\mathrm i}|t-t'|}\
\left[\bm{R}_{\mathrm s}(t'-t)\right]_{ab}$, acquires
a rotation matrix factor
\[
\bm{R}_{\mathrm s}(t'-t)=\bm{R}_{\mathrm s}^{-1}(t)\bm{R}_{\mathrm s}(t')
\]
even if the lab frame noise $\bm{\xi}^{\mathrm i}(t)$ is isotropic.

The term $\bm{R}_{\mathrm s}(t)$ in backaction and noise kernels presents a difficulty in obtaining a closed form 
of $G_z(t)$ via $\bm{\xi}^{\mu}$ averaging in \eqref{eq:G_z_rot}. We make progress
by keeping the LF dynamics exactly while treating $\bm{\xi}^{\mu}$ in 
$\bm{R}_{\mathrm s}$ perturbatively.
For this purpose, we let $\bm{\Omega}_{\mu}\equiv-\frac{1}{S}\bm{\xi}^{\mu}$
and $\omega_{0}=\gamma_{\mu}B_{\mathrm L}$ and write explicitly the antisymmetric cross product
matrix as
$[\bm{\Omega}_{\mathrm s}]_{\times}=[\omega_{0}\hat{\bm{z}}]_{\times}+[\bm{\Omega}_{\mu}]_{\times}$.
Since $[\omega_{0}\hat{\bm{z}}]_{\times}$ and $[\bm{\Omega}_{\mu}]_{\times}$ do not commute
unless $\bm{\Omega}_{\mu}\parallel\hat{z}$,
the exponential in $\bm{R}_{\mathrm s}$ cannot in general be factorized. We go to the interaction picture
with respect to LF and denote $\bm{R}_{0}(t)=\exp\left(t [\omega_{0}\hat{\bm{z}}]_{\times}\right)$
so that the full static rotation can be factored as
\[
	\bm{R}_{\mathrm s}(t)=\bm{R}_{0}(t)\bm{U}(t)
\]
where $\bm{U}(t)$ contains the effect of $\bm{\Omega}_{\mu}$.
By differentiating $\bm{R}_{\mathrm s}(t)$, we obtain an equation of motion
for $\bm{U}(t)$ whose formal solution can be expanded in Dyson series as
\[
	\bm{U}(t)=\mathbb{I}+\int_{0}^{t}ds_{1}[\bm{\Omega}_{\mu}^{I}(s_1)]_{\times}\
	+\int_{0}^{t}ds_{1}\int_{0}^{s_1}ds_{2}[\bm{\Omega}_{\mu}^{I}(s_1)]_{\times}[\bm{\Omega}_{\mu}^{I}(s_2)]_{\times}\
	+\cdots,
\]
where $\bm{\Omega}_{\mu}^{I}(t)\equiv\bm{R}_{0}^{-1}(t)\bm{\Omega}_{\mu}$.
Inserting the above into the rotation matrix factor 
$\bm{R}_{\mathrm s}(t-t')=\bm{U}^{-1}(t)\bm{R}^{-1}_{0}(t)\bm{R}_{0}(t')\bm{U}(t')
=\bm{U}^{-1}(t)\bm{R}_{0}(t'-t)\bm{U}(t')$ yield to first order
\[
	\bm{R}_{\mathrm s}(t'-t)=\bm{R}_{0}(t'-t)+\bm{R}_{0}(t'-t)\int_{0}^{t}ds[\bm{\Omega}_{\mu}^{I}(s)]_{\times}\
	+\int_{0}^{t'}ds_{1}[\bm{\Omega}_{\mu}^{I}(s_1)]_{\times}\bm{R}_{0}(t'-t)+\mathcal{O}((\bm{\Omega}_{\mu}^{I}(s))^2)
\]
Since the muon static field $\bm{\xi}_{\mu}$ is Gaussian (and so is $\bm{\Omega}_{\mu}$),
only the first term of $\bm{R}_{\mathrm s}(t-t')$ is non-trivial after averaging.
Hence, $\bm{R}_{\mathrm s}(t'-t)\approx\bm{R}_{0}(t'-t)$ and similarly $\bm{R}_{\mathrm s}(-\tau)\approx\bm{R}_{0}(-\tau)$.
In terms of the complex transverse
variable $u=m_x +i m_y$, $u^{*}=m_x -i m_y$ and 
complex transverse noise $\eta_{+}=\eta_x + i \eta_y$ with
$\eta_{-}=\eta_{+}^{*}$, the Dyson reduction
yields a pure LF phase $e^{-i\omega_{0}t}$ and
the resulting linearized equation reads
\begin{equation}
	\dot{u}(t)=i\Phi_{z}(t)u(t)+\frac{i}{S}\int_{0}^{t}d\tau\Gamma(\tau)e^{-i\omega_{0}\tau}u(t-\tau)
	-\frac{i}{S}\eta_{+}(t).
	\label{eq:trans_sde}
\end{equation}
Here, to linear order ($m_{z}\simeq 1$), 
$\Phi_{z}(t)=\omega_{z}(t)-\eta_{z}(t)/S$, where $\omega_{z}(t)\equiv \int_{0}^{t} d\tau\Gamma(\tau)$.
Denoting $\theta(t)\equiv\int_{0}^{t}ds\Phi_{z}(s)$, the local term $\Phi_{z}(t)u(t)$ 
can be removed exactly by the phase transformation
\[
	v(t)=e^{-i\theta(t)}u(t) \qquad \varphi(t)=e^{-i\theta(t)}\eta_{+}(t),
\]
giving
\begin{equation}
	\dot{v}(t)=\frac{i}{S}\int_{0}^{t}d\tau\tilde{\Gamma}(\tau)v(t-\tau)-\frac{i}{S}\varphi(t) 
	\label{eq:SDE_v}
\end{equation}
where the memory kernel 
$\tilde{\Gamma}(\tau)=\Gamma(\tau)e^{-i[\theta(t)-\theta(t-\tau)]}e^{-i\omega_{0}\tau}$
acquires an extra phase factor.
Keeping this factor exactly would lead to
stochastic equation with a random (multiplicative-noise) memory kernel, for which a closed
solution is not available. We therefore employ a standard averaged-kernel (self-averaging)
closure and replace the phase factor by its Gaussian average 
$e^{i[\theta(t)-\theta(t-\tau)]}\to e^{-i\omega_{\mathrm{eff}} \tau}\mathcal{D}(\tau)$
which restores time translational invariance and yields a deterministic dressed convolution kernel
\[
	\tilde{\Gamma}(\tau)=S^{2}\Lambda_{\mathrm i}e^{-i(\nu_{\mathrm i}+\omega_{\mathrm{eff}})\tau}\
	\mathcal{D}(\tau)\Theta(\tau).
\] 
Here,
$\mathcal{D}(\tau)=\langle e^{i[\theta_{\mathrm{fl}}(t)-\theta_{\mathrm{fl}}(t-\tau)]}\rangle$ with
$\theta_{\mathrm{fl}}(t)=-\frac{1}{S}\int_{0}^{t}ds\eta_{z}(s)$, and
$\omega_{\mathrm{eff}}=\omega_{0}+\omega_{z}$. 
To this end, we split $\theta(t)=\omega_{\mathrm{eff}}t + \theta_{\mathrm{fl}}(t)$
and assume independence during averaging.
Because $\eta_{z}$ is Gaussian and stationary, the averaged phase factor $\mathcal{D}(\tau)$
depends only on $\tau$ and evaluates (see Appendix \ref{app:phase_fluctuation_kernel}) to
\[
	\mathcal{D}(\tau)=\exp\left[-\frac{2\Delta_{\mathrm i}^{2}}{\nu_{\mathrm i}}
		\left(\nu_{\mathrm i}\tau -1+e^{-\nu_{\mathrm i}\tau}\right) \right].
\]
Interestingly, the phase-diffusion factor $\mathcal{D}(\tau)$ itself takes
the Abragam\cite{abragam} (Gaussian-Markov) form, reflecting that it is a characteristic
function of Gaussian accumulated phase.
Similarly, the transformed noise correlator becomes
\[
	\langle \varphi(t)\varphi^{*}(t')\rangle =\
2S^{2}\Delta_{\mathrm i}^{2}e^{-\nu_{\mathrm i}|t-t'|}e^{+i\omega_{\mathrm{eff}}(t-t')}\
\mathcal{D}(t-t').
\]
Let us denote $d\equiv\frac{2\Delta_{\mathrm i}^{2}}{\nu_{\mathrm i}^{2}}$ and
rewrite 
\begin{equation}
	\mathcal{D}(\tau)=e^{d}\sum_{n=0}^{\infty}\frac{(-d)^{n}}{n!}e^{-(d+n)\nu_{\mathrm i}\tau}
\label{eq:D_series}.
\end{equation}
To obtain a closed analytic propagator we approximate the dressed retarded kernel
as a finite exponential (rational) representation by truncating the series,
$n=0,1,\dots,N$, so that the dressed kernel reads
\[
	\tilde{\Gamma}(\tau)=\sum_{n=0}^{N}\kappa_{n}e^{-\beta_{n}\tau},
\]
where $\kappa_{n}=S^{2}\Lambda_{\mathrm i}e^{d}\
(-d)^{n}/{n!}$ and $\beta_{n}= [d + (n+1)]\nu_{\mathrm i} + i\omega_{\mathrm{eff}}$.
This step is purely technical and is done for the purpose of obtaining a rational Green's function
with finite number of poles in the succeeding Laplace transform solutions.
The truncation order $N$ is chosen such that $\tilde{\Gamma}(\tau)$ is converged 
over the relevant $\mu$SR time/frequency window.

Performing the Laplace transform of \eqref{eq:SDE_v} yields
\begin{equation}
	\tilde{v}(s)=\tilde{\mathcal G}(s)\tilde{u}(0)-\frac{i}{S}\tilde{\mathcal G}(s)\tilde{\varphi}(s),
	\label{eq:u_Laplace}
\end{equation}
where the causal Laplace space Green's function is a rational function ($M=N+1$)
\[
	\tilde{\mathcal G}(s)=\left(s-\frac{i}{S}\sum_{n=0}^{M-1}\frac{\kappa_{n}}{s+\beta_{n}}\right)^{-1}.
\]
For convenience, we express it as a ratio of polynomials
\[
	\tilde{\mathcal G}(s)=\frac{Q(s)}{P(s)},
\]
where $Q(s)=\prod_{n=0}^{M-1}\left(s + \beta_{n} \right)$ and 
$P(s)=sQ(s)-\frac{i}{S}\sum_{n=0}^{M-1}\kappa_{n}\prod_{m=0;m\neq n}^{M-1}\left(s+\beta_{n}\right)$.
The poles $r_k$ are the roots of $P(s)=0$, and the time-domain retarded
propagator follows as a finite sum of exponentials,
\begin{equation}
	\mathcal G(t)=\sum_{k=0}^{M} B_k e^{r_k t}\Theta(t)
	\label{eq:GF_series}
\end{equation}
with residues $B_k=\mathrm{Res}[\tilde{\mathcal G}(s),s=r_k]$.
Concretely, for a special case $N=0$ ($M=1$) the retarded kernel exactly has
one exponential and the roots of the quadratic denominator of $\tilde{G}(s)$ gives
the two poles ($k=+,-$) 
\[
	r_{\pm} = \frac{-\beta_{0}\pm\sqrt{\beta_{0}^{2}+\frac{4i}{S}\kappa_{0}}}{2}
\]
and the residues 
\[
	B_{\pm} = \frac{r_{\pm}+\beta_{0}}{r_{\pm}-r_{\mp}}.
\]
From the time-domain solution of \eqref{eq:u_Laplace} with $u(0)=0$ 
we obtain ($\langle |u(t)|^2 \rangle=\langle |v(t)|^2 \rangle$)
\begin{equation}
	\langle |u(t)|^2 \rangle =\frac{1}{S^2}\int_{0}^{t} dt_{1}\int_{0}^{t}dt_{2}
	\mathcal G(t-t_{1})\mathcal G^{*}(t-t_{2})\langle\varphi(t_{1})\varphi^{*}(t_{2})\rangle,
\label{eq:uu_tdom}
\end{equation}
Inserting \eqref{eq:GF_series} into \eqref{eq:uu_tdom} with $u(0)=0$
and evaluating the double time integrals yield the closed form
\begin{equation}
	\langle |u(t)|^2 \rangle =2\Delta_{\mathrm i}^{2}e^{d}\sum_{m=0}^{\infty}\frac{(-d)^{m}}{m!}
	\sum_{k,k'=0}^{M}B_{k}B_{k'}^{*}F_{kk'}(t;\lambda_{m},\omega_{\mathrm{eff}}).
\label{eq:uu_tdom1}
\end{equation}
where 
\[
	F_{kk'}(t;\lambda_{m},\omega_{\mathrm{eff}})=2\mathrm{Re}
	\left\{ \frac{e^{\left(r_{k}+r_{k'}^{*} \right) t}}{r_{k'}^{*}-\lambda_{m}+i\omega_{\mathrm{eff}}}\
	\left[\frac{1-e^{-\left(r_{k}+\lambda_{m}-i\omega_{\mathrm{eff}}\right) t}}{r_{k}+\lambda_{m}-i\omega_{\mathrm{eff}}}\
	-\frac{1-e^{-\left(r_{k}+r_{k'}^{*}\right) t}}{r_{k} + r_{k'}^{*}}\
	 \right]\right\}.
\]
Here, the $m$-sum originates from the exponential decomposition of the
effective transverse correlator
\[
\langle\varphi(t_{1})\varphi^{*}(t_{2})\rangle = 
2S^{2}\Delta_{\mathrm i}^{2}e^{+i\omega_{\mathrm{eff}}(t_1-t_2)}
e^{d}\sum_{m=0}^{\infty}\frac{(-d)^m}{m!}\,e^{-\lambda_m|t_1-t_2|},
\]
where
\[
\lambda_m=[d+(m+1)]\nu_{\mathrm i}.
\]

We now use two properties that follow directly from our setup.
First, the dynamic bath/noise that drives $\bm{m}(t)$ is statistically
independent of the static muon background field $\bm{\xi}^{\mu}$ entering
$\bm{R}_{\mathrm s}(t)$ so joint averages in \eqref{eq:G_z_rot} factorize into
$\langle\cdots\rangle=\langle\cdots\rangle_{\bm{\xi}^{\mu}}
\langle\cdots\rangle_{\mathrm{dyn}}$. Second, 
for an initially polarized ensemble along $+\hat{z}$ one has
$m_{x}(0)=m_{y}(0)=0$, and the linearized transverse dynamics is driven by
zero-mean noise, hence $\langle m_{x}(t)\rangle_{\mathrm{dyn}}
=\langle m_{y}(t)\rangle_{\mathrm{dyn}}=0$ at all times.
Consequently, the $zx$ and $zy$ contributions vanish and 
the polarization function \eqref{eq:G_z_rot} factorizes as 
\begin{equation}
  G_{z}(t)
  \;=\;
  G_{\mathrm{stat}}(t)\,G_{\mathrm{dyn}}(t).
  \label{eq:Gz_factorized}
\end{equation}
Here, 
$G_{\mathrm{dyn}}(t)\equiv\langle m_{z}(t) \rangle \approx 1-\frac{1}{2}\langle |u(t)|^{2} \rangle$,
and 
\begin{equation}
  G_{\mathrm{stat}}(t)\;\equiv\;
  \big\langle [\bm{R}_{\mathrm s}(t)]_{zz}\big\rangle_{\bm{\xi}^{\mu}}
  \;=\;
  \Big\langle \big(\bm{R}_{\mathrm s}(t)\hat{z}\big)_{z}\Big\rangle_{\bm{\xi}^{\mu}},
  \label{eq:Gstat_def}
\end{equation}
which is exactly the static KT/LF-KT polarization \eqref{eq:G_z_kt} derived in
Sec.~\ref{sec:static_KT} upon identifying the static field distribution
with $\bm{\xi}\to\bm{\xi}^{\mu}/S$ and field width $\Delta\to\Delta_{\mu}$. 
As established above, the transverse dynamics is linear
and driven by a Gaussian noise. Therefore $u(t)$ is a Gaussian
functional of the noise, and the standard  
(second-cumulant) resummation gives
\begin{equation}
  G_{\mathrm{dyn}}(t) \simeq
  \exp\!\left[-\frac{1}{2}\langle |u(t)|^{2} \rangle\right].
  \label{eq:G_dyn}
\end{equation}
\eqref{eq:Gz_factorized}, together with Equations \ref{eq:Gstat_def}, \ref{eq:G_dyn} and \ref{eq:uu_tdom1}
constitute the analytical form \eqref{eq:LLG}.

In the ZF and no backaction ($\Lambda_{\mathrm i}=0$) 
limit the retarded propagator reduces to
$\mathcal G(t)=1$ (equivalently $\tilde{\mathcal G}(s)=1/s$), and the transverse variance becomes
\begin{equation}
  \langle |u(t)|^2\rangle
  =
  4\Delta_{\mathrm i}^{2}\,e^{d}
  \sum_{m=0}^{\infty}\frac{(-d)^{m}}{m!}
  \left[
    \frac{t}{\lambda_{m}}
    -\frac{1-e^{-\lambda_{m}t}}{\lambda_{m}^{2}}
  \right].
  \label{eq:uu_ZF_nomem_series}
\end{equation}
It is useful to introduce the standard Abragam building block
\begin{equation}
  f(t,\lambda)\equiv
  \frac{1}{\lambda^{2}}\big(\lambda t-1+e^{-\lambda t}\big),
  \label{eq:Abragam_block}
\end{equation}
so that we can write \eqref{eq:uu_ZF_nomem_series} compactly as
\begin{equation}
	G_{\mathrm{dyn}}(t) 
	= \exp\left[-2\Delta_{\mathrm i}^{2}
\sum_{m=0}^{\infty} w_{m}\, f(t,\lambda_{m})
	\right]
  \label{eq:uu_generalized_Abragam}
\end{equation}
where $w_{m}\equiv e^{d}\frac{(-d)^{m}}{m!}$ and $\sum_{m=0}^{\infty} w_{m}=1$.
For comparison, the conventional Abragam (Gaussian-Markov) result\cite{abragam,keren_1994} 
corresponds to a single correlation rate $\lambda\to\nu$ and reads
\begin{equation}
	G_{\mathrm{Abragam}}(t) 
	= \exp\left[-2\Delta_{\mathrm i}^{2}f(t,\nu)
	\right]
	=\exp\left[-\frac{2\Delta_{\mathrm i}^{2}}{\nu^{2}}
  	\left(\nu t-1+e^{-\nu t}\right)
\right].
    \label{eq:uu_Abragam_single}
\end{equation}
\eqref{eq:uu_generalized_Abragam} therefore provides a natural
generalized Abragam form. Here the longitudinal phase diffusion modifies the
effective correlator from a single exponential to an exponential mixture, which
in turn produces a superposition of Abragam kernels with rates
$\lambda_m$. The coefficients $w_m$ originate from an exact series
representation of the dephasing factor and alternate in sign for $d>0$.
Hence they should be viewed as expansion weights rather than a positive probability
distribution. In particular, truncating \eqref{eq:uu_generalized_Abragam}
at low $m$ is generally unreliable when $d$ is large, even though the full
series resums to a smooth correlator.

In principle, \eqref{eq:uu_tdom1} can be
evaluated in closed form by brute pole/residue double sum.
However, in practice, as mentioned above, its direct evaluation is numerically fragile in the
quasi-static regime $d\gg 1$. The coefficients
$w_m$ alternate in sign and large cancellations are required to
recover the smooth correlator. This means that any finite truncation becomes numerically unstable.
In addition, implementing \eqref{eq:uu_tdom1} requires locating the poles
$r_k$ of the rational Laplace-space Green's function and computing residues
$B_k$. This entails root finding of high-order polynomials which can become
ill-conditioned as parameters vary.

Therefore, we evaluate instead the same analytical approximation in the time domain by
approximating the dressed exponential memory factor with a finite mixture of decaying exponentials.
Their nonnegative weights are obtained by a nonnegative least-squares (NNLS) fit on
the time grid. This in turn allows a stable auxiliary-variable (Markovian embedding) propagation of the
$\mathcal G(t)$ without explicit pole extraction.
Finally, $\langle|u(t)|^2\rangle$ is computed from
$\mathcal G(t)$ and the dressed correlator by time-domain convolution (implemented
efficiently via FFT/recurrence).

Under the rotating-frame and small-angle approximations described above, the longitudinal
polarization admits a compact factorized form and a closed expression for the transverse
variance. While the non-Markovian analysis is carried out mainly by MC evaluation of spin SDE 
\eqref{eq:LLG}, the derived closed expressions provide a transparent baseline on how $(\Delta_{\mathrm i}$,
$\nu_{\mathrm i})$ and $\Lambda$ reshape the $\mu$SR line shape accross regimes.

\section{Numerical Simulations}
To evaluate the muon spin polarization by MC we integrate the stochastic equation
\eqref{eq:LLG} numerically. 
%\paragraph*{Markovian embedding.}
To avoid history integrals we introduce auxiliary “memory” variables for each
exponential kernel,
\[
  \bm u_{(\zeta)}(t) \equiv \int_{-\infty}^{t}\!dt'\,
  e^{-\nu_{\zeta}(t-t')}\,\bm n(t'),
  \qquad
  \dot{\bm u}_{(\zeta)} = -\nu_{\zeta}\,\bm u_{(\zeta)} + \bm n,
  \quad \bm u_{(\zeta)}(0)=\bm 0,
\]
so that
\[
  \int_{0}^{\infty}\! d\tau\,\boldsymbol{\Gamma}(\tau)\cdot\bm n(t-\tau)
  =
  \boldsymbol{\Lambda}_{(\mu)}\cdot\bm u_{(\mu)}(t)
  +
  \boldsymbol{\Lambda}_{(\mathrm i)}\cdot\bm u_{(\mathrm i)}(t).
\]
For the colored noise we use independent Ornstein-Uhlenbeck (OU) processes
for the two bath channels.  Writing the Hubbard-Stratonovich fields as
$\boldsymbol{\xi}(t)=\boldsymbol{\xi}_{(\mu)}(t)+\boldsymbol{\xi}_{(\mathrm i)}(t)$,
we sample
\[
  d\boldsymbol{\xi}_{(\zeta)}(t)
  = -\nu_{\zeta}\,\boldsymbol{\xi}_{(\zeta)}(t)\,dt
  + \sqrt{2\nu_{\zeta}}\,\bm L_{(\zeta)}\,d\bm W_{(\zeta)}(t),
  \qquad
  \bm L_{(\zeta)}\bm L_{(\zeta)}^{\mathsf T}=S^{2}\boldsymbol{\Delta}^{2}_{(\zeta)},
\]
with independent Wiener increments $d\bm W_{(\mu)}$ and $d\bm W_{(\mathrm i)}$.
With these definitions \eqref{eq:LLG} becomes the local system
\begin{equation}
  \dot{\bm n}(t) = -\,\boldsymbol{\Omega}(t)\times\bm n(t),
  \qquad
  \boldsymbol{\Omega}(t)
  = \gamma_{\mu}\bm B 
  - \sum_{\zeta}\left[\frac{\boldsymbol{\xi}_{(\zeta)}(t)}{S}
  - \frac{\boldsymbol{\Lambda}_{(\zeta)}\cdot\bm u_{(\zeta)}(t)}{S}\right]
  \label{eq:Em_SDE}
\end{equation}
%\paragraph*{Time stepping and estimator.}
Assuming \(\boldsymbol{\Omega}(t)\) is quasi-constant on \([t_n,t_n+\Delta t]\),
we update \(\bm n\) with a Rodrigues rotation:
\[
  \bm n_{n+1}
  = \bm n_{n}\cos\theta_{n}
    - (\hat{\boldsymbol{\Omega}}_{n}\times \bm n_{n})\sin\theta_{n}
    + \hat{\boldsymbol{\Omega}}_{n}\big(\hat{\boldsymbol{\Omega}}_{n}\!\cdot\!\bm n_{n}\big)\,(1-\cos\theta_{n}),
\]
where \(\theta_{n}=\|\boldsymbol{\Omega}_{n}\|\Delta t\) and
\(\hat{\boldsymbol{\Omega}}_{n}=\boldsymbol{\Omega}_{n}/\|\boldsymbol{\Omega}_{n}\|\).
The auxiliary variables \(\bm u_{(\mu)}\), \(\bm u_{(\mathrm i)}\) and
\(\boldsymbol{\xi}_{(\mu)}\), \(\boldsymbol{\xi}_{(\mathrm i)}\) are advanced with
Euler--Maruyama using the same \(\Delta t\).
The polarization is estimated by ensemble averaging over \(N\) trajectories
\[
  G_{z}(t_n)= N^{-1}\sum_{k=1}^{N}\,[n_{z}^{(k)}]_n.
\]
%\paragraph*{Anisotropy and isotropic specialization.}
The scheme is fully tensorial: anisotropy and cross-correlations enter through
\( \boldsymbol{\Lambda}_{(\mu,\mathrm i)} \) and
\( \boldsymbol{\Delta}^{2}_{(\mu,\mathrm i)} \).
For presentation purposes, we specialize below to the isotropic case,
\(
\Delta^{2}_{(\mu)\alpha\beta}=\Delta_{\mu}^{2}\delta_{\alpha\beta}
\)
and
\(
\Delta^{2}_{(\mathrm i)\alpha\beta}=\Delta_{\mathrm i}^{2}\delta_{\alpha\beta},
\)
so that
\(
\boldsymbol{\Delta}^{2}_{(\mu)}=\Delta_{\mu}^{2}\bm I
\)
and
\(
\boldsymbol{\Delta}^{2}_{(\mathrm i)}=\Delta_{\mathrm i}^{2}\bm I,
\)
and we take
\(
\boldsymbol{\Lambda}_{(\mu)}=\Lambda_{\mu}\bm I
\),
\(
\boldsymbol{\Lambda}_{(\mathrm i)}=\Lambda_{\mathrm i}\bm I
\),
with $\bm{I}$ the $3\times3$ identity in Cartesian spin space.

For simplicity, we parametrize the retarded backaction of the combined environment by a
single friction scale, $\Lambda=\Lambda_{\mu}=\Lambda_{\mathrm i}$. This
choice assumes that the dominant memory-producing local reorganization is
shared by the background and ion-modulated sectors, whereas their distinct
contributions to the $\mu$SR signal are carried mainly by their field widths
and correlation times. Practically, 
this avoids introducing additional parameter
that the $\mu$SR data can meaningfully constrain.
We integrate \eqref{eq:Em_SDE} with a fixed time step $\Delta t= 10^{-3}$,
propagating $N=5\times10^{4}$ independent trajectories with initial conditions
$\bm{n}=(0,0,1)$, $\bm{u}_{(\zeta)}(0)=\bm 0$ and
$
  \boldsymbol{\xi}_{(\zeta)}(0)\sim\mathcal{N}(\bm{0},S^{2}\Delta_{\zeta}^{2}\bm{I}),
$
for $\zeta$ channel independently.
Here, $\mathcal{N}$ denotes a normal (Gaussian) distribution with the stated covariance.
If static field inhomogeneity exists, it can be included by drawing, for each trajectory, a time-independent
local field
$\bm{B}\sim\mathcal{N}(0,\Delta_{\mathrm{st}}^{2}\bm{I})$, and holding
it fixed throughout the integration; $\Delta_{\mathrm{st}}^{2}$ is the static KT width.

\subsection{Markovian limit and benchmarks}
\paragraph*{Purely dynamic ion limit}
\begin{figure}
  \centering
  \mbox{
	  \subfloat[\label{a}]{\includegraphics[scale=0.42]{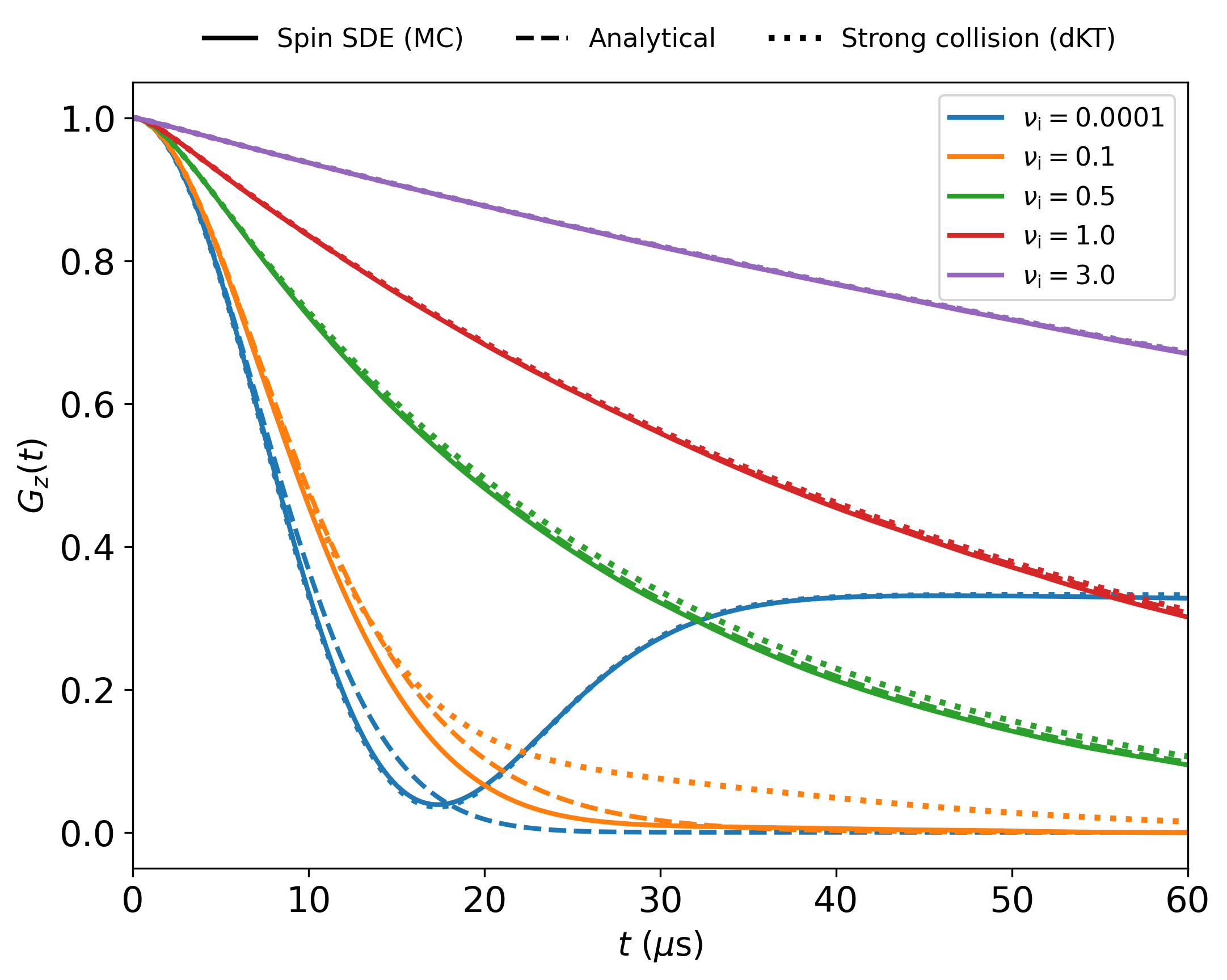}\label{fig:ZF_dKT}}\quad
	  \hspace{0.8em}%
	  \subfloat[\label{b}]{\includegraphics[scale=0.42]{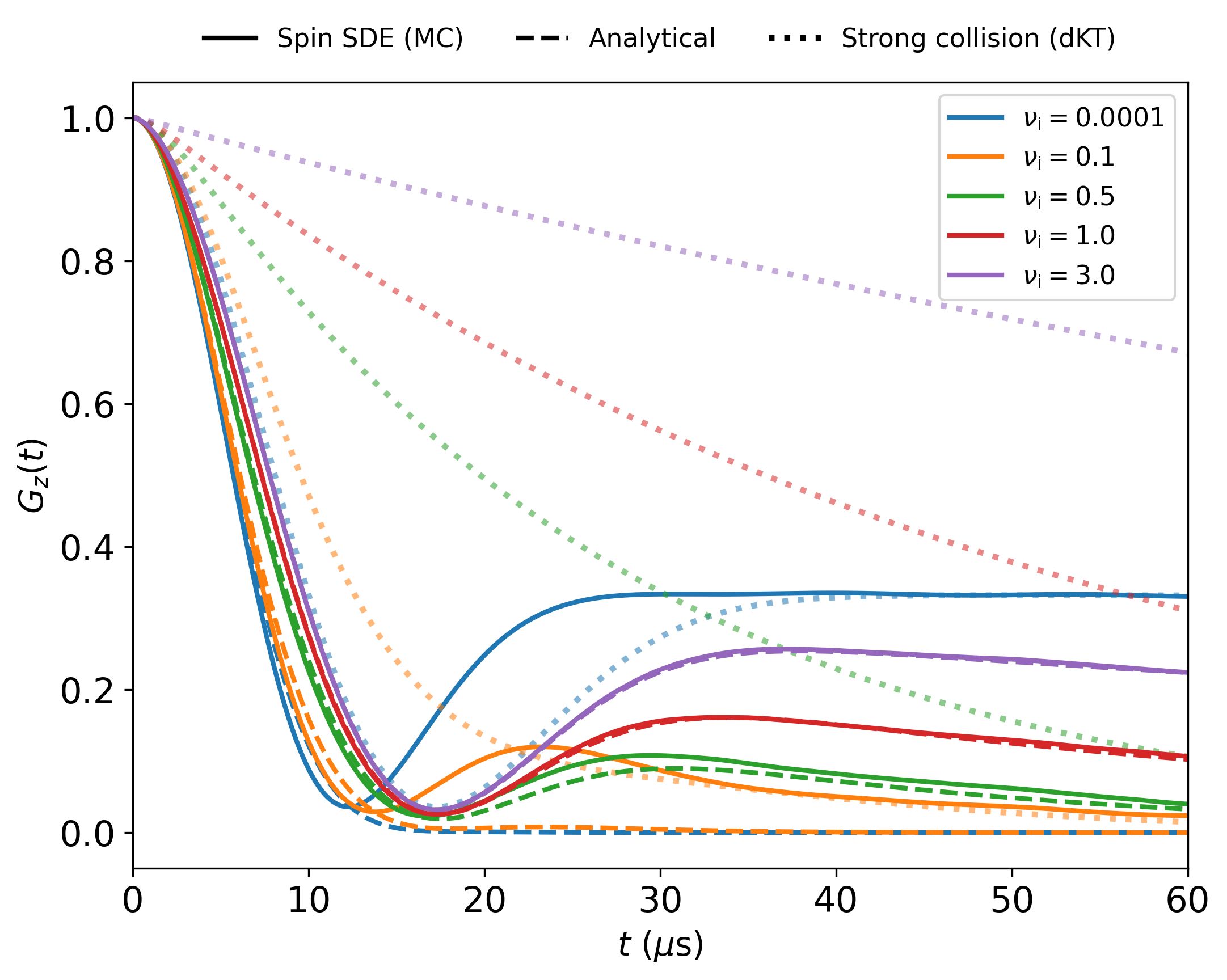}\label{fig:ZF_IK}}\quad
	
  }
  \mbox{
	  \subfloat[\label{c}]{\includegraphics[scale=0.42]{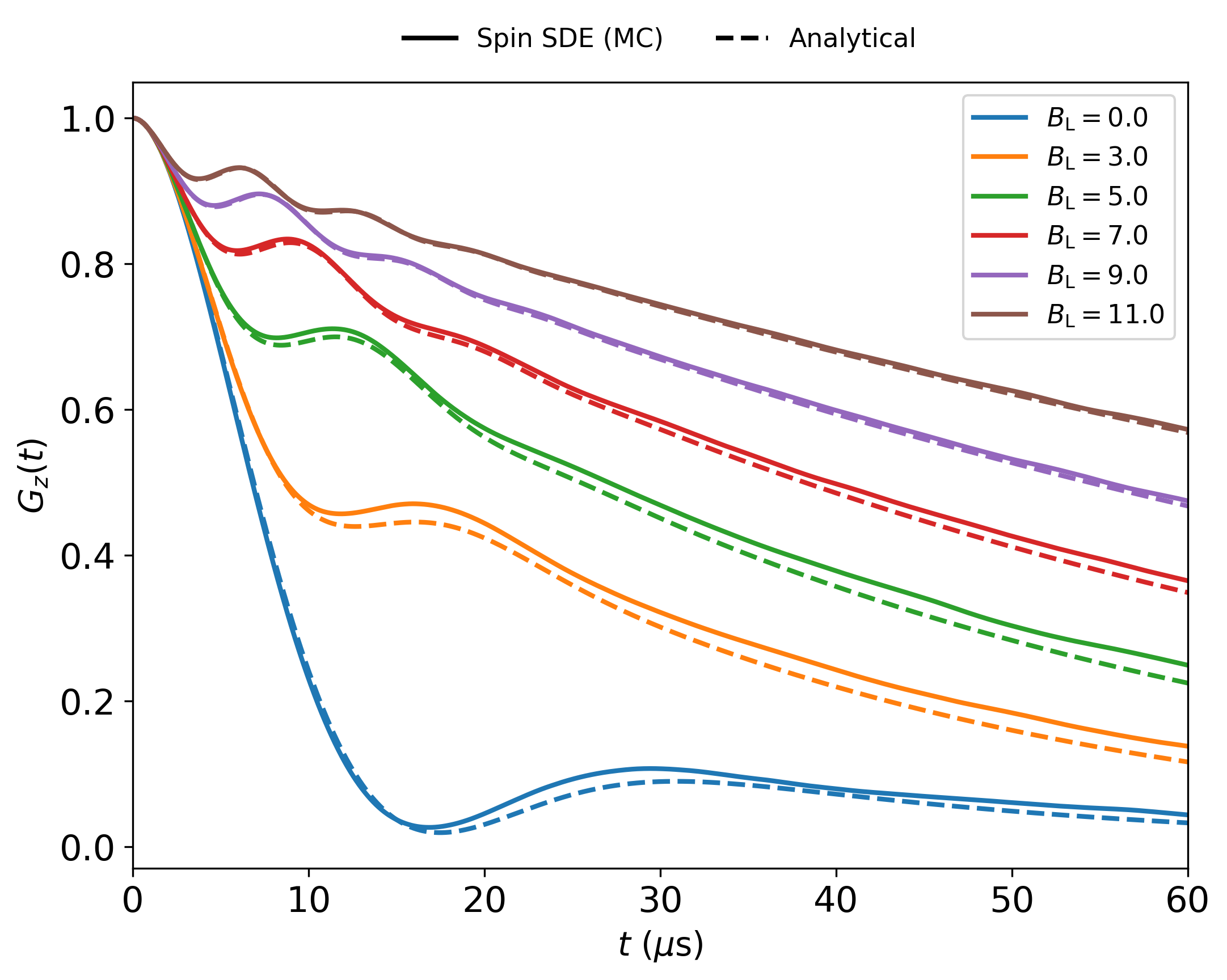}\label{fig:LF_IK}}\quad
	  \hspace{0.8em}%
	  \subfloat[\label{d}]{\includegraphics[scale=0.42]{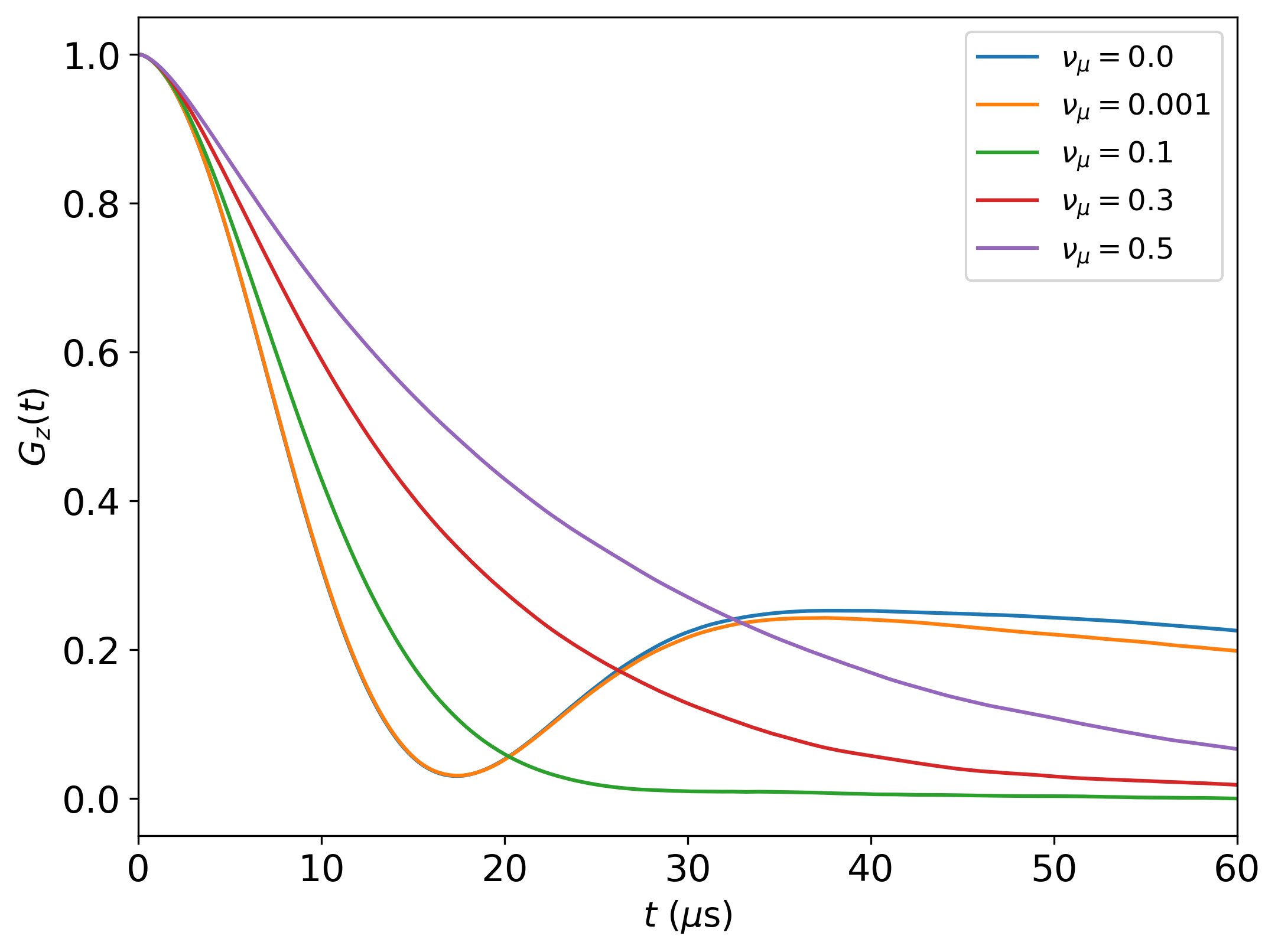}\label{fig:nu_mu_eff}}\quad
  }
  \captionsetup{justification=raggedright,singlelinecheck=false}
  \caption{Muon polarization $G_{z}(t)$ versus time for:
    (a) ZF, ion-dynamics only ($\nu_{\mu}=\Delta_{\mu}=0$);
    (b) ZF, static muon ($\nu_{\mu}=0$) with $\Delta_{\mu}=\Delta_{\mathrm i}$;
    (c) LF scans at different $B_{\mathrm L}$ (Gauss) for a static muon
        ($\nu_{\mu}=0$, $\Delta_{\mu}=\Delta_{\mathrm i}$) with
        $\nu_{\mathrm i}=0.5~\mu\mathrm{s}^{-1}$;
    (d) ZF, motional-narrowing regime with $\Delta_{\mu}=\Delta_{\mathrm i}$ and
        $\nu_{\mathrm i}=3.0~\mu\mathrm{s}^{-1}$ (see panel for the $\nu_{\mu}$ values).
	In all panels, $\Delta_{\mathrm i}=0.1~\mu\mathrm{s}^{-1}$ and $\alpha=0.0~\mu\mathrm{s}^{-2}$.}
  \label{fig:sim_res}
\end{figure}

Figure~\ref{fig:ZF_dKT} benchmarks the ion-only limit
($\nu_\mu=\Delta_\mu=0$), i.e.\ the standard dynamic
Kubo-Toyabe (dKT) problem, by comparing our spin SDE
(solid) to the strong-collision dKT formula (dotted) and to the
analytical reduction in \eqref{eq:Gz_factorized} (dashed).
In the quasi-static regime $\nu_{\mathrm i}\ll \Delta_{\mathrm i}$
(e.g.\ $\nu_{\mathrm i}=10^{-4}\,\mu\mathrm{s}^{-1}$), the fluctuating
ion field is effectively frozen over the $\mu$SR time window and the
polarization approaches the static KT line shape: a pronounced minimum
at $t\sim O(1/\Delta_{\mathrm i})$ followed by recovery to the
characteristic $1/3$ tail.
With increasing ion hopping, the minimum is progressively filled in and
the long-time tail is suppressed, reflecting the onset of longitudinal
relaxation induced by time-dependent fields.
In the fast-fluctuation (motional-narrowing) limit
$\nu_{\mathrm i}\gg \Delta_{\mathrm i}$, the field is efficiently
averaged and $G_z(t)$ becomes nearly exponential, with a relaxation rate
that scales as $\propto \Delta_{\mathrm i}^2/\nu_{\mathrm i}$
(hence weaker relaxation for faster hopping).
Over the full range of $\nu_{\mathrm i}$ shown, the spin SDE results are
in near quantitative agreement with the strong-collision dKT curve,
validating the numerical implementation in this Markovian benchmark.
By contrast, the analytical curves deviate markedly in the
quasi-static limit: it misses the static KT recovery and instead
over-depolarizes at long times.
This failure is expected because the reduction leading to
\eqref{eq:Gz_factorized} yields a dynamic factor $G_{\mathrm{dyn}}(t)$
that is independent of the static field $\bm{\xi}^{\mu}$ after Dyson series
truncation. In the current ZF and no backaction case, $G_{\mathrm{dyn}}(t)$
is essentially an Abragam-type function \eqref{eq:uu_generalized_Abragam}.
For $\nu_\mu=\Delta_\mu=0$, $G_{\mathrm{stat}}=1$ and the spectra
follow $G_{\mathrm{dyn}}(t)$ leading to over-depolarization at longer times
even when $\nu_{\mathrm i} \to 0$\cite{keren_1994}.

\paragraph*{Static muon with quenched background field}
Figure~\ref{fig:ZF_IK} benchmarks our spin SDE simulations against the
Ito-Kadono (IK) setting~\cite{Ito2024} for a \emph{static} muon
($\nu_\mu=0$) with comparable static and ion-modulated second moments,
$\Delta_\mu=\Delta_{\mathrm i}=0.1~\mu\mathrm{s}^{-1}$.
This corresponds to the intermediate case $Q=1/2$ in the IK
parameterization, where $Q=0$ and $Q=1$ denote purely static and purely
dynamic limits, respectively.
For reference, we also display the strong-collision dKT curves (faded
dotted) evaluated in the dynamic ion-only limit. Since
$\Delta_\mu\neq 0$ here, these curves are not meant as a quantitative
comparison but simply indicate the behavior of the purely dynamic
component.
For any finite $\nu_{\mathrm i}$, ion motion generates additional
longitudinal relaxation and progressively suppresses the $1/3$ tail.
The long-time behavior is non-monotonic in $\nu_{\mathrm i}$.
At intermediate hopping the relaxation is strongest, while in the
fast-hopping regime $\nu_{\mathrm i}\gg \Delta_{\mathrm i}$ the
ion-induced relaxation is motional-narrowed
so the decay slows down and $G_z(t)$ approaches the static-KT envelope
set by $\Delta_\mu$.
In this regime (roughly $\nu_{\mathrm i}\gtrsim \Delta_{\mathrm i}$),
the analytical function tracks the spin SDE semi-quantitatively,
capturing the crossover to a weak, nearly exponential long-time decay.
In the quasi-static limit, the analytical curves again
exhibit pronounced over-depolarization within the simulated time window, in clear
disagreement with the spin-SDE results. As in the previous case, this reflects the
breakdown of the approximations underlying the factorized form $G_z\simeq
G_{\mathrm{stat}}G_{\mathrm{dyn}}$. Although $\Delta_\mu$ is finite, the factorized
ansatz assigns the dominant long-time decay to the Abragam-like dephasing factor
$G_{\mathrm{dyn}}(t)$. When $\nu_\mathrm{i}\ll\Delta_\mu$ this dephasing becomes excessively
strong and drives $G_z(t)$ to decay much faster than observed, rather than approaching
the correct quasi-static KT envelope.

\paragraph*{Longitudinal field effects when $\Delta_{\mu}\neq 0$}

The LF response by varying the applied field
$B_{\mathrm L}$ in the same IK setting ($Q=1/2$) at fixed
$\nu_{\mathrm i}=0.5~\mu\mathrm{s}^{-1}$  is shown in Figure~\ref{fig:LF_IK}.
Solid and dashed curves again denote the spin SDE simulations and
analytical functions, respectively.
Increasing $B_{\mathrm L}$ progressively decouples the muon polarization
from the transverse quasi-static components. This is characterized by the reduction 
of the KT-like and increase of the long-time polarization.
The monotonic increase of the late-time polarization with
$B_{\mathrm L}$ and the suppression of the ZF minimum reproduce the
qualitative LF trends reported by IK for the intermediate-$Q$ case.
Within the Gaussian phase-diffusion closure, $B_{\mathrm L}$ enters not only via
$\omega_0$ in $G_{\mathrm{stat}}(t)$ but also through
$\omega_{\mathrm{eff}}=\omega_0+\omega_z$ in the dressed kernel and dressed
noise correlator. In other words, as an $\omega_{\mathrm{eff}}$-dependent shift of the
poles of the transverse Green's function, which produces LF decoupling.
Across the full field range shown, the analytical function captures
the monotonic decoupling trend and reproduces the overall magnitude and
field dependence of the long-time polarization.
Residual discrepancies with spin SDE are most visible at low fields (notably near ZF),
where quasi-static components and the neglected $\Phi_{z}$ dressing are
expected to be most important. At larger $B_{\mathrm L}$ the agreement
improves as the external field stabilizes the longitudinal axis and
reduces sensitivity to the transverse quasi-static distribution.

\paragraph*{Muon hopping effects}
Finally, \figref{fig:nu_mu_eff} illustrates the effect of muon hopping on the ZF
polarization in the fast-ion regime ($\nu_{\mathrm i}\gg \Delta_{\mathrm i}$).
Keeping $\Delta_{\mu}=\Delta_{\mathrm i}=0.1~\mu\mathrm{s}^{-1}$ and
$\nu_{\mathrm i}=3.0~\mu\mathrm{s}^{-1}$ fixed, we vary the muon hopping rate
$\nu_{\mu}$.
For $\nu_{\mu}\approx 0$ the muon background component is effectively static,
leading to a partial recovery at long times, while the ion-modulated component
produces only weak additional relaxation due to motional narrowing.
As $\nu_{\mu}$ becomes comparable to the field scale ($\nu_{\mu}\sim\Delta$),
the static component is converted into a fluctuating field with correlation
time $\tau_{\mu}\sim 1/\nu_{\mu}$, which strongly suppresses the KT tail recovery.
Upon further increasing $\nu_{\mu}$, the muon motion itself enters the
motional-narrowing regime and the initial relaxation becomes slower again,
consistent with the reduction of the effective relaxation rate in the fast
fluctuation limit.
Overall, muon diffusion removes the residual static contribution and drives
$G_{z}(t)$ toward complete depolarization at long times, with the strongest
relaxation occurring at intermediate $\nu_{\mu}$.
This implies that if muon hopping becomes appreciable at high $T$, our model predicts that it will
further shorten the field correlation time and thus modify the ZF relaxation in
a way that is qualitatively similar to motional narrowing. We emphasize however that, this trend
is not unique to muon diffusion and can also result from other fast fluctuating
field sources.

\subsection{Non-Markovian backaction and memory effects}
\label{subsec:alpha_results}

\begin{figure}
  \centering
  \mbox{
	  \subfloat[\label{a}]{\includegraphics[scale=0.42]{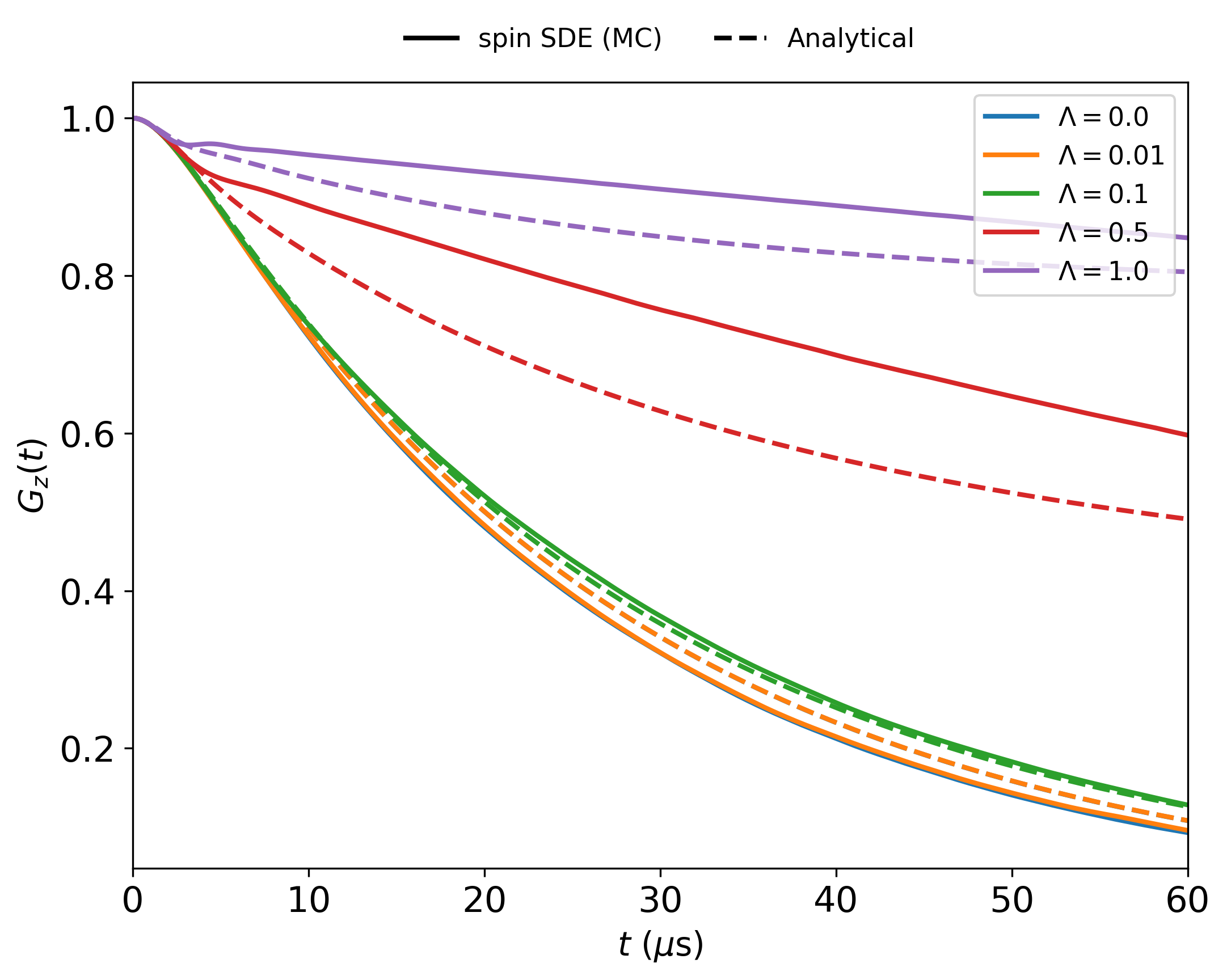}\label{fig:Lambda_nui_05_Dmu_0}}\quad
	  \hspace{0.8em}%
	  \subfloat[\label{b}]{\includegraphics[scale=0.42]{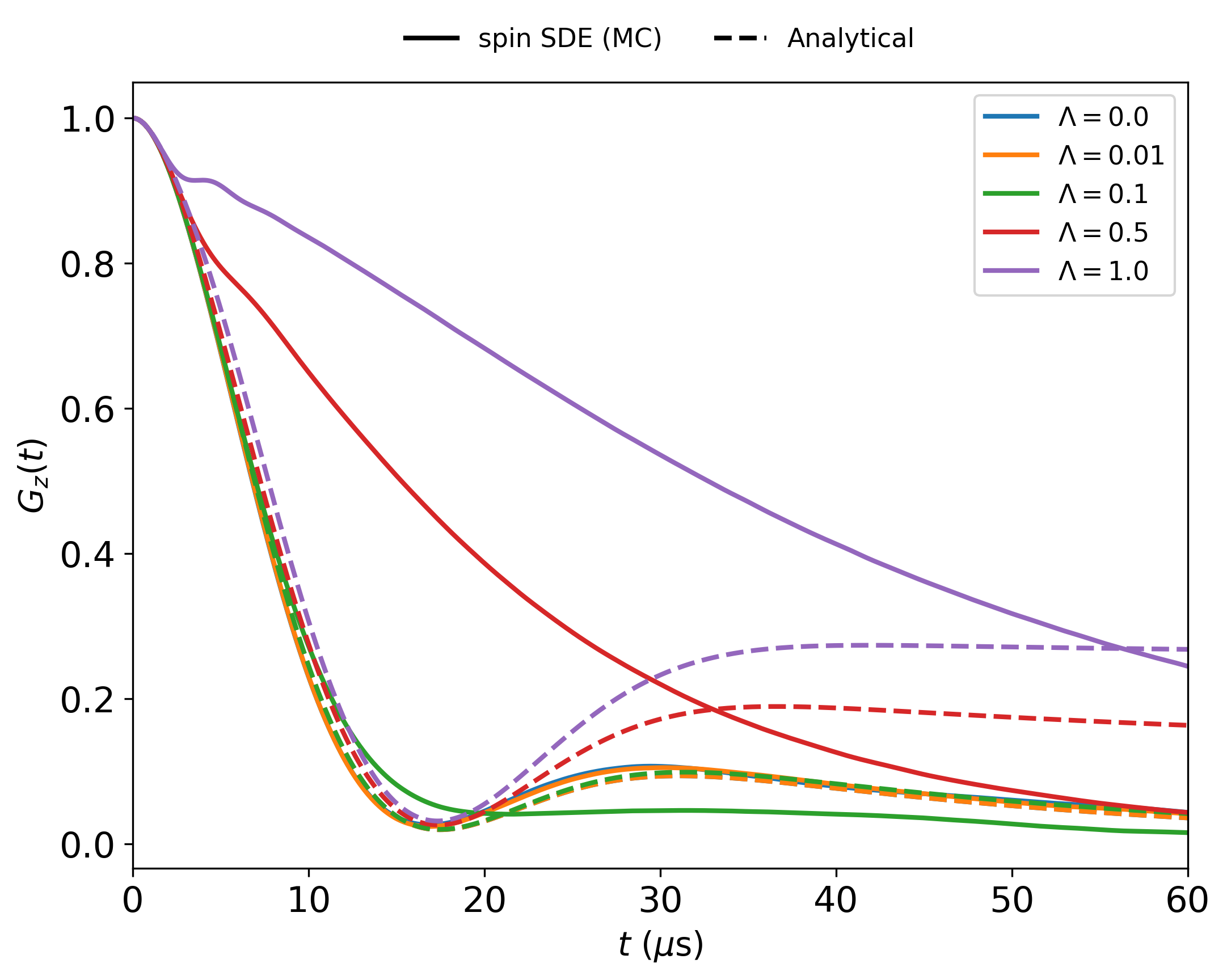}\label{fig:Lambda_nui_05_Dmu_01}}\quad
  }
  \mbox{
	  \subfloat[\label{c}]{\includegraphics[scale=0.42]{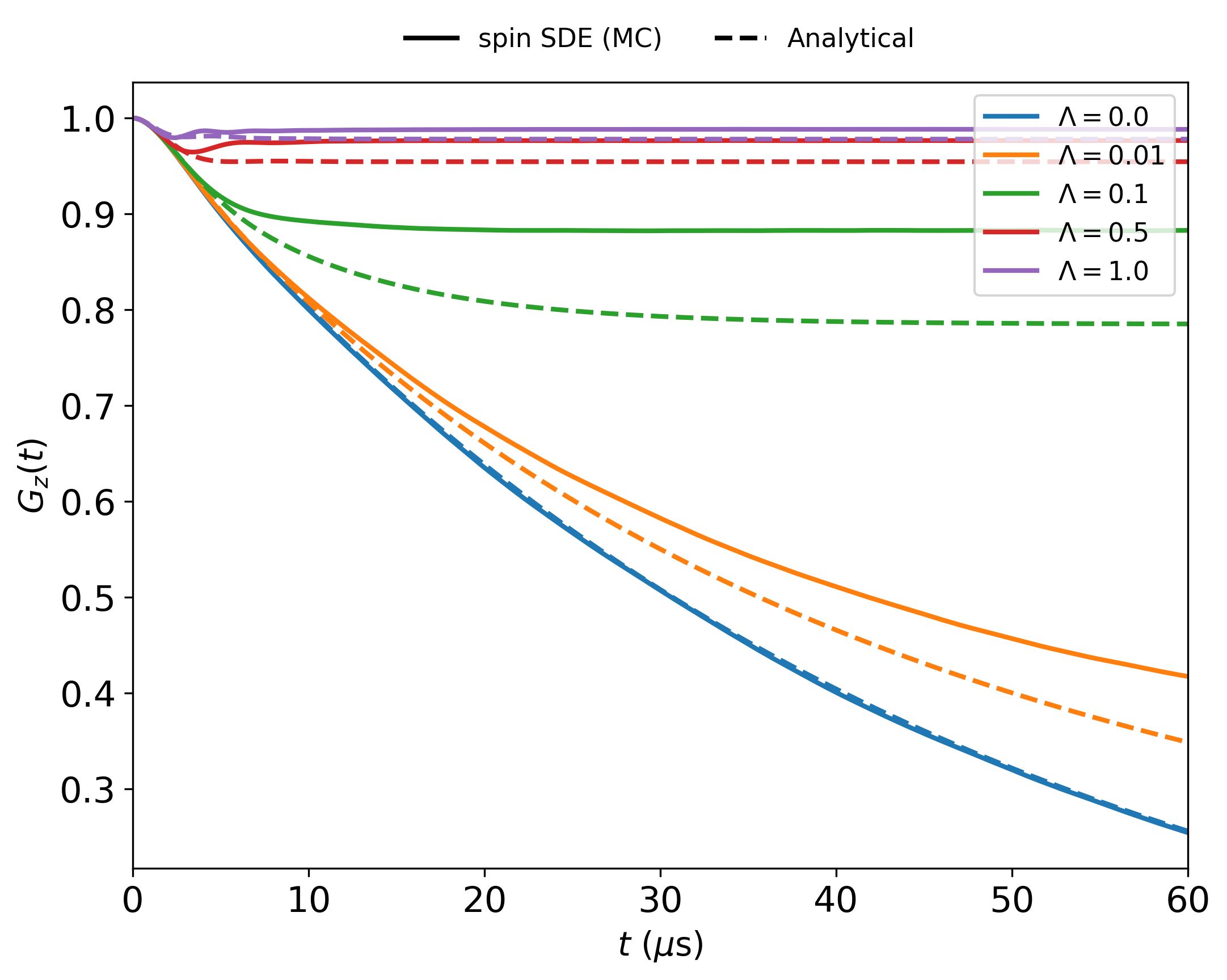}\label{fig:Lambda_LF_memory}}\quad
	  \hspace{0.8em}%
	  \subfloat[\label{d}]{\includegraphics[scale=0.42]{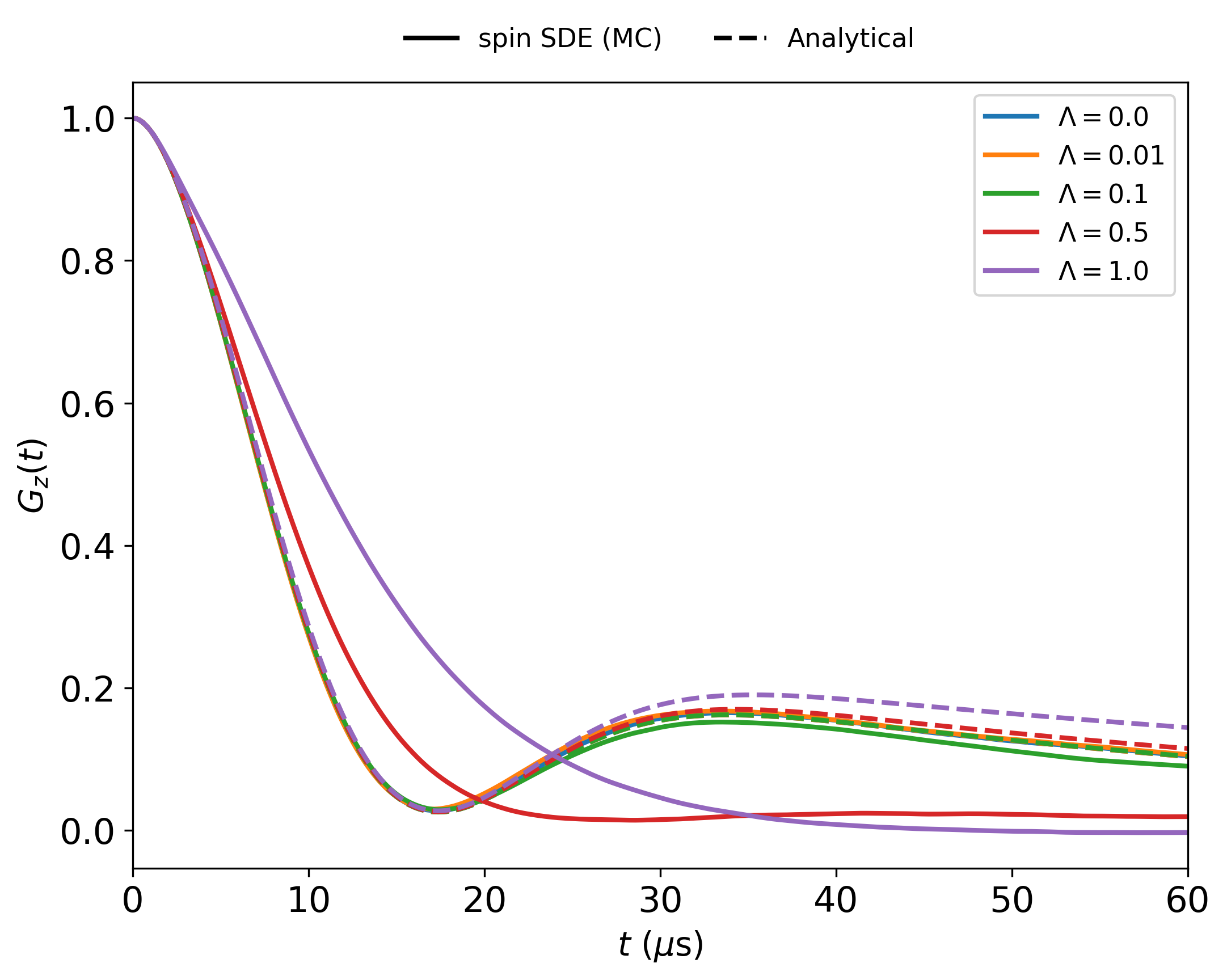}\label{fig:Lambda_ZF_mn}}\quad
  }
  \captionsetup{justification=raggedright,singlelinecheck=false}
  \caption{
	  Effect of the backaction (retarded-torque) strength $\Lambda$ (in units of $\mu s^{-2}$) on the longitudinal
polarization $G_z(t)$ computed from the spin SDE Monte Carlo (solid curves).
Dashed curves show the analytical reduction.
All panels use an ion-modulated width $\Delta_{\mathrm{i}}=0.1~\mu\mathrm{s}^{-1}$, and
$\Lambda$ is varied as indicated in the legend.
(a) Purely dynamical limit in ZF: $\Delta_\mu=0$, with an
intermediate fluctuation rate $\nu_{\mathrm{Li}}=0.5~\mu\mathrm{s}^{-1}$.
(b) Mixed static-dynamic case in ZF: $\Delta_\mu=0.1~\mu\mathrm{s}^{-1}$ and
$\nu_{\mathrm{Li}}=0.5~\mu\mathrm{s}^{-1}$.
(c) Longitudinal-field response in the dynamical limit: $B_L=5$~G, $\Delta_\mu=0$, and
$\nu_{\mathrm{Li}}=0.5~\mu\mathrm{s}^{-1}$.
(d) Motional-narrowing regime in ZF with a quenched component:
$\Delta_\mu=\Delta_{\mathrm{Li}}=0.1~\mu\mathrm{s}^{-1}$ and
$\nu_{\mathrm{Li}}=1.0~\mu\mathrm{s}^{-1}$.
}
  \label{fig:alpha_eff}
\end{figure}

\paragraph*{Intermediate $\nu_{\mathrm i}$ with ZF}
Figures~\ref{fig:Lambda_nui_05_Dmu_0} and \ref{fig:Lambda_nui_05_Dmu_01}
compare the spin SDE results with the
analytical reduction at fixed intermediate
$\nu_{\mathrm i}=0.5~\mu\mathrm{s}^{-1}$ value, for purely dynamic regime ($\Delta_\mu=0$) and 
static--dynamic fields interplay ($\Delta_\mu=\Delta_{\mathrm i}=0.1~\mu\mathrm{s}^{-1}$).
When $\Delta_\mu=0$, the muon does not experience an independent static field
distribution and all relaxation originates from the dynamical ion-fluctuation
bath and the memory/backaction channel.
In this regime the analytical model and the spin SDE agree quantitatively for
small $\Lambda$ ($\Lambda\lesssim 0.1~\mu\mathrm{s}^{-2}$), both in the 
early-time dip and in the long-time tail.
As $\Lambda$ increases, the spin SDE exhibits a clear motional-narrowing-like
trend. The depolarization is progressively suppressed and $G_z(t)$ remains closer
to unity over the entire time window.
The analytical curves capture the qualitative stabilization for moderate
$\Lambda$, but for large $\Lambda$ they underestimate the degree of narrowing.
This deviation reflects the breakdown of the averaged-kernel closure when the
longitudinal phase dressing becomes strong and dynamically correlated with the
transverse mode.
For a finite $\Delta_\mu$ 
the polarization now exhibits the characteristic early-time
static dephasing associated with a Gaussian field distribution 
even at $\Lambda=0$. In the full spin SDE, increasing $\Lambda$ not only suppresses the dynamical
relaxation channel but also partially averages the effective static field
distribution along the stochastic trajectory.
As a result, for $\Lambda\gtrsim 0.5$ the spin SDE curves show a dramatic stabilization.
The long-time decay is substantially reduced compared to the $\Lambda=0$
reference, and the overall envelope is markedly flatter.
By contrast, the analytical reduction displays a much weaker $\Lambda$-dependence.
Although it reproduces the small-$\Lambda$ behavior well, it retains a pronounced
KT-like recovery structure and fails to capture the strong dynamical averaging
observed in the spin SDE at large $\Lambda$.
The effective 
factorization of $G_z(t)$ into static and dynamic components
with further approximate closure, neglects these
trajectory-dependent correlations.
Consequently, it cannot fully describe the $\Lambda$-induced dynamical narrowing
of the effective static width.

\paragraph*{LF response with intermediate $\nu_{\mathrm i}$}
Figure~\ref{fig:Lambda_LF_memory} illustrates the LF response in the
ion-only setting ($\Delta_\mu=\nu_\mu=0$) at an intermediate hopping rate
$\nu_{\mathrm i}=0.5~\mu\mathrm{s}^{-1}$ with $B_{\mathrm L}=5$~G and
$\Delta_{\mathrm i}=0.1~\mu\mathrm{s}^{-1}$.
The spin SDE curves exhibit a clear memory-induced
decoupling characterized by the suppression of the long-time relaxation 
and high-polarization plateau with 
increasing the retarded-torque strength $\Lambda$.
The analytical reduction reproduces this trend and is quantitatively
accurate in the weak-$\Lambda$ regime, where the retarded torque
acts as a perturbative correction and the self-averaged kernel closure
is satisfactory. At larger $\Lambda$ the self-averaged $\Phi_z$ dressing becomes uncontrolled and the
Gaussian closure misses nonlinear feedback between retarded torque and fluctuations,
so the analytical model over-decouples compared with spin SDE.

\paragraph*{Motional narrowing $\nu_{\mathrm i}$ with ZF}
In the intermediate and motional-narrowing regimes ($\nu_{\mathrm i}\gtrsim\Delta_{\mathrm i}$)
shown in Figures~\ref{fig:ZF_dKT} and \ref{fig:ZF_IK},
the analytic reduction provides a useful closed-form interpolation; however, its
$\Lambda$-dependence is not uniformly controlled. This behavior is depicted in \figref{fig:Lambda_ZF_mn}.
For small enough $\Lambda$ that the retarded
torque produces only a perturbative correction over the bath correlation time,
the dashed curves closely track the spin SDE across the full time
window. It reproduces both the early-time KT-like depolarization and the subsequent
crossover to the dynamical tail.
In this regime the small-angle linearization and Gaussian (second-cumulant)
resummation for the transverse mode remain valid, and the self-averaged kernel
closure introduces only linewidth-level changes.
At larger $\Lambda$, however, the agreement becomes progressively poorer: the full
spin SDE exhibits clear motional-narrowing behavior
driven by nonlinear feedback between retarded torque and fluctuations, whereas the
analytical function captures only part of the resulting $\Lambda$-dependence.
This limitation becomes apparent when $\Delta_\mu\neq 0$ in which the static background
field couples to the dynamical response through rotation dressing of both memory and
noise kernels, and joint averages no longer factorize.
Consequently, the analytical curves can underestimate the $\Lambda$-induced narrowing
seen in spin SDE simulations, even when $\nu_{\mathrm i}\gtrsim\Delta_{\mathrm i}$.

\paragraph*{Quasi-static $\nu_{\mathrm i}$ with ZF}
Similarly as already shown in Figures~\ref{fig:ZF_dKT} and \ref{fig:ZF_IK},
in the quasi-static regime $\nu_{\mathrm i}\ll\Delta_{\mathrm i}$,
the ion field is effectively frozen over the experimental time
window, so the polarization is governed by a \emph{static} field average rather
than by a motional-narrowing cumulant.
The spin SDE curves therefore approach the static KT form (dip and recovery to
a long-time plateau) determined by the combined quasi-static width,
$\Delta_{\rm eff}=\sqrt{\Delta_\mu^2+\Delta_{\mathrm i}^2}$.
the dependence on
$\Lambda$ manifests primarily through coherent precession/stiffening effects
of the retarded torque rather than through a simple exponential relaxation rate.
In contrast, the analytical reduction through its simplifications,
becomes sensitive to the specific
(realization-dependent) static field configuration leading to
over-depolarization and incorrect long-time behavior compared to spin SDE, and the
$\Lambda$-dependence is not captured reliably.

These comparisons define the regime of validity of the analytical reduction.
It provides a fast, transparent, and quantitatively accurate description from 
intermediate fluctuation rates into the motional-narrowing regime when the 
spin–bath coupling is weak to moderate (small $\Lambda$). When $\Lambda$ is large 
and static broadening becomes dynamically entangled with the fluctuating bath,
the averaged-kernel closure breaks down and the full spin SDE is the controlled reference.
\begin{figure}[t]
  \centering
  \mbox{
    \subfloat[\label{fig:mem_nu_3}]{
      \includegraphics[width=0.32\linewidth]{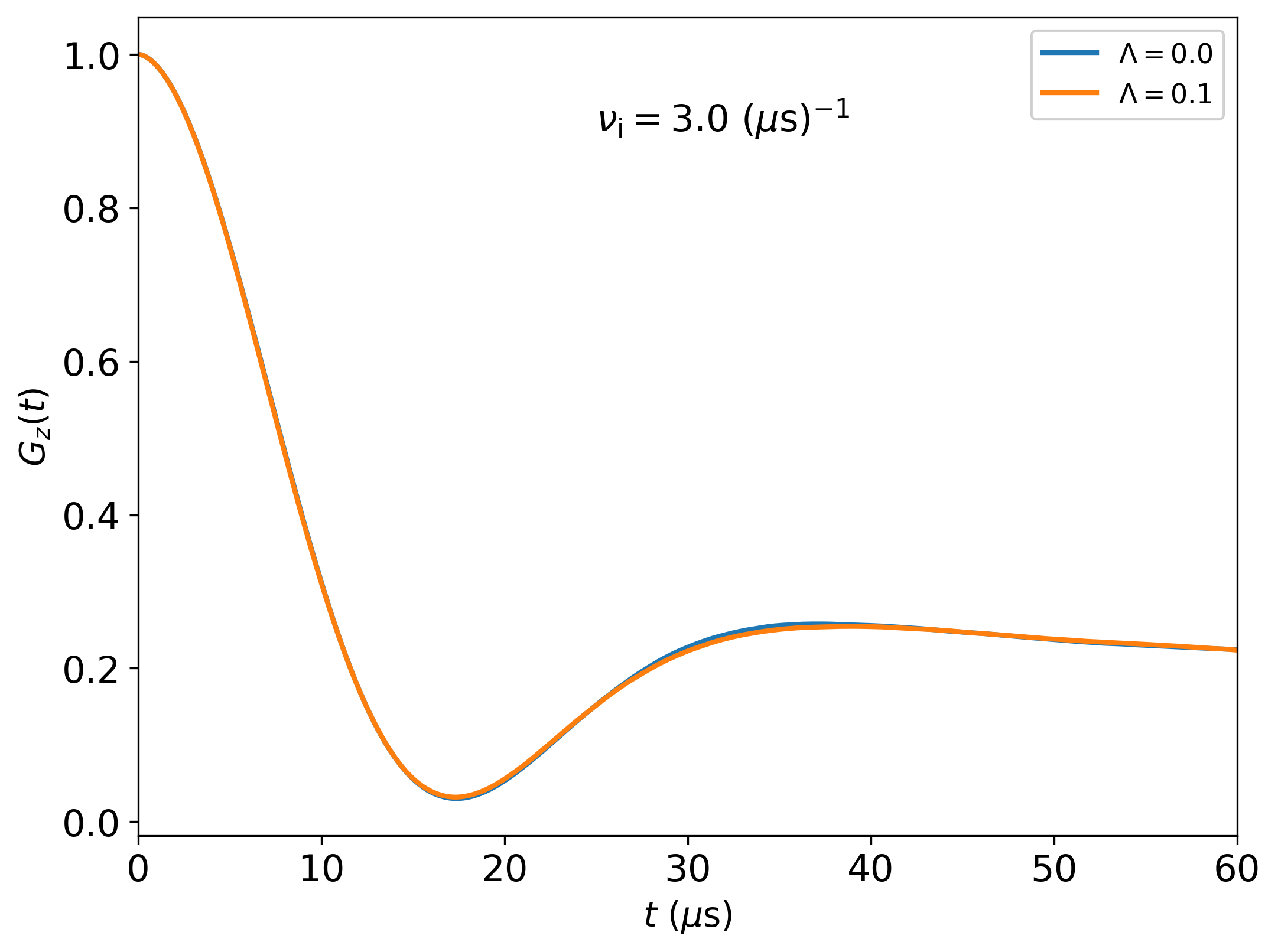}
    }\hspace{0.8em}
    \subfloat[\label{fig:mem_nu_05}]{
      \includegraphics[width=0.32\linewidth]{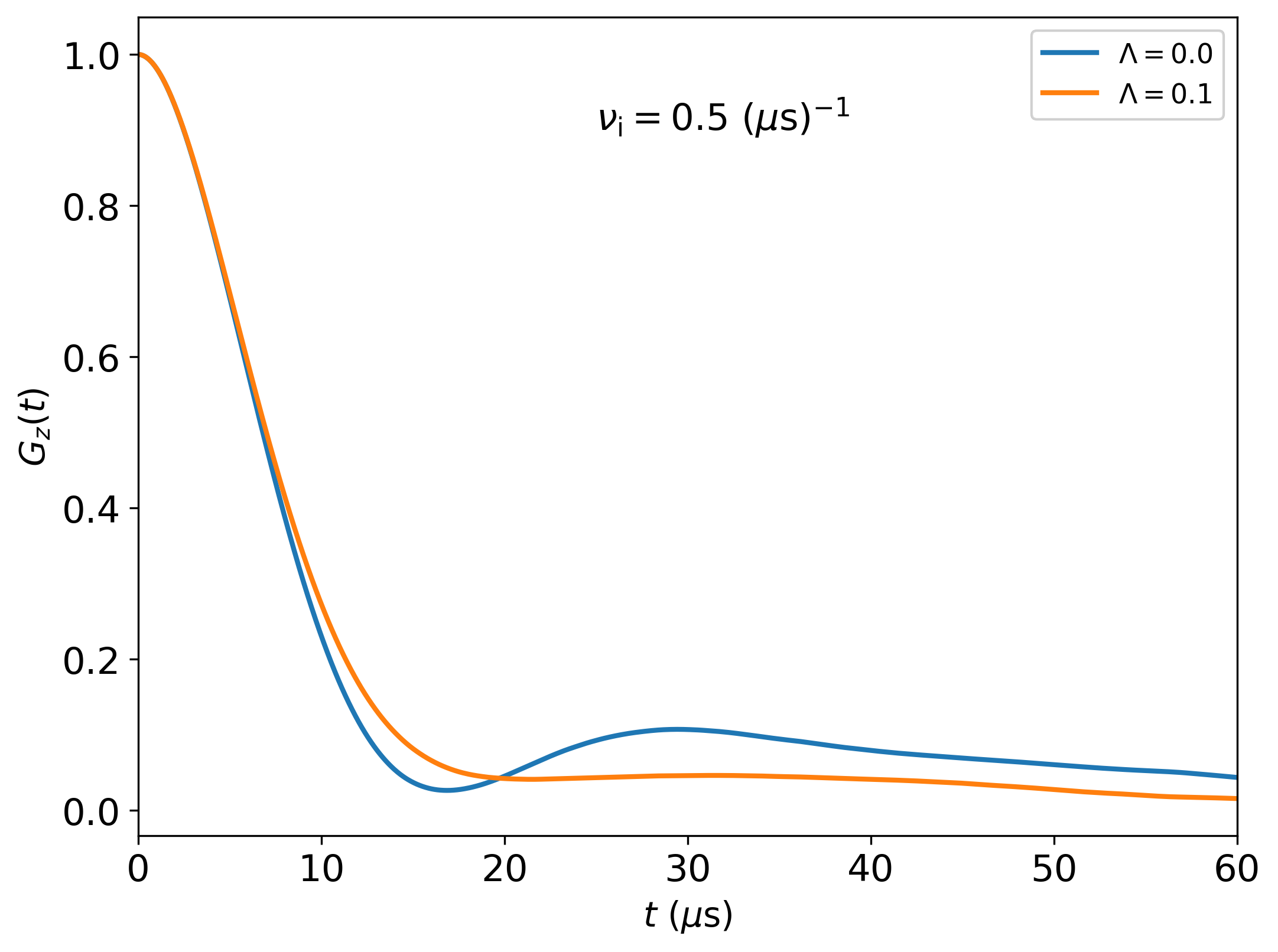}
    }\hspace{0.8em}
    \subfloat[\label{fig:mem_nu_1e4}]{
      \includegraphics[width=0.32\linewidth]{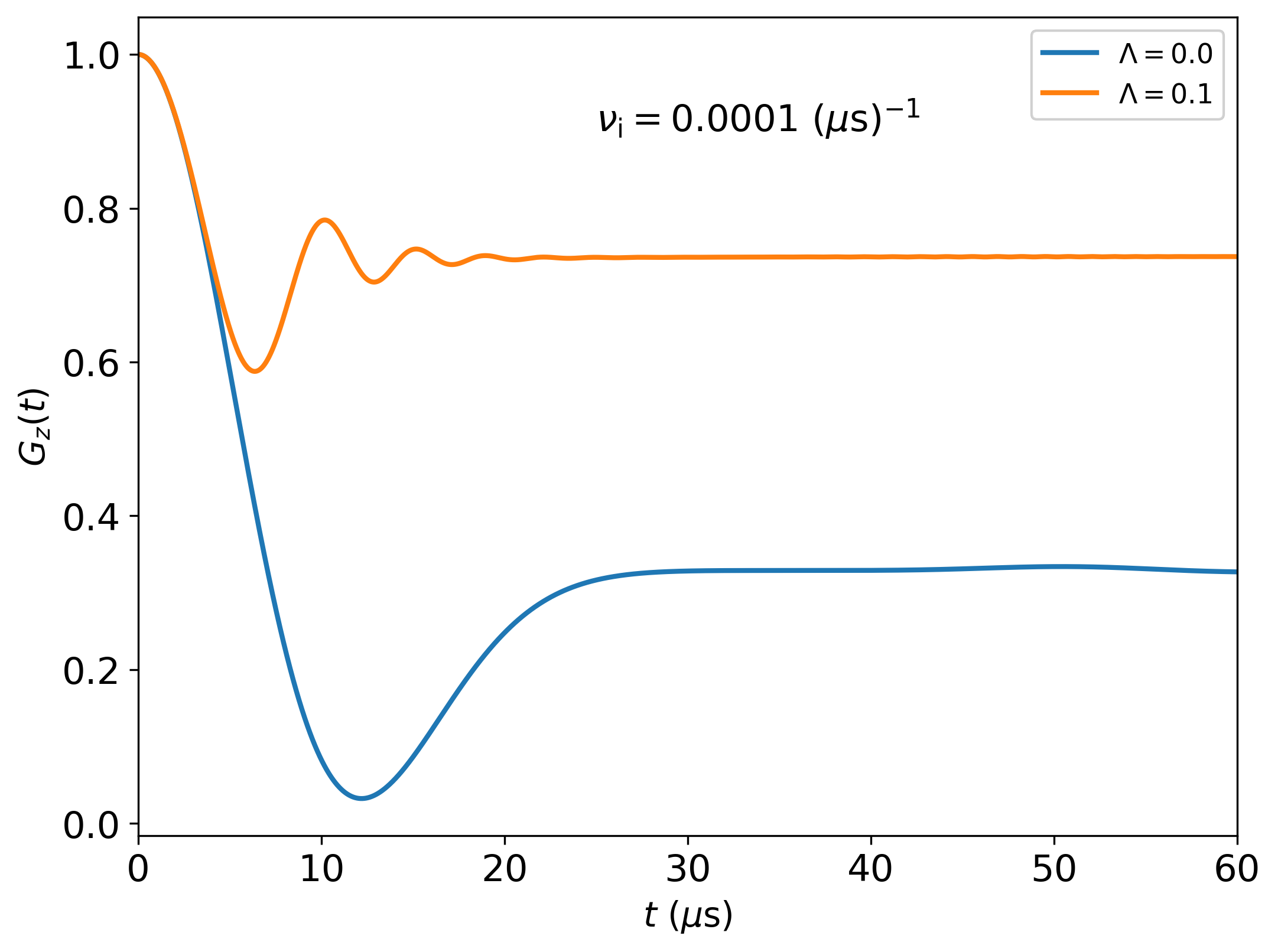}
    }
  }
  \captionsetup{justification=raggedright,singlelinecheck=false}
  \caption{Sensitivity of the spin-SDE polarization to the retarded-torque
	  strength $\Lambda$ (in units of $\mu s^{-2}$) across fluctuation regimes set by the ion correlation rate
  $\nu_{\mathrm i}$ (MC results only).
  Panels (a)--(c) compare $G_z(t)$ for $\Lambda=0$ and $\Lambda=0.1$ at fixed
  $\Delta_{\mathrm i}$ while varying $\nu_{\mathrm i}$:
  (a) fast-fluctuation (short-memory) regime, $\nu_{\mathrm i}=3.0~\mu\mathrm{s}^{-1}$;
  (b) crossover regime, $\nu_{\mathrm i}=0.5~\mu\mathrm{s}^{-1}$;
  (c) quasi-static (long-memory) regime, $\nu_{\mathrm i}=10^{-4}~\mu\mathrm{s}^{-1}$.
  The impact of $\Lambda$ is non-monotonic in $\nu_{\mathrm i}$: it is weak when the
  bath is fast (a), most visible in the intermediate crossover regime (b), and can
  produce qualitatively different behavior when the bath is effectively frozen
  over the $\mu$SR window (c).}
  \label{fig:mem_nu_scan}
\end{figure}

Finally, Figure~\ref{fig:mem_nu_scan} illustrates that the influence of the retarded-torque
parameter $\Lambda$ depends strongly on the fluctuation regime set by
$\nu_{\mathrm i}$, even when $\Lambda$ and $\Delta_{\mathrm i}$ are held fixed.
This behavior can be understood directly from the Markovian embedding of the
exponential kernel ($\bm{u}_{\mathrm i}(t)\to \bm{u}(t)$),
\begin{equation}
	\bm{u}(t)\equiv \int_{0}^{t}ds\
	e^{-\nu_{\mathrm i}(t-s)}\bm{n}(s),
  \qquad
  \dot{\bm u}=-\nu_{\mathrm i}\bm u+\bm n,
\end{equation}
for which the backaction torque enters as an effective field
$\propto \Lambda\,\bm u(t)$.

In the fast-fluctuation regime (\figref{fig:mem_nu_3}), $\nu_{\mathrm i}$ is large
enough that $\bm u(t)$ closely follows the instantaneous spin direction,
$\bm u(t)\simeq \bm n(t)/\nu_{\mathrm i}$ (up to corrections of $\mathcal{O}(1/\nu_{\mathrm i}^{2})$). 
For an isotropic kernel this leading contribution is
nearly parallel to $\bm n(t)$ and therefore produces little torque,
$\bm n\times(\Lambda\,\bm u)\approx 0$.  Consequently, $G_z(t)$ shows only a
very weak dependence on $\Lambda$ in this limit, consistent with the near-overlap of
the $\Lambda=0$ and $\Lambda=0.1$ curves.
At intermediate $\nu_{\mathrm i}$ (\figref{fig:mem_nu_05}), $\bm u(t)$ is no
longer simply proportional to $\bm n(t)$, and instead represents a weighted
running average of the spin direction over a finite recent time interval of order
$\mathcal{O}(1/\nu_{\mathrm i})$.  In this regime the vector $\bm u(t)$ typically develops a
component not parallel to $\bm n(t)$, so the retarded-torque term generates a
nontrivial additional precession/damping contribution.  As a result, switching on
$\Lambda$ visibly reshapes the late-time relaxation of $G_z(t)$, making $\Lambda$
most identifiable in practice in this crossover regime.
In the quasi-static regime (\figref{fig:mem_nu_1e4}), $\nu_{\mathrm i}$ is so small
that the exponential weight is essentially unity over the $\mu$SR window and
$\bm u(t)\approx \int_0^t ds\,\bm n(s)$.  The backaction term then depends on
the accumulated history of the spin direction and can generate qualitatively
distinct line shapes (including high-$G_z$ plateaus and oscillatory structure) even
for moderate $\Lambda$.  This implies that in $\mu$SR fitting, the low-$T$ spectra in this regime can place strong
constraints on $\Lambda$ because the effect of turning on $\Lambda$ is highly visible.

\section{Application to \texorpdfstring{$\mathrm{LiCoO_2}$}{LiCoO2}}

We now apply our generalized-kernel formalism to Li-ion diffusion in the
prototypical layered cathode $\mathrm{LiCoO_{2}}$.  Sugiyama \emph{et al.}\
\cite{Sugiyama2009} studied Li-ion dynamics in $\mathrm{Li_{x}CoO_{2}}$
($x=0.75$ and $0.50$) by ZF and LF $\mu$SR.  Their spectra were analyzed
using the dynamic Kubo--Toyabe (dKT) function, from which the temperature
dependence of the effective field width $\Delta(T)$ and fluctuation rate
$\nu(T)$ was extracted up to $T\simeq 400~\mathrm{K}$.
To enable efficient fitting over the
full parameter space, we precomputed a multidimensional numerical table
\[
  G_{z}(\Delta_{\mu},\Delta_{\mathrm i},\nu_{\mu},\nu_{\mathrm i},\alpha,B_{\mathrm L};t)
\]
by numerically integrating the spin SDE 
\eqref{eq:LLG} over a grid of parameters and interpolating between grid
points during fitting.
\begin{figure}
	\centering
  \mbox{
    \subfloat[\label{a}]{\includegraphics[scale=0.42]{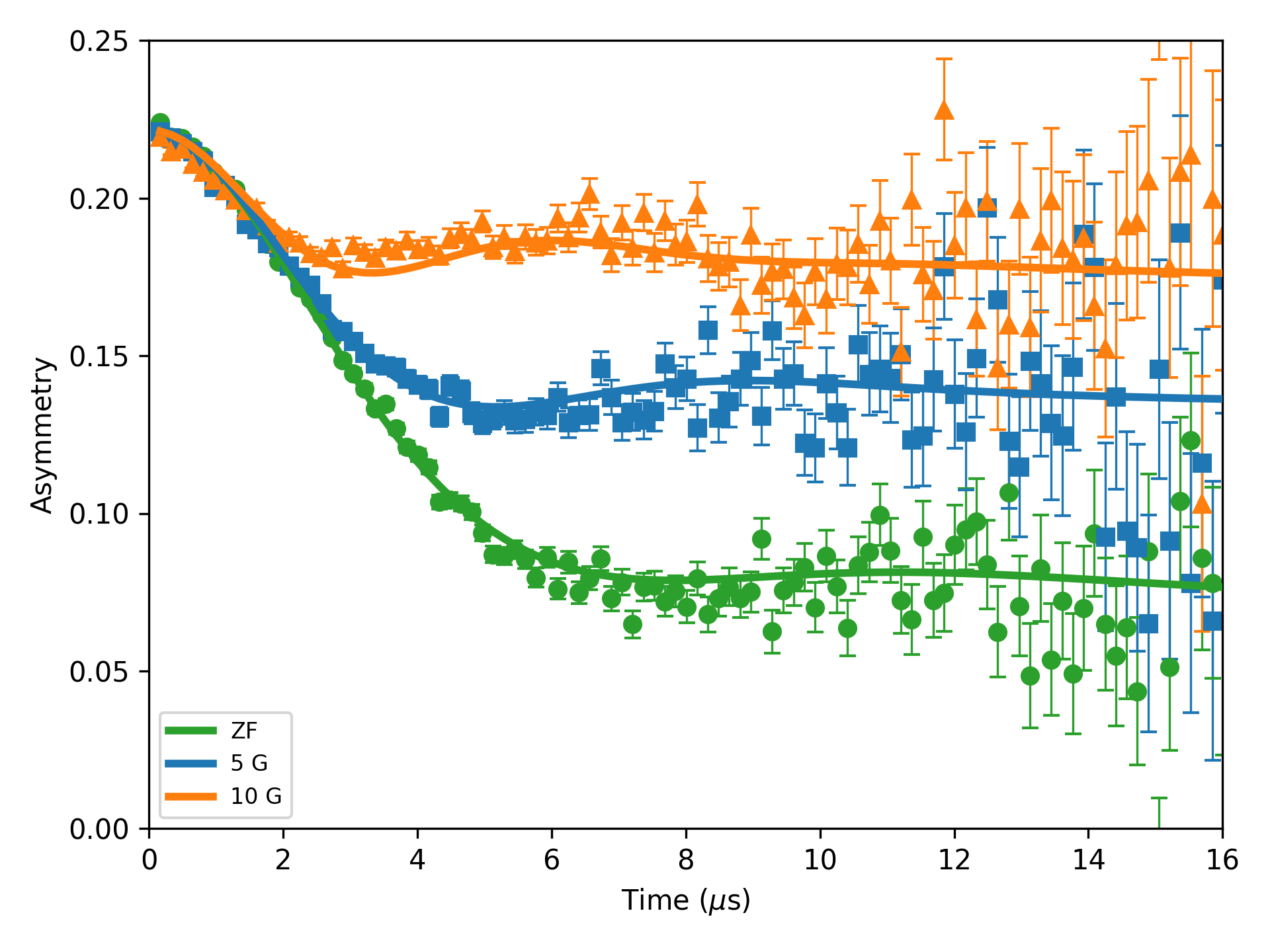}\label{fig:fit_G_T100K}}\quad
    \hspace{0.8em}%
    \subfloat[\label{b}]{
      \begin{overpic}[scale=0.50]{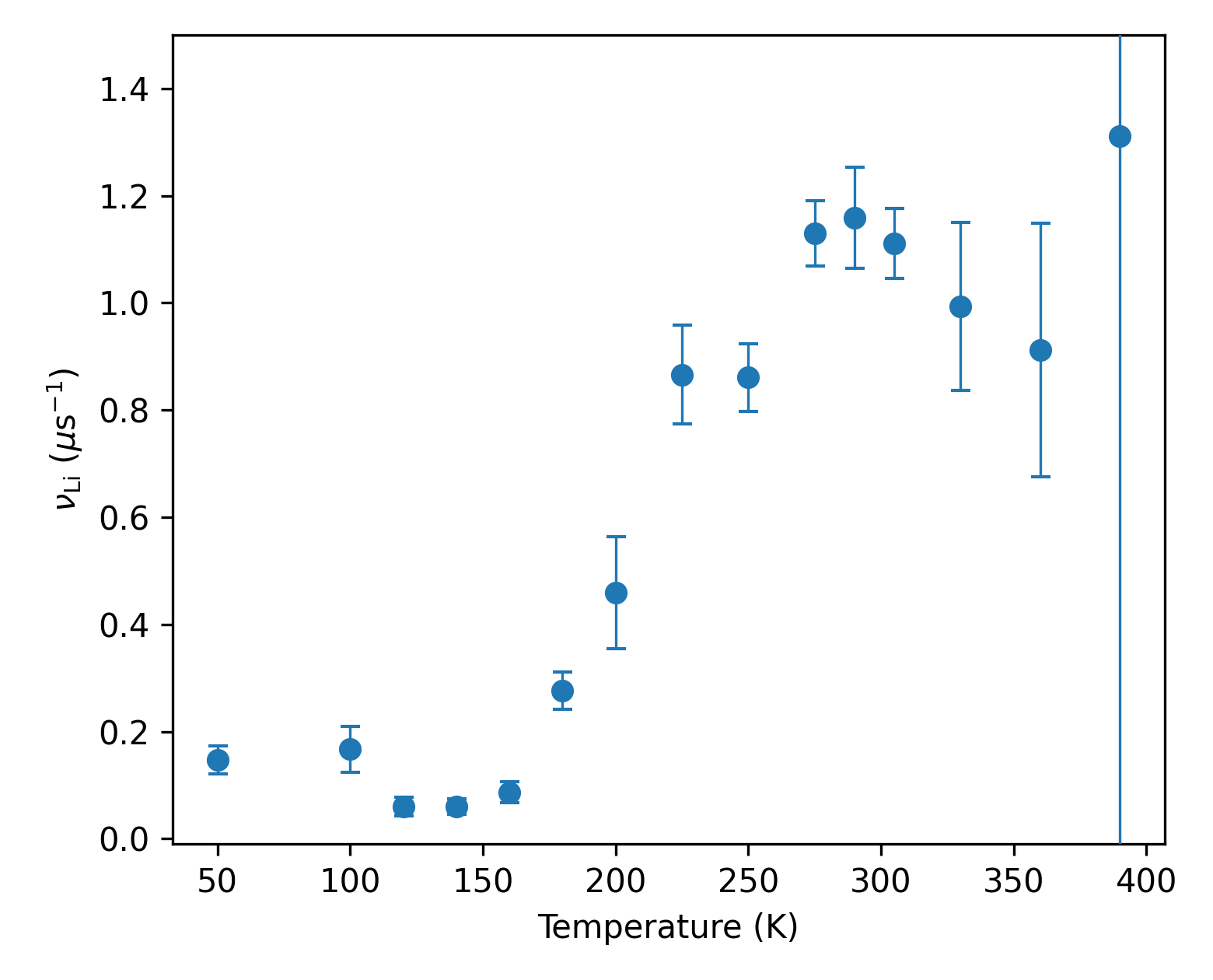}
        \put(15,50){\includegraphics[width=0.15\linewidth]{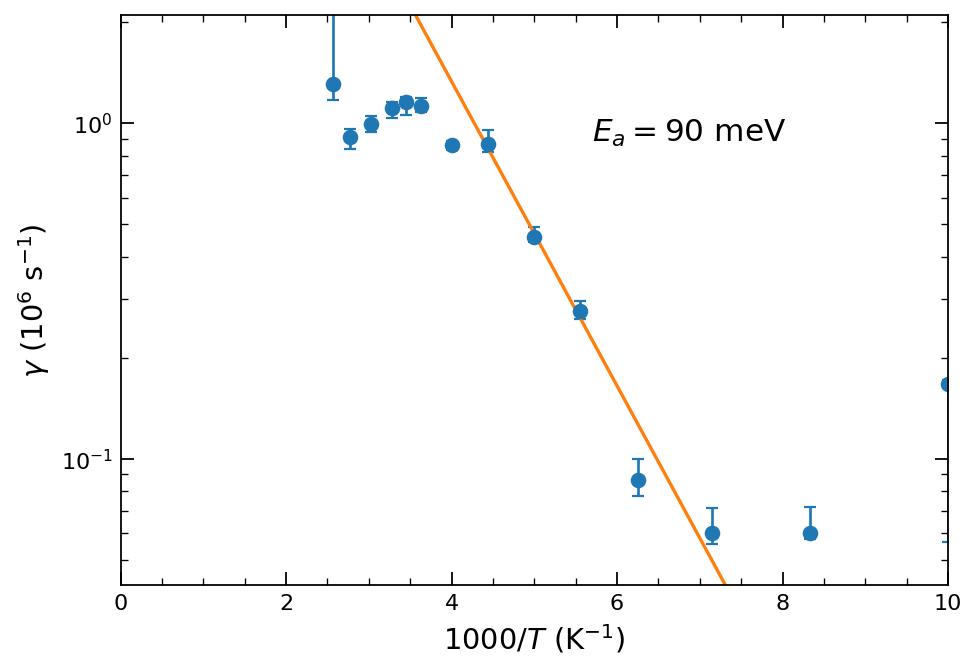}}
      \end{overpic}
      \label{fig:fit_nu_Li}
    }\quad
  }
  \mbox{
    \subfloat[\label{c}]{\includegraphics[scale=0.50]{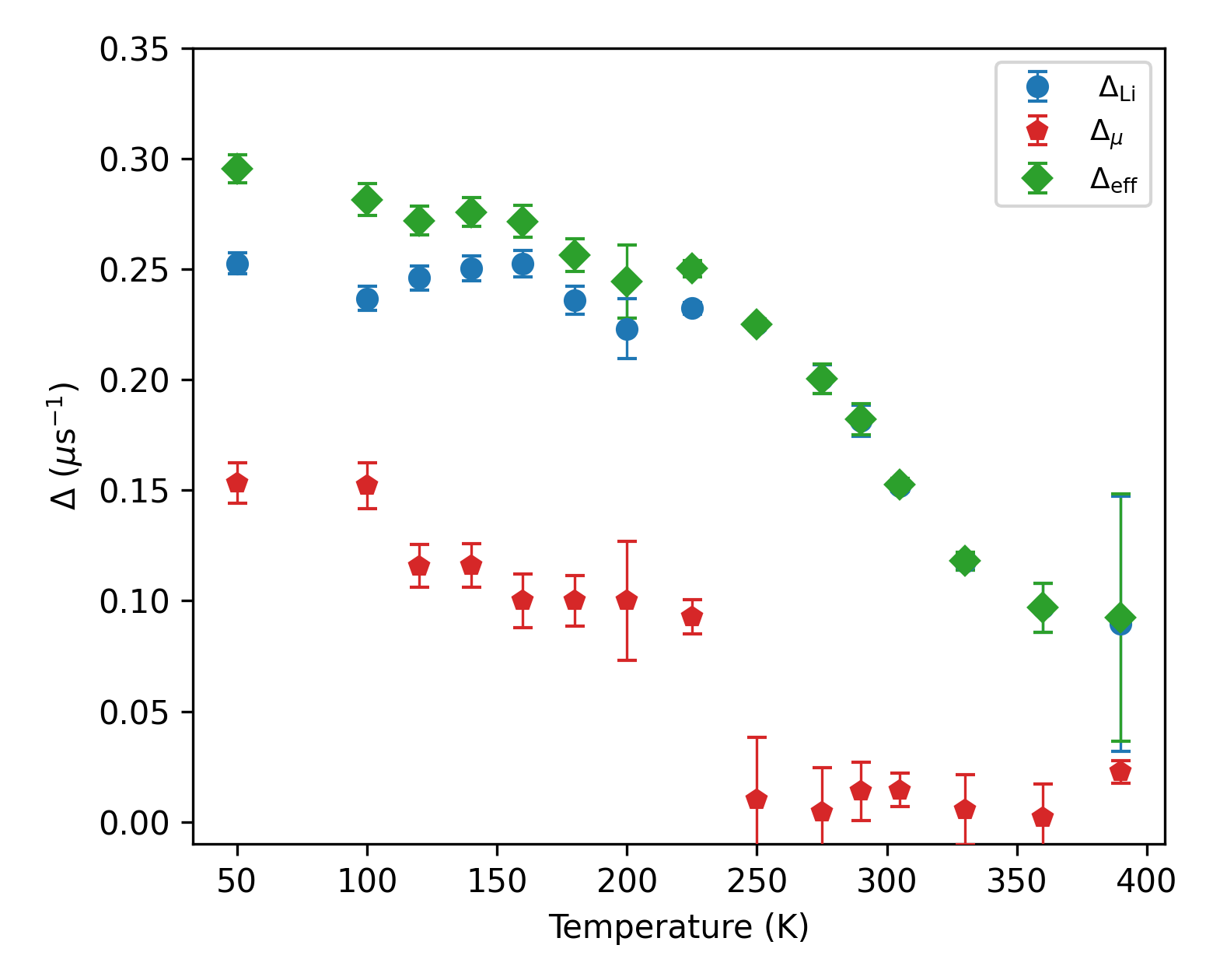}\label{fig:fit_Delta}}\quad
    \hspace{0.8em}%
    \subfloat[\label{d}]{\includegraphics[scale=0.50]{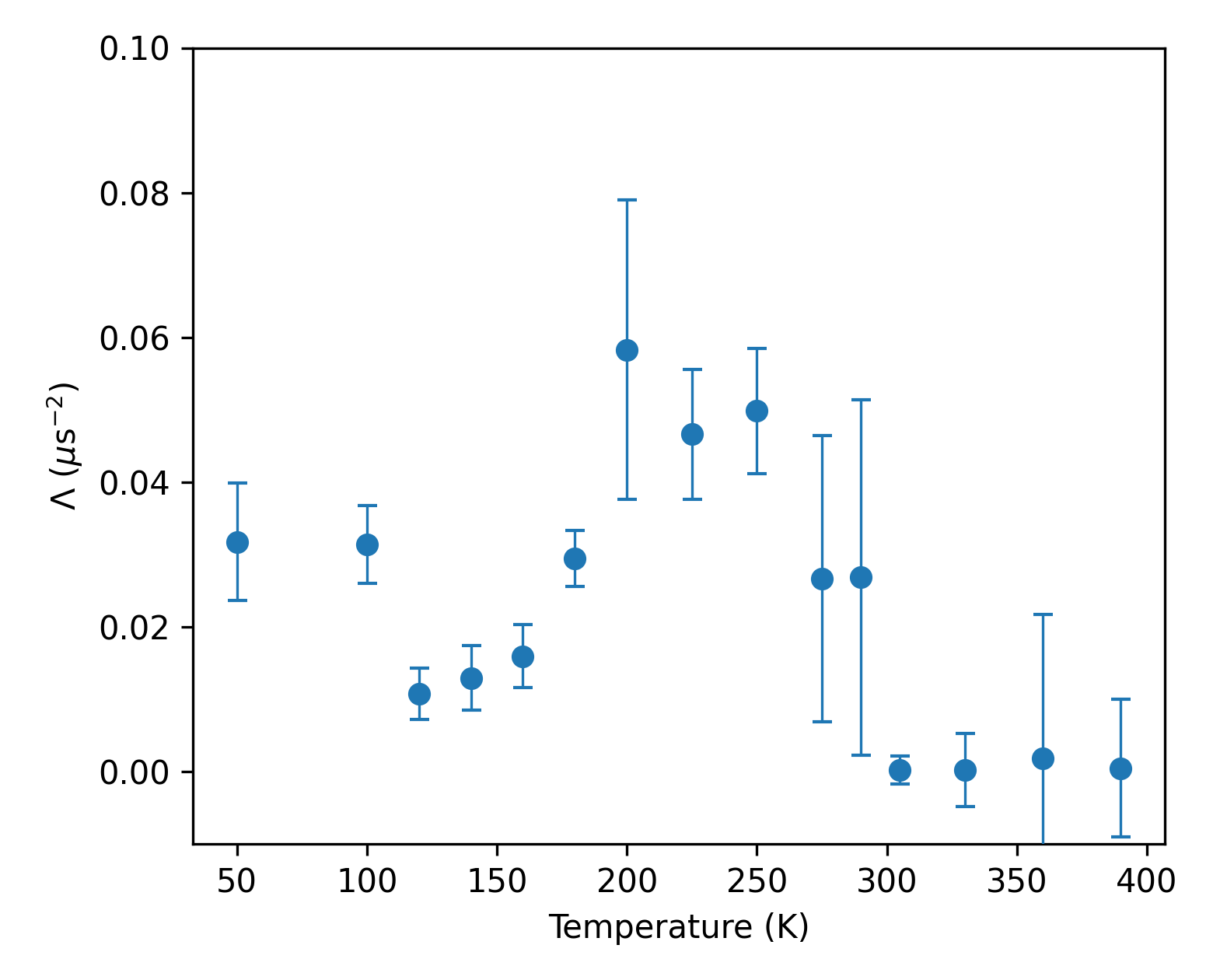}\label{fig:fit_Lambda}}\quad
  }
  \captionsetup{justification=raggedright,singlelinecheck=false}
  \caption{
Spin SDE description of $\mu$SR spectra of $\mathrm{Li}_{0.73}\mathrm{CoO}_2$ and
temperature dependence of the fitted parameters (global ZF/weak-LF analysis).
(a) Representative simultaneous fit of ZF, 5~G, and 10~G asymmetry spectra at a fixed
temperature using a single parameter set. The fit reproduces the KT-like early-time
dephasing (quenched width $\Delta_\mu$, with $\nu_\mu=0$), the systematic LF
decoupling, and the remaining dynamical relaxation at long times.
(b) Extracted Li-driven fluctuation rate $\nu_{\mathrm{Li}}(T)$.
(c) Extracted field-width parameters: dynamical width $\Delta_{\mathrm{Li}}(T)$, quenched
(static) width $\Delta_\mu(T)$, and $\Delta_{\mathrm{eff}}(T)\equiv
\sqrt{\Delta_{\mathrm{Li}}^2+\Delta_\mu^2}$.
(d) Extracted backaction (memory) strength $\Lambda(T)$.
The data indicate a crossover from a low-$T$ regime requiring a substantial quenched
component to a higher-$T$ regime dominated by Li-driven dynamics, with backaction effects
most pronounced in the intermediate-temperature window. 
}
  \label{fig:LiCoO2_2x2}
\end{figure}

Figure~\ref{fig:LiCoO2_2x2} summarizes the performance of the spin SDE framework
against experimental ZF and weak-LF spectra of $\mathrm{Li}_{0.73}\mathrm{CoO}_2$
and the corresponding temperature dependence of the fitted parameters.
Throughout this application we set $\nu_\mu=0$, so the muon-background channel
acts as a quenched Gaussian distribution characterized by a single static width
$\Delta_\mu$; no additional independent
static broadening is introduced.

In \figref{fig:fit_G_T100K} we show a representative simultaneous fit at $T=100$~K to the ZF, 5~G,
and 10~G asymmetry spectra using a single parameter set shared across the three
fields. The spin SDE model reproduces the principal
field-dependent line-shape features: (i) the pronounced KT-like early-time
depolarization in ZF governed by the quenched width $\Delta_\mu$, (ii) the
systematic weak-LF decoupling in which increasing $B_{\mathrm L}$ suppresses the ZF minimum
and raises the late-time polarization, and (iii) the remaining incomplete LF
recovery indicating a residual dynamical relaxation channel.
The ability of a single parameter set to reproduce the ZF and LF spectra
provides a nontrivial consistency check of the spin SDE description beyond single-curve fits.

The corresponding extracted ion-driven fluctuation rate $\nu_{\mathrm{Li}}(T)$
depicted in \figref{fig:fit_nu_Li} increases strongly with temperature.
The onset of a rapid rise in $\nu_{\mathrm{Li}}$ around the intermediate-$T$
window signals that Li-induced field fluctuations enter the $\mu$SR time window.
At higher temperatures, $\nu_{\mathrm{Li}}$ approaches the $\sim O(1~\mu\mathrm{s}^{-1})$
scale with larger uncertainties, consistent with the progressive approach toward
the motional-narrowing regime in which the spectra become less sensitive to
further increases in the correlation rate.
To further assess whether the extracted Li-driven fluctuation rate follows
thermally activated kinetics, we show in the inset of the $\nu_{\mathrm{Li}}(T)$
panel an Arrhenius representation, plotting $\nu_{\mathrm{Li}}$ versus $1000/T$
on a logarithmic scale.
Over the temperature window where $\nu_{\mathrm{Li}}$ exhibits a clear monotonic
increase and the uncertainties remain moderate, the data are approximately
consistent with activated behavior,
\begin{equation}
  \nu_{\mathrm{Li}}(T)=\nu_{0}\exp\!\left(-\frac{E_{a}}{k_{B}T}\right),
  \label{eq:nu_arrhenius}
\end{equation}
as indicated by the near-linear trend in $\ln \nu_{\mathrm{Li}}$ vs.\ $1/T$.
A representative fit (solid line in the inset) yields an activation energy of
order $E_{a}\simeq 90~\mathrm{meV}$ consistent with the previous results\cite{Sugiyama2009}.
\figref{fig:fit_Delta} shows the fitted widths $\Delta_{\mathrm{Li}}(T)$ (dynamic ion-modulated
width), $\Delta_\mu(T)$ (quenched/static width), and their quadrature combination
$\Delta_{\mathrm{eff}}(T)\equiv\sqrt{\Delta_{\mathrm{Li}}(T)^2+\Delta_\mu(T)^2}$.
At low temperatures the spectra require a substantial quenched component
$\Delta_\mu$, consistent with the KT-like ZF minimum and strong weak-LF decoupling.
With increasing temperature, the fitted $\Delta_\mu$ decreases while the dynamics
encoded by $(\Delta_{\mathrm{Li}},\nu_{\mathrm{Li}})$ becomes increasingly dominant.
The decrease of $\Delta_{\mathrm{eff}}$ with $T$ reflects the combined effect of
the reduced need for a quenched component as the spectra become more dynamic
and correlations between $\Delta_{\mathrm{Li}}$ and $\nu_{\mathrm{Li}}$ in the
fast-fluctuation regime.

The extracted backaction (retarded-torque) strength $\Lambda(T)$ is shown in
\figref{fig:fit_Lambda}.  Its temperature dependence is naturally interpreted in
terms of the fluctuation regimes set by the fitted $\nu_{\mathrm{Li}}(T)$.
At high temperatures, where $\nu_{\mathrm{Li}}$ is large and the ion-driven field
is rapidly fluctuating, the retarded kernel becomes effectively short-ranged on
the $\mu$SR window and the polarization becomes only weakly sensitive to
$\Lambda$ (cf.\ the fast-fluctuation benchmark in
Fig.~\ref{fig:mem_nu_3}).  Accordingly, $\Lambda$ is only weakly identifiable in
this regime and the fit returns values consistent with $\Lambda\simeq 0$ within
uncertainties.  In contrast, in the intermediate-$T$ window where
$\nu_{\mathrm{Li}}\sim \Delta_{\mathrm{Li}}$, the retarded-torque channel
produces a visible reshaping of the line shape, improving the global consistency across
ZF and weak-LF spectra.
This is illustrated by the benchmark in \figref{fig:mem_nu_05}.
For the same $(\Delta_{\mathrm i},\nu_{\mathrm i})$, the spin SDE curves with
$\Lambda=0$ and $\Lambda=0.1$ separate clearly at intermediate and late times.
In other words, in this crossover regime a finite $\Lambda$ leaves a clear
signature in the line shape.
At low temperatures, the spectra are dominated by quasi-static dephasing from the
quenched width $\Delta_\mu$ together with slow Li-driven fluctuations (small $\nu_{\mathrm{Li}}$).  
In \figref{fig:mem_nu_1e4} finite $\Lambda$ generates pronounced line shape stiffening
and oscillatory signatures in this quasi-static regime which are not visible in
the low-$T$ spectra. Therefore, the data
restrict $\Lambda$ to remain small in this regime, although it need not vanish
identically. 

Overall, these trends indicate that a retarded-torque (backaction) channel
is operative in $\mathrm{Li}_{0.73}\mathrm{CoO}_2$ on the $\mu$SR time window.
Its effect is most clearly visible in
the crossover regime compared to the fast-fluctuation and 
quasi-static limits. Thus, $\Lambda(T)$ should be interpreted as a genuine
non-Markovian feedback scale that becomes observable only when the fluctuation
dynamics falls in the intermediate regime.

\section{Summary and Conclusions}

In this work we developed an open-quantum-system description of muon spin
relaxation ($\mu$SR) that treats the implanted muon as a quantum spin coupled to
a fluctuating magnetic environment with intrinsic temporal correlations.
Starting from a Schwinger-Keldysh spin-coherent-state path integral, integrating
out the bath degrees of freedom yields an influence functional specified by two
kernels: a retarded kernel that produces a causal history-dependent torque
(backaction) and a Keldysh kernel that sets the colored-noise correlations.
The resulting stochastic equation of motion, \eqref{eq:LLG}, provides an
effective non-Markovian extension of standard KT modeling. In this framework,
the stochastic fluctuations are tied to the modeled symmetrized local-field
correlator, while retarded memory is incorporated through an effective causal
kernel, allowing static broadening, dynamical fluctuations, and backaction to
be analyzed beyond the strong-collision approximation.

For the exponential memory kernel used here, the dynamics admits an efficient
Markovian embedding. We implemented this using auxiliary memory variables
together with Ornstein-Uhlenbeck sampling of the colored noise, enabling
stable and scalable Monte Carlo evaluation of the polarization function
$G_z(t)$. Systematic benchmarks show that the spin SDE reproduces the
established quasi-static to motional-narrowing crossover in the ion-only
(dynamic KT) setting, captures the expected longitudinal-field decoupling
trends, and quantifies the effect of muon hopping when it becomes appreciable.
Furthermore, within the present effective description, increasing backaction 
strength $\Lambda$ can suppress
depolarization over the experimental time window in a manner reminiscent of
motional narrowing.

To complement the numerical approach, we derived a closed analytical reduction
for a controlled parameter regime.
In the static-muon setting ($\nu_\mu=0$), a rotating-frame, small-angle
treatment yields a tractable transverse dynamics whose cumulant resummation
recovers the Abragam form in the Markov limit and produces a practical
interpolation into the intermediate regime.
We emphasized, however, that this reduction has a sharply defined domain of
validity. It is reliable when the fluctuations are not quasi-static
($\nu_{\mathrm i}\gtrsim \Delta_{\mathrm i}$) and the backaction is weak to
moderate, but it breaks down in the quasi-static limit and in the strong-memory
regime, particularly when a quenched width is present.
In these cases, the static background enters the dynamics through rotation
dressing of the noise and memory kernels, and the approximations that restore a
deterministic convolution kernel are
no longer quantitatively valid.
Accordingly, the full spin SDE serves as the baseline description
whenever static-dynamic entanglement or strong backaction becomes important.

Finally, we applied the framework to $\mathrm{Li}_{0.73}\mathrm{CoO}_2$ using a
global analysis of ZF and weak-LF spectra (5~G and 10~G) within the static muon assumption.
The resulting fits reproduce the ZF KT-like early-time depolarization, the
systematic weak-field decoupling, and the residual dynamical relaxation at long
times with a single parameter set at each temperature.
The extracted $\nu_{\mathrm{Li}}(T)$ increases strongly with temperature and is
consistent with activated behavior over an intermediate-temperature window,
with a representative activation energy of order $E_a\simeq 90$~meV. Further,
$\Lambda(T)$ is most clearly required in the crossover regime where the spectra
retain enough structure to distinguish memory-induced stabilization from a
simple rescaling of $(\Delta_{\mathrm{Li}},\nu_{\mathrm{Li}})$.

Overall, this work establishes a practical, fit-ready route to incorporate
non-Markovian backaction and colored field correlations into quantitative
$\mu$SR modeling.
The framework is readily extendable to anisotropic kernels, additional bath
channels, and more realistic correlators for correlated ionic motion, providing
a systematic basis for $\mu$SR studies of ion-driven magnetic fluctuations in
functional materials.

% Create the reference section using BibTeX:
\section{Acknowlegments}
We are grateful to Jun Sugiyama, Kazuki Ohishi, Masatoshi Hiraishi, and
Izumi Umegaki for providing the experiment $\mu$SR spectra and valuable discussions.
This research is supported by the New Energy and Industrial Technology Development Organization 
(NEDO) project, MEXT as “Program for Promoting Researches on the Supercomputer Fugaku” 
(Fugaku battery \& Fuel Cell Project) 
(Grant No. JPMXP1020200301, Project No.: hp220177, hp210173, hp200131), 
Digital Transformation Initiative for Green Energy Materials (DX-GEM)
and JSPS Grants-in-Aid for Scientific Research (Young Scientists) No. 19K15397. 
Some calculations were done using the supercomputing facilities of the 
Institute for Solid State Physics, The University of Tokyo.

\appendix

\section{Intermediate scattering function from correlated Li--$\mu$ hopping}
\label{app:ISF_QME_CME}
We consider a muon on lattice sites $a$ (positions $\bm R_a$) and Li on
lattice sites $i$ (positions $\bm r_i$), with joint basis
$|a,i\rangle$ and positional Hamiltonian
\begin{equation}
  \hat H_{\rm pos}
  =
  \hat H_\mu + \hat H_{\rm Li} + \hat H_{\mu{\rm Li}},
\end{equation}
where $\hat H_\mu$ and $\hat H_{\rm Li}$ are tight-binding hopping Hamiltonians
and $\hat H_{\mu{\rm Li}}$ allows (in general) correlated energetics and/or
correlated moves.
Introduce a phonon bath $\hat H_B$ and a system--bath coupling that induces
\emph{incoherent} jumps between joint configurations,
\begin{equation}
  \hat H
  =
  \hat H_{\rm pos} + \hat H_B + \hat H_{SB},
  \qquad
  \hat H_{SB}
  =
  \sum_{(a,i)\neq(b,j)}
  \Bigl(
    |b,j\rangle\langle a,i| \,\hat B_{b j, a i}
    + \text{H.c.}
  \Bigr),
\end{equation}
with bath operators $\hat B_{b j, a i}=t_{bj,ai}\hat{X}_{bj,ai}$ (linear in phonon operators in the
usual case).  Under a Born--Markov approximation, tracing out the bath yields
a Lindblad-type quantum master equation (QME)\cite{Lindblad1976} with jump operators
$\hat L_{b j \leftarrow a i}=\sqrt{W_{b j, a i}}\;|b,j\rangle\langle a,i|$,
\begin{equation}
  \dot{\hat\rho}
  =
  \sum_{(a,i)\neq(b,j)}
  \left(
    \hat L_{b j \leftarrow a i}\,\hat\rho\,\hat L_{b j \leftarrow a i}^\dagger
    -
    \frac12
    \left\{
      \hat L_{b j \leftarrow a i}^\dagger \hat L_{b j \leftarrow a i},\hat\rho
    \right\}
  \right),
\end{equation}
with rates given by bath correlation functions,
\begin{equation}
  W_{b j, a i}
  =
  |t_{bj,ai}|^{2}\int_{-\infty}^{+\infty} dt\;
  e^{+i\omega_{b j, a i} t}\,
  \bigl\langle
    \hat X_{b j, a i}(t)\,\hat X_{b j, a i}^\dagger(0)
  \bigr\rangle_B ,
\end{equation}
where $\omega_{b j, a i}$ is the energy difference between the two joint
configurations
Let the joint populations be
\begin{equation}
  P_{a i}(t)\equiv \langle a,i|\hat\rho(t)|a,i\rangle,
  \qquad
  \sum_{a,i}P_{ai}(t)=1.
\end{equation}
Projecting the QME onto populations\cite{BreuerPetruccione2002} 
gives the classical master equation (CME)\cite{vanKampen2007} on the product space, 
\begin{equation}
  \dot P_{a i}(t)
  =
  \sum_{b,j}
  \Bigl[
    W_{a i, b j}\,P_{b j}(t)
    -
    W_{b j, a i}\,P_{a i}(t)
  \Bigr].
  \label{eq:CME_joint_general}
\end{equation}
Next we define the joint intermediate scattering function (ISF)
\begin{equation}
  F(\bm q_\mu,\bm q_{\rm Li};t)
  \equiv
  \Bigl\langle
    e^{i\bm q_\mu\cdot[\bm R_\mu(t)-\bm R_\mu(0)]
      +i\bm q_{\rm Li}\cdot[\bm r_{\rm Li}(t)-\bm r_{\rm Li}(0)]}
  \Bigr\rangle .
  \label{eq:joint_ISF_def}
\end{equation}
In the Poisson limit we assume homogeneity and symmetric
rates such that the total escape rate from any joint state $(a,i)$ is
state-independent,
\begin{equation}
  \nu
  \equiv
  \sum_{(b,j)\neq(a,i)} W_{b j, a i},
  \qquad \text{independent of $(a,i)$}.
  \label{eq:nu_joint}
\end{equation}
Then the number of jumps $N(t)$ is Poisson distributed,
\begin{equation}
  \mathbb P[N(t)=n]
  =
  e^{-\nu t}\frac{(\nu t)^n}{n!}.
  \label{eq:Poisson_joint}
\end{equation}
Each transition $(a,i)\to(b,j)$ produces a \emph{joint} displacement
\begin{equation}
  \delta\bm R_\mu
  =
  \bm R_b-\bm R_a,
  \qquad
  \delta\bm r_{\rm Li}
  =
  \bm r_j-\bm r_i,
\end{equation}
drawn, in the Poisson (time-homogeneous) limit, from a stationary single-jump
distribution
\begin{equation}
  p(\delta\bm R_\mu,\delta\bm r_{\rm Li})
  =
  \sum_{(b,j)\neq(a,i)}
  \frac{W_{b j, a i}}{\nu}\;
  \delta_{\delta\bm R_\mu,\bm R_b-\bm R_a}\,
  \delta_{\delta\bm r_{\rm Li},\bm r_j-\bm r_i},
  \label{eq:p_joint}
\end{equation}
Conditioned on exactly $N(t)=n$ jumps, the total displacement is a sum of $n$
independent draws from $p$, hence
\begin{equation}
  \Bigl\langle
    e^{i\bm q_\mu\cdot\Delta\bm R_\mu(t)
      +i\bm q_{\rm Li}\cdot\Delta\bm r_{\rm Li}(t)}
  \Bigr\rangle_{N=n}
  =
  \Bigl[\alpha(\bm q_\mu,\bm q_{\rm Li})\Bigr]^n,
\end{equation}
with the single-jump characteristic function
\begin{equation}
  \alpha(\bm q_\mu,\bm q_{\rm Li})
  \equiv
  \sum_{\delta\bm R_\mu,\delta\bm r_{\rm Li}}
  p(\delta\bm R_\mu,\delta\bm r_{\rm Li})\,
  e^{i\bm q_\mu\cdot\delta\bm R_\mu
    +i\bm q_{\rm Li}\cdot\delta\bm r_{\rm Li}} .
  \label{eq:alpha_joint}
\end{equation}

Averaging over the Poisson distribution \eqref{eq:Poisson_joint} yields\cite{ChudleyElliott1961}
\begin{align}
  F(\bm q_\mu,\bm q_{\rm Li};t)
  &=
  \sum_{n=0}^{\infty}
  e^{-\nu t}\frac{(\nu t)^n}{n!}\,
  \Bigl[\alpha(\bm q_\mu,\bm q_{\rm Li})\Bigr]^n
  \nonumber\\
  &=
  \exp\!\Bigl\{-\nu t\bigl[1-\alpha(\bm q_\mu,\bm q_{\rm Li})\bigr]\Bigr\}.
  \label{eq:ISF_joint_Poisson_final}
\end{align}
Muon-only and Li-only ISFs are recovered by setting the other wavevector to
zero, e.g.\ $F_\mu(\bm q,t)=F(\bm q,\bm 0;t)$.

If muon and Li jumps are uncorrelated, the transition rates separate into
muon-only and Li-only parts and lose conditional dependence,
\begin{equation}
  W_{b j, a i}
  =
  W^{(\mu)}_{b a}\,\delta_{j i}
  +
  W^{({\rm Li})}_{j i}\,\delta_{b a},
  \label{eq:W_uncorr}
\end{equation}
so that the CME generator is a direct sum and the two counting processes are
independent Poisson processes with rates
\begin{equation}
  \nu_\mu=\sum_{b\neq a}W^{(\mu)}_{b a},
  \qquad
  \nu_{\rm Li}=\sum_{j\neq i}W^{({\rm Li})}_{j i}.
\end{equation}
Repeating the Poisson argument separately gives
\begin{equation}
  F(\bm q_\mu,\bm q_{\rm Li};t)
  =
  \exp\!\Bigl\{
    -\nu_\mu t\bigl[1-\alpha_\mu(\bm q_\mu)\bigr]
    -\nu_{\rm Li} t\bigl[1-\alpha_{\rm Li}(\bm q_{\rm Li})\bigr]
  \Bigr\},
  \label{eq:ISF_uncorr_final}
\end{equation}
where $\alpha_\mu$ and $\alpha_{\rm Li}$ are the corresponding single-hop
characteristic functions, defined as in \eqref{eq:alpha_joint} but with
the appropriate single-species hop distributions.  In particular,
$F(\bm q_\mu,\bm q_{\rm Li};t)=F_\mu(\bm q_\mu,t)\,F_{\rm Li}(\bm q_{\rm Li},t)$.
Explicit forms used in the main text are recovered by making the substitution
$\alpha_\mu(\bm q_\mu)\to\alpha(\bm q)$ for the muon background and
$\alpha_{\rm Li}(\bm q_{\rm Li})\to\lambda(\bm q)$ for the ion-modulated ISF's,
respectively.

\section{Derivation of the phase-diffusion factor}
\label{app:phase_fluctuation_kernel}
Let's start with the Gaussian closure
\begin{equation}
	\mathcal{D}(\tau)=\langle e^{i[\theta_{\mathrm{fl}}(t)-\theta_{\mathrm{fl}}(t-\tau)]}\rangle,
  \label{eq:Gaussian_closure}
\end{equation}
where the term $\theta_{\mathrm{fl}}(t)=-\frac{1}{S}\int_{0}^{t}ds\eta_{z}(s)$ comes from the noise 
of $\Phi_{z}$. Because the noise is stationary, it depends only on $\tau$ not on $t$. Writing
the difference explicitly as a zero-mean Gaussian random variable
\[
	X\equiv\theta_{\mathrm{fl}}(t)-\theta_{\mathrm{fl}}(t-\tau)=-\frac{1}{S}\int_{t-\tau}^{t}ds\eta_{z}(s),
\]
we express \eqref{eq:Gaussian_closure} as
\begin{equation}
	\mathcal{D}(\tau)=\langle e^{iX}\rangle=\exp\left(-\frac{1}{2}\langle X^{2}\rangle\right).
  \label{eq:Gaussian_closure1}
\end{equation}
Using $\langle\eta_{z}(s)\eta_{z}^{*}(s')\rangle=2S^{2}\Delta_{\mathrm i}^{2}e^{-\nu_{\mathrm i}|s-s'|}$,
\begin{equation}
	\langle X^{2}\rangle=2\Delta_{\mathrm i}^{2}\int_{t-\tau}^{t}ds\int_{t-\tau}^{t}ds'e^{-\nu_{\mathrm i}|s-s'|}.
  \label{eq:X_ave}
\end{equation}
The double integral can be evaluated by triangle splitting which results in
\begin{equation}
	\mathcal{D}(\tau)=\exp\left\{-\frac{2\Delta_{\mathrm i}^{2}}{\nu_{\mathrm i}^{2}}
	\left[\nu_{\mathrm i}\tau - \left(1 - e^{-\nu_{\mathrm i}\tau}\right) \right]\right\}.
  \label{eq:Gaussian_closure2}
\end{equation}

\bibliographystyle{apsrev4-2}
\bibliography{Ref_PI.bib}

@article{Takahashi2020,
  author = {Takahashi, Hideaki and Tanimura, Yoshitaka},
  title = {Open Quantum Dynamics Theory of Spin Relaxation: Application to $\mu$SR and Low-Field NMR Spectroscopies},
  journal = {Journal of the Physical Society of Japan},
  volume = {89},
  number = {6},
  pages = {064710},
  year = {2020},
  doi = {10.7566/JPSJ.89.064710},
  url = {https://doi.org/10.7566/JPSJ.89.064710}
}

@article{Ito2024,
  author = {Ito, Takashi U. and Kadono, Ryosuke},
  title = {Distinguishing Ion Dynamics from Muon Diffusion in Muon Spin Relaxation},
  journal = {Journal of the Physical Society of Japan},
  volume = {93},
  number = {4},
  pages = {044602},
  year = {2024},
  doi = {10.7566/JPSJ.93.044602},
  url = {https://doi.org/10.7566/JPSJ.93.044602}
}

@article{Hayano1979,
  title = {Zero- and low-field spin relaxation studied by positive muons},
  author = {Hayano, R. S. and Uemura, Y. J. and Imazato, J. and Nishida, N. and Yamazaki, T. and Kubo, R.},
  journal = {Phys. Rev. B},
  volume = {20},
  number = {3},
  pages = {850--859},
  year = {1979},
  month = aug,
  publisher = {American Physical Society},
  doi = {10.1103/PhysRevB.20.850},
  url = {https://link.aps.org/doi/10.1103/PhysRevB.20.850}
}

@article{Moller2013,
  doi = {10.1088/0031-8949/88/06/068510},
  url = {https://doi.org/10.1088/0031-8949/88/06/068510},
  year = {2013},
  month = dec,
  publisher = {IOP Publishing},
  volume = {88},
  number = {6},
  pages = {068510},
  author = {M{\"o}ller, J. S. and Bonf{\`a}, P. and Ceresoli, D. and Bernardini, F. and Blundell, S. J. and Lancaster, T. and De Renzi, R. and Marzari, N. and Watanabe, I. and Sulaiman, S. and Mohamed-Ibrahim, M. I.},
  title = {Playing quantum hide-and-seek with the muon: localizing muon stopping sites},
  journal = {Physica Scripta}
}

@article{Blundell2004,
  author = {Blundell, Stephen J.},
  title = {Muon-Spin Rotation Studies of Electronic Properties of Molecular Conductors and Superconductors},
  journal = {Chemical Reviews},
  volume = {104},
  number = {11},
  pages = {5717--5736},
  year = {2004},
  doi = {10.1021/cr030632e},
  note = {PMID: 15535666},
  url = {https://doi.org/10.1021/cr030632e}
}

@article{Uemura1985,
  title = {Muon-spin relaxation in AuFe and CuMn spin glasses},
  author = {Uemura, Y. J. and Yamazaki, T. and Harshman, D. R. and Senba, M. and Ansaldo, E. J.},
  journal = {Phys. Rev. B},
  volume = {31},
  number = {1},
  pages = {546--563},
  year = {1985},
  month = jan,
  publisher = {American Physical Society},
  doi = {10.1103/PhysRevB.31.546},
  url = {https://link.aps.org/doi/10.1103/PhysRevB.31.546}
}

@inbook{Kubo-Toyabe,
  author = {Kubo, Ryogo and Toyabe, Toru},
  title = {Magnetic field fluctuations and nuclear magnetic relaxation},
  booktitle = {Magnetic Resonance and Relaxation},
  editor = {Blinc, R.},
  publisher = {North-Holland},
  address = {Amsterdam},
  year = {1967},
  pages = {810}
}

@article{Sulaiman1994,
  author = {Sulaiman, Shukri B. and Sahoo, N. and Srinivas, Sudha and Hagelberg, F. and Das, T. P. and Torikai, E. and Nagamine, K.},
  title = {Theory of location and associated hyperfine properties of the positive muon in La$_2$CuO$_4$},
  journal = {Hyperfine Interactions},
  volume = {84},
  number = {1},
  pages = {87--103},
  year = {1994},
  doi = {10.1007/BF02060647},
  url = {https://doi.org/10.1007/BF02060647}
}

@article{Williams2016,
  doi = {10.1088/0953-8984/28/7/076001},
  url = {https://doi.org/10.1088/0953-8984/28/7/076001},
  year = {2016},
  month = jan,
  publisher = {IOP Publishing},
  volume = {28},
  number = {7},
  pages = {076001},
  author = {Williams, R. C. and Xiao, F. and Thomas, I. O. and Clark, S. J. and Lancaster, T. and Cornish, G. A. and Blundell, S. J. and Hayes, W. and Paul, A. K. and Felser, C. and Jansen, M.},
  title = {Muon-spin relaxation study of the double perovskite insulators Sr$_2$BOsO$_6$ ($B=\mathrm{Fe},\,\mathrm{Y},\,\mathrm{In}$)},
  journal = {Journal of Physics: Condensed Matter}
}

@article{Blundell2023,
  author = {Blundell, S. J. and Lancaster, T.},
  title = {DFT+\,{$\mu$}: Density functional theory for muon site determination},
  journal = {Applied Physics Reviews},
  volume = {10},
  number = {2},
  pages = {021316},
  year = {2023},
  month = jun,
  issn = {1931-9401},
  doi = {10.1063/5.0149080},
  url = {https://doi.org/10.1063/5.0149080}
}

@article{Sugiyama2011,
  title = {Magnetic and diffusive nature of LiFePO$_4$ investigated by muon spin rotation and relaxation},
  author = {Sugiyama, Jun and Nozaki, Hiroshi and Harada, Masashi and Kamazawa, Kazuya and Ofer, Oren and M\aa{}nsson, Martin and Brewer, Jess H. and Ansaldo, Eduardo J. and Chow, Kim H. and Ikedo, Yutaka and Miyake, Yasuhiro and Ohishi, Kazuki and Watanabe, Isao and Kobayashi, Genki and Kanno, Ryoji},
  journal = {Phys. Rev. B},
  volume = {84},
  number = {5},
  pages = {054430},
  year = {2011},
  month = aug,
  publisher = {American Physical Society},
  doi = {10.1103/PhysRevB.84.054430},
  url = {https://link.aps.org/doi/10.1103/PhysRevB.84.054430}
}

@article{Okabe2024,
  title = {Nanoscale dynamics of hydrogen in VO$_2$ studied by $\mu$SR},
  author = {Okabe, H. and Hiraishi, M. and Koda, A. and Matsushita, Y. and Ohsawa, T. and Ohashi, N. and Kadono, R.},
  journal = {Phys. Rev. Mater.},
  volume = {8},
  number = {2},
  pages = {024602},
  year = {2024},
  month = feb,
  publisher = {American Physical Society},
  doi = {10.1103/PhysRevMaterials.8.024602},
  url = {https://link.aps.org/doi/10.1103/PhysRevMaterials.8.024602}
}

@article{Sugiyama2009,
  title = {Li Diffusion in Li$_x$CoO$_2$ Probed by Muon-Spin Spectroscopy},
  author = {Sugiyama, Jun and Mukai, Kazuhiko and Ikedo, Yutaka and Nozaki, Hiroshi and M\aa{}nsson, Martin and Watanabe, Isao},
  journal = {Phys. Rev. Lett.},
  volume = {103},
  number = {14},
  pages = {147601},
  year = {2009},
  month = sep,
  publisher = {American Physical Society},
  doi = {10.1103/PhysRevLett.103.147601},
  url = {https://link.aps.org/doi/10.1103/PhysRevLett.103.147601}
}

@article{Sugiyama2012,
  title = {Diffusive behavior in Li$M$PO$_4$ with $M=\mathrm{Fe},\ \mathrm{Co},\ \mathrm{Ni}$ probed by muon-spin relaxation},
  author = {Sugiyama, Jun and Nozaki, Hiroshi and Harada, Masashi and Kamazawa, Kazuya and Ikedo, Yutaka and Miyake, Yasuhiro and Ofer, Oren and M\aa{}nsson, Martin and Ansaldo, Eduardo J. and Chow, Kim H. and Kobayashi, Genki and Kanno, Ryoji},
  journal = {Phys. Rev. B},
  volume = {85},
  number = {5},
  pages = {054111},
  year = {2012},
  month = feb,
  publisher = {American Physical Society},
  doi = {10.1103/PhysRevB.85.054111},
  url = {https://link.aps.org/doi/10.1103/PhysRevB.85.054111}
}

@article{Ashton2014,
  author = {Ashton, Thomas E. and Vidal Laveda, Josefa and MacLaren, Donald A. and Baker, Peter J. and Porch, Adrian and Jones, Martin O. and Corr, Serena A.},
  title = {Muon studies of Li$^+$ diffusion in LiFePO$_4$ nanoparticles of different polymorphs},
  journal = {J. Mater. Chem. A},
  year = {2014},
  volume = {2},
  number = {17},
  pages = {6238--6245},
  publisher = {The Royal Society of Chemistry},
  doi = {10.1039/C4TA00543K},
  url = {http://dx.doi.org/10.1039/C4TA00543K}
}

@article{Laveda2018,
  author = {Vidal Laveda, Josefa and Johnston, Beth and Paterson, Gary W. and Baker, Peter J. and Tucker, Matthew G. and Playford, Helen Y. and Jensen, Kirsten M. {\O}. and Billinge, Simon J. L. and Corr, Serena A.},
  title = {Structure--property insights into nanostructured electrodes for Li-ion batteries from local structural and diffusional probes},
  journal = {J. Mater. Chem. A},
  year = {2018},
  volume = {6},
  number = {1},
  pages = {127--137},
  publisher = {The Royal Society of Chemistry},
  doi = {10.1039/C7TA04400C},
  url = {http://dx.doi.org/10.1039/C7TA04400C}
}

@article{Ohishi2022,
  author = {Ohishi, Kazuki and Igarashi, Daisuke and Tatara, Ryoichi and Umegaki, Izumi and Koda, Akihiro and Komaba, Shinichi and Sugiyama, Jun},
  title = {Operando Muon Spin Rotation and Relaxation Measurement on LiCoO$_2$ Half-Cell},
  journal = {ACS Applied Energy Materials},
  volume = {5},
  number = {10},
  pages = {12538--12544},
  year = {2022},
  doi = {10.1021/acsaem.2c02175},
  url = {https://doi.org/10.1021/acsaem.2c02175}
}

@article{Sugiyama2013,
  author = {Sugiyama, Jun and Mukai, Kazuhiko and Harada, Masashi and Nozaki, Hiroshi and Miwa, Kazutoshi and Shiotsuki, Taishi and Shindo, Yohei and Giblin, Sean R. and Lord, James S.},
  title = {Reactive surface area of the Li$_x$(Co$_{1/3}$Ni$_{1/3}$Mn$_{1/3}$)O$_2$ electrode determined by $\mu^+$SR and electrochemical measurements},
  journal = {Phys. Chem. Chem. Phys.},
  year = {2013},
  volume = {15},
  number = {25},
  pages = {10402--10412},
  publisher = {The Royal Society of Chemistry},
  doi = {10.1039/C3CP51662H},
  url = {http://dx.doi.org/10.1039/C3CP51662H}
}

@book{AltlandSimons2010,
  author    = {Altland, Alexander and Simons, Ben D.},
  title     = {Condensed Matter Field Theory},
  edition   = {2},
  year      = {2010},
  publisher = {Cambridge University Press},
  address   = {Cambridge}
}

@book{Kamenev2011,
  author    = {Kamenev, Alex},
  title     = {Field Theory of Non-Equilibrium Systems},
  year      = {2011},
  publisher = {Cambridge University Press},
  address   = {Cambridge}
}

@article{Edwards1976,
doi = {10.1088/0305-4608/6/10/022},
url = {https://doi.org/10.1088/0305-4608/6/10/022},
year = {1976},
month = {oct},
publisher = {},
volume = {6},
number = {10},
pages = {1927},
author = {S F Edwards and P W Anderson},
title = {Theory of spin glasses. II},
journal = {Journal of Physics F: Metal Physics},
}

@inbook{abragam,
    author        = "Abragam, Anatole",
    title         = "{$\mathrm{The}$ principles of nuclear magnetism;
    $\mathrm{Reprint}$ with corrections}",
    publisher     = "Clarendon Press",
    address       = "Oxford, U.K.",
    series        = "International series of monographs on physics",
    year          = "1989",
      url           = "https://cds.cern.ch/record/109195",
}

@book{kalvius,
  author ={G. M. Kalvius and D. R. Noakes and O. Hartmann},
  title ={Handbook on the Physics and Chemistry of Rare Earths}, 
  editor ={K. A. Gschneidner and Jr. L. Eyring and G. H. Lander},
  publisher = {North-Holland},
  year = 2001,
  address ={Amsterdam, Holland},
  volume = 32,
  chapter = 206,
  pages ={55-451} 
}

@book{yaouanc,
  author ={A. Yaouanc and P. Dalmas de R{\'e}otier},
  title = {Muon Spin Rotation, Relaxation, and Resonance, 
  Application to Condensed Matter}, 
  publisher = {Oxford University Press},
  year = 2011,
  address ={New York}
}

@InBook{chandran_16,
author = "C. Vinod Chandran and P. Heitjans",
editor = "Graham A. Webb",
series = "Annu. Rep. \text{NMR} Spectrosc.",
publisher = "Academic Press",
volume = "89",
pages = "1 - 102",
year = "2016",
title = "\text{Chapter One -} \text{Solid}-\text{State} \text{NMR} \text{Studies} of \text{Lithium} \text{Ion} \text{Dynamics} \text{Across} \text{Materials Classes}",
issn = "0066-4103",
doi = "https://doi.org/10.1016/bs.arnmr.2016.03.001",
url = "http://www.sciencedirect.com/science/article/pii/S0066410316300126",
}

@article{heitjans_03,
	doi = {10.1088/0953-8984/15/30/202},
	url = {https://doi.org/10.1088/0953-8984/15/30/202},
	year = 2003,
	month = {jul},
	publisher = {{IOP} Publishing},
	volume = {15},
	number = {30},
	pages = {R1257--R1289},
	author = {Paul Heitjans and Sylvio Indris},
	title = {Diffusion and ionic conduction in nanocrystalline ceramics},
	journal = {J. Phys.: Condens. Matter},
}

@article{keren_1994,
  author = {Keren, Amit},
  journal = {Phys. Rev. B},
  volume = {50},
  issue = {14},
  pages = {10039--10042},
  numpages = {0},
  year = {1994},
  month = {Oct},
  publisher = {American Physical Society},
  title = {Generalization of the $\mathrm{Abragam}$ relaxation function to a longitudinal field},
  doi = {10.1103/PhysRevB.50.10039},
  url = {https://link.aps.org/doi/10.1103/PhysRevB.50.10039}
}

@book{hempelmann_2000,
    author = {Hempelmann, Rolf},
    title = {Quasielastic Neutron Scattering and Solid State Diffusion},
    publisher = {Oxford University Press},
    year = {2000},
    month = {11},
    isbn = {9780198517436},
    doi = {10.1093/acprof:oso/9780198517436.001.0001},
    url = {https://doi.org/10.1093/acprof:oso/9780198517436.001.0001},
}

@article{mizushima_80,
  author = 	 {K. Mizushima and P. C. Jones and P. J. Wiseman and J. B. Goodenough},
  title = 	 {\text{Li}$_x$\text{CoO}$_2 (0<x\leq1)$: \text{A} NEW CATHODE MATERIAL FOR BATTERIES OF HIGH ENERGY DENSITY},
  journal = 	 {Mat. Res. Bull.},
  year = 	 1980,
  volume = 	 {15},
  pages = 	 {783-789}}

@article{padhi_1997,
doi = {10.1149/1.1837571},
url = {https://doi.org/10.1149/1.1837571},
year = {1997},
month = {apr},
publisher = {The Electrochemical Society, Inc.},
volume = {144},
number = {4},
pages = {1188},
author = {Padhi, A. K. and Nanjundaswamy, K. S. and Goodenough, J. B.},
title = {Phospho‐olivines as {P}ositive‐{E}lectrode {M}aterials for {R}echargeable {L}ithium {B}atteries},
journal = {Journal of The Electrochemical Society},
}

@article{amatucci_2003,
doi = {10.1149/1.1566965},
url = {https://doi.org/10.1149/1.1566965},
year = {2003},
month = {mar},
publisher = {The Electrochemical Society, Inc.},
volume = {150},
number = {5},
pages = {L9},
author = {Amatucci, G. and Tarascon, J.-M.},
title = {Optimization of {Insertion} {Compounds} {Such} as {LiMn}$_2${O}$_4$ for {Li}-{Ion} {Batteries} [Journal of The Electrochemical Society, 149, K31 (2002)]},
journal = {Journal of The Electrochemical Society},
}

@article{ohta2025,
author = {Ohta, Hiroto and Sugiyama, Jun},
title = {The Role of {DFT} Calculations in $\mu^+${SR} Studies of Battery Materials},
journal = {The Journal of Physical Chemistry C},
volume = {129},
number = {41},
pages = {18406-18416},
year = {2025},
doi = {10.1021/acs.jpcc.5c03880},
URL = {https://doi.org/10.1021/acs.jpcc.5c03880},
eprint = {https://doi.org/10.1021/acs.jpcc.5c03880}
}

@article{ChudleyElliott1961,
  author  = {Chudley, C. T. and Elliott, R. J.},
  title   = {Neutron Scattering from a Liquid on a Jump Diffusion Model},
  journal = {Proceedings of the Physical Society},
  volume  = {77},
  pages   = {353--361},
  year    = {1961}
}

@book{BreuerPetruccione2002,
  author    = {Breuer, Heinz-Peter and Petruccione, Francesco},
  title     = {The Theory of Open Quantum Systems},
  publisher = {Oxford University Press},
  year      = {2002}
}

@book{vanKampen2007,
  author    = {van Kampen, N. G.},
  title     = {Stochastic Processes in Physics and Chemistry},
  edition   = {3},
  publisher = {North-Holland},
  year      = {2007}
}

@article{Lindblad1976,
  author  = {Lindblad, G{\"o}ran},
  title   = {On the generators of quantum dynamical semigroups},
  journal = {Communications in Mathematical Physics},
  volume  = {48},
  pages   = {119--130},
  year    = {1976}
}

@article{Willwater2022,
  title = {Muon spin rotation and relaxation study on ${\mathrm{Nb}}_{1\ensuremath{-}y}{\mathrm{Fe}}_{2+y}$},
  author = {Willwater, J. and Eppers, D. and Kimmel, T. and Sadrollahi, E. and Litterst, F. J. and Grosche, F. M. and Baines, C. and S\"ullow, S.},
  journal = {Phys. Rev. B},
  volume = {106},
  issue = {13},
  pages = {134408},
  numpages = {12},
  year = {2022},
  month = {Oct},
  publisher = {American Physical Society},
  doi = {10.1103/PhysRevB.106.134408},
  url = {https://link.aps.org/doi/10.1103/PhysRevB.106.134408}
}

@article{Radcliffe1971,
  author  = {Radcliffe, J. M.},
  title   = {Some properties of coherent spin states},
  journal = {J. Phys. A: Gen. Phys.},
  volume  = {4},
  pages   = {313--323},
  year    = {1971}
}

@article{Vilao2017,
  title = {Role of the transition state in muon implantation},
  author = {Vil\~{a}o, R. C. and Vieira, R. B. L. and Alberto, H. V. and Gil, J. M. and Weidinger, A.},
  journal = {Phys. Rev. B},
  volume = {96},
  issue = {19},
  pages = {195205},
  numpages = {7},
  year = {2017},
  month = {Nov},
  publisher = {American Physical Society},
  doi = {10.1103/PhysRevB.96.195205},
}

@article{Vilao2025,
	title={{Muonium reaction in MgO: A showcase for the final steps of ion implantation}},
	author={Rui C. Vil\~{a}o and Ali Roonkiani and Apostolos G. Marinopoulos and Helena V. Alberto and João M. Gil and Ricardo B. L. Vieira and Robert Scheuermann and Alois Weidinger},
	journal={SciPost Phys. Core},
	volume={8},
	pages={056},
	year={2025},
	publisher={SciPost},
	doi={10.21468/SciPostPhysCore.8.3.056},
	url={https://scipost.org/10.21468/SciPostPhysCore.8.3.056},
}

@article{Dehn2021,
  title = {Local Electronic Structure and Dynamics of Muon-Polaron Complexes in ${\mathrm{Fe}}_{2}{\mathrm{O}}_{3}$},
  author = {Dehn, M. H. and Shenton, J. K. and Arseneau, D. J. and MacFarlane, W. A. and Morris, G. D. and Maign\'e, A. and Spaldin, N. A. and Kiefl, R. F.},
  journal = {Phys. Rev. Lett.},
  volume = {126},
  issue = {3},
  pages = {037202},
  numpages = {6},
  year = {2021},
  month = {Jan},
  publisher = {American Physical Society},
  doi = {10.1103/PhysRevLett.126.037202},
}

@article{Dehn2020,
  title = {Observation of a Charge-Neutral Muon-Polaron Complex in Antiferromagnetic ${\mathrm{Cr}}_{2}{\mathrm{O}}_{3}$},
  author = {Dehn, M. H. and Shenton, J. K. and Holenstein, S. and Meier, Q. N. and Arseneau, D. J. and Cortie, D. L. and Hitti, B. and Fang, A. C. Y. and MacFarlane, W. A. and McFadden, R. M. L. and Morris, G. D. and Salman, Z. and Luetkens, H. and Spaldin, N. A. and Fechner, M. and Kiefl, R. F.},
  journal = {Phys. Rev. X},
  volume = {10},
  issue = {1},
  pages = {011036},
  numpages = {18},
  year = {2020},
  month = {Feb},
  publisher = {American Physical Society},
  doi = {10.1103/PhysRevX.10.011036},
}

\end{document}